\documentclass[lineno]{jfm}
\usepackage{graphicx}
\usepackage{newtxtext}
\usepackage{newtxmath}
\usepackage{natbib}
\usepackage{hyperref}
\usepackage[labelfont={small}, font={normal}]{subfig}
\usepackage{xcolor}
\usepackage[normalem]{ulem}
\usepackage{amsmath}
\newcommand{\stkout}[1]{\ifmmode\text{\sout{\ensuremath{#1}}}\else\sout{#1}\fi}

\hypersetup{
	colorlinks = true,
	urlcolor  = blue,
	citecolor = black,
}

\newcommand{\RomanNumeralCaps}[1]
\linenumbers

\title{Sedimentation and shear-induced dynamics of spheroids in fluids with spatial viscosity variations}
\author{Arjun Sharma\aff{1}
	\corresp{\email{asharm1@sandia.gov, as3833@cornell.edu}},
	Peter A. Bosler\aff{1},
	Rama Govindarajan\aff{2},
	\and Donald L. Koch\aff{3} }
\affiliation{\aff{1} Center for Computing Research, Sandia National Laboratories, Albuquerque, NM, 87185, USA
	\aff{2} International Centre for Theoretical Sciences, Tata Institute of Fundamental Research, Bengaluru 560089, India
	\aff{3} Robert Frederick Smith School of Chemical and Biomolecular Engineering, Cornell University, Ithaca, NY, 14853, USA}

\begin{document}
	\maketitle
	\begin{abstract}
		A generalized reciprocal theorem is used to relate the force and torque induced on  a particle in an inertia-less fluid {with small variation in viscosity} to integrals involving Stokes flow fields and the spatial dependence of viscosity. These resistivity expressions are analytically evaluated using spheroidal harmonics and then used to obtain the mobility of the spheroid during sedimentation, and in linear flows, of a fluid with linear viscosity stratification. The coupling between the rotational and translational motion induced by stratification rotates the spheroid's centerline, creating a variety of rotational and translational dynamics dependent upon the particle's aspect ratio, $\kappa$, and the component of the stratification unit vector in the gravity direction, $d_g$. Spheroids with $0.55\lessapprox\kappa\lessapprox2.0$ exhibit the largest variety of settling behaviors. Interestingly, this range covers most microplastics and typical microorganisms. One of the modes include a stable orientation dependent only on $\kappa$ and $d_g$, but independent of initial orientation, thus allowing for the potential control of settling angles and sedimentation rates. In a simple shear flow, cross-streamline migration occurs due to the stratification-induced force generated on the particle. Similarly, a particle no longer stays at the stagnation point of a uniaxial extensional flow. While fully analytical results are obtained for spheroids, numerical simulations provide a source of validation. These simulations also provide additional insights into the stratification-induced force- and torque-producing mechanisms through the stratification-induced stress, which is not accessed in the reciprocal theorem-based analytical calculations.
	\end{abstract}
	
	
	
	\section{Introduction}\label{sec:Intro}
	{Numerous natural and engineering scenarios involve the motion of dispersed biological (micro-organisms) or artificial particles in fluids with non-uniform viscosity. The variation in viscosity experienced by the particles can change their trajectory relative to that in a uniform viscosity scenario.} 
	For example, spatio-temporally varying temperature and dissolved nutrients in the ocean lead to viscosity stratification which may influence the sedimentation of dead organic matter or marine snow, and micro-plastics. The motion of these particles, that are responsible both for oceanic pollution and for the helpful absorption of 30\% of anthropogenic CO$_2$ \citep{gruber2019oceanic}, could also be affected by the interaction between background shear and variable viscosity. The viscosity of the mucus present in airways increases from the cilium layers to the air-mucus interface \citep{barton1967analytical}, likely playing a role in the movement of bacteria (\textit{Escherichia Coli}) and other pathogens through cellular surfaces. As fluid motion near a wall may be described as simple shear flow and that exiting from or entering a pore as an extensional flow, these particles experience locally linear flows. A variety of particle shapes ranging from spheres and spheroids to irregularly shaped micro-plastics are encountered in the aforementioned instances. Irregular shapes are most often approximated as spheroids, which enables analytical treatment {for accessing a range of shapes by changing the aspect ratio of the particle. Spheroids have been previously used to explain experiments on micro-organisms such as \textit{Paramecium caudatum} \citep{keller1977porous}, \textit{Escherichia Coli} \citep{bai2006dielectric} and a system of sedimenting phytoplankton and secreted mucus \citep{chajwa2024hidden}.} Therefore, in this paper, we study freely suspended spheroids in fluids with viscosity gradient, under gravity and in linear flow, and elucidate previously unexplored mechanisms of rotation and translation. This is achieved through analytical calculations of stratification-induced forces and torques, supplemented by insights from pressure and viscous stresses obtained via numerical simulations. These findings can be valuable in designing desirable intervention strategies for a variety of applications such as ocean farming to enhance long term carbon sequestration provided by marine snow \citep{jones2022climate}, devising effective micro-plastic extraction strategies \citep{van2015microplastics}, targeted drug delivery \citep{xie2020bacteria} and understanding the role of viscosity stratification in pathogenicity of bacteria.
	
	In the absence of fluid and particle inertia, and in fluid of constant physical properties, a fore-aft and axisymmetric particle does not experience a force when placed at the center of a linear flow, or a torque when fixed in a uniform flow \citep{kim2013microhydrodynamics}. As a result, when such a particle is freely suspended, in linear flows it does not translate relative to the local imposed fluid velocity and upon free sedimentation it maintains its initial orientation while settling at an orientation dependent velocity. In simple shear flow, the axis of symmetry of the spheroid undergoes non-uniform periodic trajectories known as Jeffery orbits \citep{jeffery1922motion}, and the exact orbit chosen is dependent on initial conditions. As indicated by numerous theoretical, computational and experimental studies, mechanisms such as fluid inertia \citep{subramanian2005inertial,dabade2016effect,di2024influence}, viscoelasticity \citep{gauthier1971particle,leal1975slow,bartram1975particle,harlen1993simple,iso1996orientation,iso1996orientationb,gunes2008flow,d2014bistability,dabade2015effects,d2022numerical,sharma2023rotation} or stratification in the fluid's density (see review by \cite{more2023motion}) may break this strong dependence on initial conditions by inducing an additional force and/ or torque.
	
Viscosity variation in the underlying fluid may provide another mechanism to break the aforementioned symmetry in particle motion. {Even though the mechanisms that result in viscosity variation of a liquid are long studied, its effect on the particle motion is a relatively new topic and previous studies have focused either on sedimenting particles or those fixed in uniform flow. The viscosity may be stratified due to several effects such as temperature \citep{seeton2006viscosity} and salt concentration \citep{jiang2003new}. Increasing temperature lowers the viscosity of liquids by creating increased agitation of molecules resulting in smaller coherent groups that exhibit lower resistance \citep{batchelor2000introduction}. Solute concentration in an electrolyte alters viscosity as an additional restoring force is required to overcome interionic attraction and thermal movements in a sheared cloud of ions \citep{stokes1965viscosity}. Furthermore, due to ion solvent interaction the dependence of viscosity on solute concentration may change with temperature from an increasing function with concentration to a decreasing function \citep{stokes1965viscosity,kaminsky1957ion}. Phytoplankton alter the viscosity of their sorroundings by as much as 40 times that of seawater \citep{guadayol2021microrheology} by secreting mucus clouds that take varying anisotropic shapes around the organism \citep{chajwa2024hidden}. The presence of artificial particles can also create viscosity disturbances in the surrounding fluid for example by changing the surrounding temperature. This may then influence the particles' motion relative to that in a uniform viscosity scenario.} 
	
\cite{oppenheimer2016motion} study the forces and torques on a hot sphere fixed in a uniform flow of a fluid with temperature-sensitive viscosity. Their theory is restricted to small P\'eclet number where temperature is transported primarily through diffusion, creating variations in fluid viscosity near the particle surface. When the sphere is uniformly heated, viscosity variation created due to the monopole thermal moment lowers the hydrodynamic drag on the particle. This could be an explanation for increased diffusivity of heated gold nanoparticles in water, as observed experimentally by \cite{rings2010hot}. As another case, \cite{oppenheimer2016motion} considered one hemisphere of the sphere to be maintained at a different temperature from the other, and found that the asymmetric viscosity distribution created by the dipole thermal moments leads to a torque on the sphere. {\cite{ziegler2022hydrodynamic} considered two spherical particles which perturb the viscosity due to their temperature (also for small P\'eclet number). Similar to \cite{oppenheimer2016motion} they find the self mobility of a hot particle to increase and that for a cold particle to decrease. However, the additional cooling or heating by the neighboring particle alters this mobility in a manner that scales inversely with the separation between particles, $s$. Furthermore, unlike the single particle case, uniformly heated spheres lead to a coupling between their translational and rotational motion at $\mathcal{O}(s^{-2})$.}

Using a regular perturbation expansion and the reciprocal theorem, \cite{datt2019active} found that a sphere of radius $l$ moving at a velocity $\mathbf{u}$ relative to a fluid with small ambient viscosity gradient $\nabla \eta$, experiences a torque $2\pi l^3\mathbf{u}\times\nabla \eta$. Through a combination of this technique and numerical integration, \cite{anand2024sedimentation} performed these calculations for a spheroidal particle. They found that the torque induced due to small linear viscosity gradients leads to a change in the spheroid's orientation, which alters its translation path and speed. However, their semi-analytical results are restricted to $\kappa<0.5$ and $\kappa>2$, where $\kappa$ is the particle's aspect ratio. {\cite{gong2024active} have extended this analysis to include fully analytical expressions for not just passive but active prolate spheroids. They find the effect of viscosity gradients on the reorientation of a swimmer towards lower viscosity is reduced as it becomes more slender.} In the limit of large aspect ratio, \cite{kamal2023resistive} formulate a resistive force theory to study the effect of small viscosity gradients on a slender fiber and a thin ring (small cross sectional length relative to its circumference) fixed in uniform flow and in rotational and uniaxial extensional flow. On these slender filaments, viscosity stratification leads to a torque in uniform flow and force in rotational and extensional flows. The latter two cases are the only results available for the effect of viscosity stratification on particles in linear flows.

While micro-plastics found in marine and freshwater environments have a wide range of length to width ratios, most of the micro-plastic population is within the $0.5<\kappa <2$ range \citep{kooi2021characterizing}. Furthermore, certain micro organisms such as \textit{Paramecium} have an aspect ratio $\kappa\approx2$ \citep{kreutz2012morphological} and \textit{Escherichia Coli} have $2\lessapprox \kappa \lessapprox 4$ \citep{kaya2009characterization,liu2014real}. Therefore, the effect of viscosity stratification in the entire $\kappa$ range, which this paper provides, is beneficial. In section \ref{sec:SpheroidFixUniform}, we revisit the sedimentation of spheroids in linear viscosity stratification. Our calculations of stratification-induced torque on a sedimenting particle are entirely analytical. From this, we obtain richer orientation dynamics behavior within the $0.55\lessapprox \kappa \lessapprox 2.0$ regime than outside. Five types of orientation trajectories are discovered that can be represented by different regions in the $\kappa-d_g$ phase space, where $d_g$ is the projection of the viscosity gradient unit vector along gravity. The force due to viscosity gradients in linear flows valid for the entire $\kappa$ range is also obtained analytically. Cross-stream migration due to stratification in simple shear flow (section \ref{sec:LinearFlows}), may inspire novel particle sorting strategies in microfluidics applications. {In our study, viscosity variation arises solely from the prescribed ambient conditions and is not influenced by the presence of the particle. An example of such a scenario is when a fluid is subjected to an ambient temperature variation, the particle is small or the fluid thermal conductivity is large (small P\'eclet number), and the particle has the same thermal conductivity as the fluid. Otherwise, one must account for the transport of the scalar (such as temperature or solute concentration) and appropriate boundary conditions at the particle surface and the external boundaries.}

	The rest of the paper is organised as follows. Section \ref{sec:Formulation} describes the governing equations, expressions for stratification-induced force and torque for a fluid with small viscosity variation, and other mathematical details. Beyond section \ref{sec:Formulation} we focus on a fluid with temporally constant, and spatially linear, viscosity stratification. We consider fore-aft and axisymmetric particles, a class that includes biconcave discs such as red blood cells, bispherical objects, dumbbells, rings, spheroids, etc. in section \ref{sec:GeneralParticle} and provide the vector equations governing the rotation and translation of freely suspended neutrally buoyant particles in linear flows and freely settling. These equations require specific stratification-induced force and torque on a fixed particle as inputs, which may be determined either analytically (as done here for spheroids) or from specific numerical simulations (that may be done for other particle shapes in future). In section \ref{sec:SpheroidResultsFix}, these forces and torques on a spheroid (obtained analytically using the spheroidal harmonics formulation of \cite{dabade2015effects,dabade2016effect}) are analyzed in detail with the underlying mechanisms elucidated. Distinct modes of a spheroid's motion as a result of the viscosity stratification while sedimenting in quiescent fluid and freely suspended in linear flows are described in sections \ref{sec:SpheroidsSedimenting} and \ref{sec:SpheroidsLinear} respectively. Finally, conclusions and suggestions for future investigations are given in section \ref{sec:Conclusions}.
	
\section{Mathematical formulation and different torque/ force generating mechanisms}\label{sec:Formulation}
Consider a particle in an unbounded flow of a variable-viscosity fluid in the absence of particle and fluid inertia. The equations governing fluid mass and momentum conservation are,
\begin{equation}
	\nabla \cdot\mathbf{u}=0,\hspace{0.1in}\nabla\cdot \boldsymbol{\sigma}=0,\label{eq:MassandMomentum}
\end{equation}
where, $\mathbf{u}$ is the fluid velocity and,
\begin{equation}
	\boldsymbol{\sigma}=-p\mathbf{I}+2\eta\mathbf{e}=-p\mathbf{I}+\eta(\nabla\mathbf{u}+\nabla {\mathbf{u}}^T),
\end{equation}
is the fluid stress with $p$ and $\eta$ being the fluid pressure and viscosity, $\mathbf{I}$ the identity tensor, and $\mathbf{e}$ the strain rate. Here, spatio-temporal variations in fluid viscosity, $\eta(\mathbf{x},t)$, are allowed. The boundary conditions are no-slip on the particle surface and a prescribed flow field (uniform flow, shear flow, etc.) at the outer boundary,
\begin{equation}
	\mathbf{u}=\mathbf{u}_\text{particle}, \text{ on the particle surface, and }\mathbf{u}\rightarrow\mathbf{u}_\infty\text{ as }\mathbf{x}\rightarrow\mathbf{x}_\text{out}.\label{eq:MassandMomentumBC}
\end{equation}
The outer boundary, $\mathbf{x}_\text{out}$, may be in the far-field, at a solid wall or a periodic boundary (in case $\mathbf{x}_\text{out}$ is at a finite distance, ``$\rightarrow$'' in the above equation is replaced with an ``$=$''). The net hydrodynamic force, $\mathbf{f}$, and torque, $\mathbf{q}$, on the particle experiencing a fluid stress $\boldsymbol{\sigma}$ on its surface are,
\begin{equation}
	\mathbf{f}(\boldsymbol{\sigma})= \int_{r_p}\text{dS}\hspace{0.05in}(\mathbf{n}\cdot\boldsymbol{\sigma}), \hspace{0.2in}	\mathbf{q}(\boldsymbol{\sigma})= \int_{r_p}\text{dS}\hspace{0.05in}\mathbf{x}\times(\mathbf{n}\cdot\boldsymbol{\sigma}).\label{eq:ForceToruqeDef}
\end{equation}
We assume that, without additional approximation, viscosity
{\begin{equation}
		\eta(\mathbf{x},t)=\eta_0(t)+\beta\eta'(\mathbf{x},t),
\end{equation}}can be decomposed into a spatially constant, $\eta_0(t)$ (the fluid viscosity at the particle's center) and spatially dependent, ${\beta}\eta'(\mathbf{x},t)$, components. {Here, the variable part of viscosity $\beta \eta'$ is assume to be much smaller than the spatially constant part, $\eta_0(t)$ with $\beta\ll 1$ being a perturbation parameter.} Typically one would require numerical discretization of the governing equations to obtain the fluid stress, $\boldsymbol{\sigma}$, at the particle surface prior to evaluating $\mathbf{f}$, and $\mathbf{q}$. However, as outlined in appendix \ref{sec:Reciprocal}, {when $\beta \ll 1$,} using a generalized reciprocal theorem {following regular perturbations of the relevant flow variables in $\beta$,} the $\mathbf{f}$, and $\mathbf{q}$ acting on a particle at time $t$, placed at the origin, in a fluid with small viscosity variation is,
\begin{align}
	\begin{split}
		&\mathbf{f}(t)=	\eta_0(t)\mathbf{f}^\text{Stokes}(t)-2{\beta}\int_\text{Fluid} \mathbf{dx}\hspace{0.1in}\eta'(\mathbf{x},t)(\mathbf{e}^\text{Stokes}(\mathbf{x};t)-\mathbf{E}_\infty(t)):\nabla\mathbf{b}^\mathbf{f}(\mathbf{x}){+\mathcal{O}(\beta^2)},\\
		&\mathbf{q}(t)=	\eta_0(t)\mathbf{q}^\text{Stokes}(t)-2{\beta}\int_\text{Fluid} \mathbf{dx}\hspace{0.1in}\eta'(\mathbf{x},t)(\mathbf{e}^\text{Stokes}(\mathbf{x};t)-\mathbf{E}_\infty(t)):\nabla\mathbf{b}^\mathbf{q}(\mathbf{x}){+\mathcal{O}(\beta^2)}.\label{eq:StratificationForceTorque}
	\end{split}
\end{align}
Here, $\eta_0\mathbf{f}^\text{Stokes}(t)=\int_{r_p}\text{dS}\hspace{0.05in}(\mathbf{n}\cdot\boldsymbol{\sigma}^\text{Stokes})$ and $\eta_0\mathbf{q}^\text{Stokes}(t)=\int_{r_p}\text{dS}\hspace{0.05in}\mathbf{x}\times(\mathbf{n}\cdot\boldsymbol{\sigma}^\text{Stokes})$ are the current force and torque on the particle in the same computational domain as the complete problem but with a spatially constant viscosity $\eta_0(t)$ (with the Stokes stress, $\boldsymbol{\sigma}^\text{Stokes}=-p^\text{Stokes}+2\eta_0\mathbf{e}^\text{Stokes}$), while the volume integral terms (with $2\mathbf{E}_\infty(t)=\nabla\mathbf{u}_\infty+(\nabla\mathbf{u}_\infty)^T$ denoting the imposed strain rate) capture the entire contribution induced by viscosity variation. The two 2-tensor fields $\mathbf{b}^\mathbf{f}(\mathbf{x})$ and $\mathbf{b}^\mathbf{q}(\mathbf{x})$ are related to the auxiliary problem in the reciprocal theorem and are obtained from the solution of a Stokes problem around the particle. A vector $\mathbf{b}^\mathbf{f}\cdot\hat{\mathbf{u}}$ is the Stokes velocity field around the particle translating with $\hat{\mathbf{u}}$ in a quiescent fluid. Similarly, $\mathbf{b}^\mathbf{q}\cdot\hat{\boldsymbol{\omega}}$ corresponds to the velocity disturbance created by a particle rotating with angular velocity $-\boldsymbol{\omega}$ in a quiescent inertia-less fluid. The tensor fields $\mathbf{b}^\mathbf{f}$ and $\mathbf{b}^\mathbf{q}$ approach zero at the far-field boundaries and are equal to zero the no-slip wall surfaces. The expressions for the hydrodynamic force and torque in equation \eqref{eq:StratificationForceTorque} are valid for a general particle shape present in an inertialess fluid with small viscosity variations that may be unbounded or contained within solid and periodic boundaries. If not held fixed by an external force or torque, the particle translates and rotates to ensure that the force- and torque-free constraints are satisfied. These expressions are utilized to reveal novel particle dynamics in viscosity-stratified fluids in the next four sections of this paper. In particular, we note that in equation \eqref{eq:StratificationForceTorque}, {when the spatial variability in viscosity is small, i.e., $\beta\ll1$,} the force and torque on a particle in a viscosity stratified fluid require knowledge of only the Stokes velocity field and spatial distribution of viscosity; we will use this fact in section \ref{sec:SpheroidResultsFix} with spheroidal harmonics to analytically determine these forces and torques on spheroids fixed in uniform and linear flows. Section \ref{sec:MoreMathDetails} provides mathematical details required to isolate different mechanistic constituents of the extra force and torque due to viscosity variation.

\subsection{Mechanistic origins of stratification-induced force and torque}\label{sec:MoreMathDetails}
The derivation of the expressions in equation \eqref{eq:StratificationForceTorque} is in appendix \ref{sec:Reciprocal} and relies on an exact decomposition of the velocity and pressure fields into two components such that
\begin{equation}
	\mathbf{u}=\mathbf{u}^\text{Stokes}+\mathbf{u}^\text{Stratified},\text{ and }
	p=(1+{\beta}\eta'/\eta_0)p^\text{Stokes}+p^\text{Stratified}.
\end{equation}
The variables $\mathbf{u}^\text{Stokes}$ and $p^\text{Stokes}$ represent the velocity and pressure in a uniform viscosity fluid (Stokes flow) with viscosity equal to that at the particle's center, whereas $\mathbf{u}^\text{Stratified}$ and $p^\text{Stokes}{\beta}\eta'/\eta_0+p^\text{Stratified}$ are velocity and pressure induced by viscosity variation. The pressure component $p^\text{Stokes}{\beta}\eta'/\eta_0$ is the Stokes pressure at any position incremented by the local viscosity. The fluid stress, $\boldsymbol{\sigma}$, consists of three parts,
\begin{equation}
	\boldsymbol{\sigma}=\boldsymbol{\sigma}^\text{Stokes}+({\beta}\eta'/\eta_0)\boldsymbol{\sigma}^\text{Stokes}+\boldsymbol{\sigma}^\text{Stratified},\label{eq:TotalStressDecomposed}
\end{equation}
with its constituent stresses,
\begin{eqnarray}
	\boldsymbol{\sigma}^\text{Stokes}=-p^\text{Stokes}\mathbf{I}+\eta_0(\nabla\mathbf{u}^\text{Stokes}+(\nabla\mathbf{u}^\text{Stokes})^T),\text{ and,}\\
	\boldsymbol{\sigma}^\text{Stratified}=-p^\text{Stratified}\mathbf{I}+(\eta_0+{\beta}\eta')[\nabla\mathbf{u}^\text{Stratified}+(\nabla\mathbf{u}^\text{Stratified})^T].
\end{eqnarray}

After applying the decomposition, the governing equations are,
\begin{eqnarray}
	\nabla\cdot \mathbf{u}^\text{Stokes}=0, \hspace{0.2in}\nabla\cdot\boldsymbol{\sigma}^\text{Stokes}=0,\label{eq:StokesProblem}\\
	\nabla\cdot \mathbf{u}^\text{Stratified}=0, \hspace{0.2in}\nabla\cdot\boldsymbol{\sigma}^\text{Stratified}+\frac{\nabla\eta}{\eta_0}.\boldsymbol{\sigma}^\text{Stokes}=0,\label{eq:StratificationProblem}
\end{eqnarray}
subject to the boundary conditions
\begin{eqnarray}
	\mathbf{u}^\text{Stokes}=\mathbf{u}_\text{particle}, \text{on the particle surface, and, }\mathbf{u}^\text{Stokes}\rightarrow\mathbf{u}_\infty\text{ as }\mathbf{x}\rightarrow\mathbf{x}_\text{out},\text{ and,}\label{eq:StokesProblemBC}\\
	\mathbf{u}^\text{Stratified}=0, \text{on the particle surface, and, as }\mathbf{x}\rightarrow\mathbf{x}_\text{out},\label{eq:StratificationProblemBC}
\end{eqnarray}
where as in equation \eqref{eq:MassandMomentumBC}, the outer boundary, $\mathbf{x}_\text{out}$, may be in the far-field, at a solid wall or a periodic boundary.
Equations \eqref{eq:StokesProblem} and \eqref{eq:StokesProblemBC} governing the evolution of $\boldsymbol{\sigma}^\text{Stokes}$ are the same as the original equations \eqref{eq:MassandMomentum} and \eqref{eq:MassandMomentumBC} but in a fluid with uniform viscosity, $\eta_0(t)$. The velocity induced by the variable viscosity effect, $\mathbf{u}^\text{Stratified}$, has zero boundary conditions, equation \eqref{eq:StratificationProblemBC}, because the imposed flow and particle motion are already accounted for in the boundary conditions for Stokes velocity, $\mathbf{u}^\text{Stokes}$ in equation \eqref{eq:StokesProblemBC}. Summing equations \eqref{eq:StokesProblem} and \eqref{eq:StratificationProblem} recovers the original formulation in equation \eqref{eq:MassandMomentum}. We have not made any assumption in decomposing the original system (equation \eqref{eq:MassandMomentum} to \eqref{eq:MassandMomentumBC}) into the two components given by equations \eqref{eq:StokesProblem} to \eqref{eq:StratificationProblemBC}. This exact decomposition is possible because the original mass and momentum equations are linear in velocity and pressure. Periodic boundary conditions (not in equations \eqref{eq:StokesProblemBC} and \eqref{eq:StratificationProblemBC}) are also compatible with this decomposition.

To support the discussion and analysis of section \ref{sec:SpheroidResultsFix}, it is useful to describe the additional stratification-induced force and torque on a particle that are not present in classical Stokes flow. From equation \eqref{eq:TotalStressDecomposed} the extra stress in a variable viscosity fluid is $({\beta}\eta'(\mathbf{x})/\eta_0)\boldsymbol{\sigma}^\text{Stokes}+\boldsymbol{\sigma}^\text{Stratified}$ which leads to the {$\mathcal{O}(\beta)$} force or torque $-2{\beta}\int_\text{Fluid} \text{dV}\hspace{0.1in}\eta'(\mathbf{e}^\text{Stokes}-\mathbf{E}_\infty):\nabla\mathbf{b}$ (with $\mathbf{b}=\mathbf{b}^\mathbf{f}$ or $\mathbf{b}^\mathbf{q}$). The net force and torque, shown in equation \eqref{eq:StratificationForceTorque}, acting on the particle can be represented as, $\mathbf{f}=\eta_0\mathbf{f}^\text{Stokes}+\mathbf{f}^\text{Stratified}_\text{A}+\mathbf{f}^\text{Stratified}_\text{B}$ and $\mathbf{q}=\eta_0\mathbf{q}^\text{Stokes}+\mathbf{q}^\text{Stratified}_\text{A}+\mathbf{q}^\text{Stratified}_\text{B}$. Here, part of the force and torque arise from the Stokes stress acting on the particle surface immersed in varying viscosity fluid,
\begin{equation}
	\mathbf{f}^\text{Stratified}_\text{A}={\beta}\int_{r_p}\text{dS}\hspace{0.05in}((\eta'/\eta_0)\mathbf{n}\cdot\boldsymbol{\sigma}^\text{Stokes}), \qquad \mathbf{q}^\text{Stratified}_\text{A}={\beta}\int_{r_p}\text{dS}\hspace{0.05in}(\eta'/\eta_0)\mathbf{x}\times(\mathbf{n}\cdot\boldsymbol{\sigma}^\text{Stokes}),\label{eq:ForceTorqueA}
\end{equation}
and the final contributions come from the velocity and pressure induced by variable viscosity,
\begin{eqnarray}\begin{split}
		\mathbf{f}^\text{Stratified}_\text{B}=\int_{r_p}\text{dS}\hspace{0.05in}(\mathbf{n}\cdot\boldsymbol{\sigma}^\text{Stratified})= -2\beta\int_\text{Fluid} \text{dV}\hspace{0.1in}\eta'(\mathbf{e}^\text{Stokes}-\mathbf{E}_\infty):\nabla\mathbf{b}^\mathbf{f}-\mathbf{f}^\text{Stratified}_\text{A}{+\mathcal{O}(\beta^2)}, \\ \mathbf{q}^\text{Stratified}_\text{B}=\int_{r_p}\text{dS}\hspace{0.05in}\mathbf{x}\times(\mathbf{n}\cdot\boldsymbol{\sigma}^\text{Stratified})= -2\beta\int_\text{Fluid} \text{dV}\hspace{0.1in}\eta'(\mathbf{e}^\text{Stokes}-\mathbf{E}_\infty):\nabla\mathbf{b}^\mathbf{q}- \mathbf{f}^\text{Stratified}_\text{A}{+\mathcal{O}(\beta^2)}.\label{eq:ForceTorqueB}
	\end{split}
\end{eqnarray}
{The force and torque, $\mathbf{f}^\text{Stratified}_\text{B}$ and $\mathbf{q}^\text{Stratified}_\text{B}$, are due to the $\mathcal{O}(\beta)$ stratified stress governed by the $\mathcal{O}(\beta)$ stratified mass and momentum equations,
	\begin{equation}
		\nabla\cdot ({\mathbf{u}}^\text{Stratified})^{(1)}=0, \hspace{0.2in}\nabla\cdot({\boldsymbol{\sigma}}^\text{Stratified})^{(1)}+\frac{\mathbf{d}}{\eta_0}\cdot\boldsymbol{\sigma}^\text{Stokes}=0.\label{eq:StratificationProblemMod3}
	\end{equation}
	Here in \eqref{eq:StratificationProblemMod3} we have regularly expanded the flow variables in $\beta$,
	\begin{eqnarray}
		&{\mathbf{u}}^\text{Stratified}=\beta({\mathbf{u}}^\text{Stratified})^{(1)}+\mathcal{O}(\beta^2),\\
		&{p}^\text{Stratified}=\beta({p}^\text{Stratified})^{(1)}+\mathcal{O}(\beta^2),\\
		&\begin{split}{\boldsymbol{\sigma}}^\text{Stratified}=& \beta\big[-(\widetilde{p}^\text{Stratified})^{(1)}\mathbf{I}+\eta_0[(\nabla{\mathbf{u}}^\text{Stratified})^{(1)}+((\nabla{\mathbf{u}}^\text{Stratified})^{(1)})^T]\big]\\&+\mathcal{O}(\beta^2).\end{split}
\end{eqnarray}}

From here on, we focus on linear viscosity stratification such that the viscosity, $\eta$, distribution is,
\begin{equation}
	\eta=\eta_0+\beta\mathbf{d}\cdot\mathbf{x},\hspace{0.1in} ||\mathbf{d}||_2=1,\hspace{0.1in} ||\nabla \eta||_2=\beta\hspace{0.1in} (\eta'(\mathbf{x},t)=\eta'(\mathbf{x})=\beta\mathbf{d}\cdot\mathbf{x}), \label{eq:LinearViscosity}
\end{equation}	
where unit vector $\mathbf{d}$ lies along the direction of viscosity stratification. For this case the parameter, $\beta$ in the regular perturbation expansion of the relevant flow variables is the magnitude of viscosity gradient.
	
	\section{Fore-aft and axisymmetric particle}\label{sec:GeneralParticle}
	This section first illustrates the structure of the stratification-induced force and torque on a fore-aft and axisymmetric particle fixed in a linear and uniform flow of a stratified fluid with small magnitude of viscosity gradient. These forces and torques, whose structure is derived in the first two sub-sections below are labeled in table \ref{tab:NewtonianCoeffsStratifiedForces}. Their explicit values may be evaluated either analytically, as demonstrated for spheroids in section \ref{sec:SpheroidResultsFix}, or numerically for more complex particle shapes such as cylinders, biconcave disks or dumbbells (that may be considered in future studies). Then the dynamics of such particles freely suspended in a variety of flow scenarios can be obtained from the upcoming ordinary differential equations (odes) defined in equations \eqref{eq:rotationratefull}, and \eqref{eq:StratificationVelocity} using the aforementioned forces and torques as input parameters. Even if only numerical evaluation of forces and torques is possible, obtaining particle motion through the solution of these odes will be computationally more efficient than alternative direct numerical simulations requiring numerical solution of the governing equations at each time step along the particle trajectory.
	
	A uniform or linear incompressible imposed flow, $\mathbf{u}_\infty$, can be obtained via a linear superposition of the 11 canonical flows in the reference frame aligned with the particle,
	\begin{eqnarray}\begin{split}
			&\mathbf{u}_\infty^\text{body}=\Sigma_{i=1}^3 a_i \hat{\mathbf{u}}^{(i)}+\mathbf{x}^\text{body}\cdot\Sigma_{i=1}^8 b_i\hat{\boldsymbol{\Gamma}}^{(i)},\\
			&\hat{\mathbf{u}}^{(1)}=\begin{bmatrix}1&0&0 \end{bmatrix},\hat{\mathbf{u}}^{(2)}=\begin{bmatrix}0&1&0 \end{bmatrix},\hat{\mathbf{u}}^{(3)}=\begin{bmatrix}0&0&1 \end{bmatrix},                  \\
			&\hat{\boldsymbol{\Gamma}}^{(1)}=\begin{bmatrix}1&0&0\\0&-1&0\\0&0&0 \end{bmatrix},
			\hat{\boldsymbol{\Gamma}}^{(2)}=\begin{bmatrix}-\frac{1}{2}&0&0\\0&-\frac{1}{2}&0\\0&0&1\end{bmatrix},
			\hat{\boldsymbol{\Gamma}}^{(3)}=\begin{bmatrix}0&1&0\\1&0&0\\0&0&0\end{bmatrix},
			\hat{\boldsymbol{\Gamma}}^{(4)}=\begin{bmatrix}0&0&1\\0&0&0\\1&0&0\end{bmatrix}, \\&     \hat{\boldsymbol{\Gamma}}^{(5)}=\begin{bmatrix}0&0&0\\0&0&1\\0&1&0\end{bmatrix},
			\hat{\boldsymbol{\Gamma}}^{(6)}=\begin{bmatrix}0&1&0\\-1&0&0\\0&0&0\end{bmatrix},
			\hat{\boldsymbol{\Gamma}}^{(7)}=\begin{bmatrix}0&0&-1\\0&0&0\\1&0&0\end{bmatrix},
			\hat{\boldsymbol{\Gamma}}^{(8)}=\begin{bmatrix}0&0&0\\0&0&1\\0&-1&0\end{bmatrix},
			\label{eq:BasisFlows}
	\end{split}\end{eqnarray}
	with appropriate values of the weights $a_i$ and $b_i$ and unit vectors, $\hat{\mathbf{u}}^{(i)}, i\in[1,3]$ and constant tensors, $\hat{\boldsymbol{\Gamma}}^{(i)}, i\in[1,8]$. Here, $\mathbf{x}^\text{body}$ is the position vector of an arbitrary point in the particle reference frame. In constant viscosity Stokes flow, the effect of a particle in any linear or uniform flow can be obtained via a linear superposition of these canonical cases. Furthermore, the $\mathcal{O}(\beta)$ (equation \eqref{eq:StratificationProblemMod2}), we observe that the stratification-induced flow is linear in the viscosity gradient, $\nabla \eta=\mathbf{d}\beta$. Hence, calculating the stratification-induced force and torque for three perpendicular viscosity gradients for each of the eleven canonical flows in equation \eqref{eq:BasisFlows} is sufficient to obtain the relevant values for an arbitrarily oriented $\mathbf{d}$ with any combination of uniform and linear imposed flow.
	
	By acknowledging the symmetry of the particle shape relative to each of the flows given in equation \eqref{eq:BasisFlows}, we observe the symmetry within the components of forces and torques acting on a fore-aft and axisymmetric particle in a constant as well as in linearly stratified viscosity fluid. For example, in a reference frame with direction 3 aligned with the fixed particle's axis of symmetry, a uniform flow of constant viscosity with velocity along direction 3, $a_3\hat{\mathbf{u}}^{(3)}$, creates a hydrodynamic force only along direction 3. The force along direction 1 and 2 is zero. There is a symmetry in the 1 and 2 directions due to axisymmetry, e.g., the force in direction 2 for flow $\hat{\mathbf{u}}^{(2)}$ is the same as that in direction 1 for $\hat{\mathbf{u}}^{(1)}$ in the particle-aligned coordinate system. The torque in the case of uniform flow is zero. Similarly, in the case of linear flows, the particle experiences a torque but no force.
	
	In the context of uniform flow relative to a particle, the hydrodynamic force and torque surface integrands shown in equation \eqref{eq:ForceToruqeDef} are even and odd in the position vector, $\mathbf{x}$, respectively. In contrast, for linear flows the force integrand is odd and the torque integrand is even. The surface integral of an integrand that is odd in $\mathbf{x}$ is zero around an axi and fore-aft symmetric particle. For a particular flow type, the stress arising from linear stratification, $(\eta'/\eta_0)\boldsymbol{\sigma}^\text{Stokes}+\boldsymbol{\sigma}^\text{Stratified}$ is odd in $\mathbf{x}$ if the stress for uniform viscosity fluid, $\boldsymbol{\sigma}^\text{Stokes}$ is even and vice versa. Therefore, due to linear stratification, an extra force (and no torque) is produced for linear flows, and an extra torque (and no force) in uniform flow. Furthermore, {the $\mathcal{O}(\beta)$ }stratification-induced forces and torques are linear in $\mathbf{d}$, the direction along which viscosity is stratified. Based on such symmetry arguments, the force and torque generated in the eleven flows listed in equation \eqref{eq:BasisFlows} in a uniform viscosity fluid are provided in the second and third columns of table \ref{tab:NewtonianCoeffsStratifiedForces} and that in a viscosity-stratified fluid are the dot product of the last two columns of this table with $\mathbf{d}^\text{body}$, i.e., the unit vector in the particle aligned frame along which viscosity increases. While we only consider linearly stratified fluid in the rest of this paper, a similar argument allows one to note that spatially quadratic variation in viscosity may lead to an additional stratification-induced force in uniform flows and torque in linear flows.
	
	The orientation of a fore-aft and axisymmetric particle can be described by a single vector: the direction of its axis of symmetry, $\mathbf{p}$. From a body-fixed coordinate system where the axis of symmetry is aligned with the 3 axis, a vector can be transformed to another reference frame using the rotation matrix,
	\begin{equation}
		\mathbf{R}=\begin{bmatrix}
			\frac{p_2}{\sqrt{p_1^2+p_2^2}} & \frac{p_1p_3}{\sqrt{p_1^2+p_2^2}} & p_1\\
			-\frac{p_1}{\sqrt{p_1^2+p_2^2}} & \frac{p_2p_3}{\sqrt{p_1^2+p_2^2}} & p_2\\
			0											 				& -\sqrt{p_1^2+p_2^2} & p_3
		\end{bmatrix},
	\end{equation}
	where the particle's orientation $\mathbf{p}=\begin{bmatrix} p_1&p_2&p_3\end{bmatrix}$ in the chosen reference frame. In section \ref{sec:Unstratified} we will use the forces and torques from table \ref{tab:NewtonianCoeffsStratifiedForces} to obtain the particle dynamics in a constant viscosity fluid. Then in sections \ref{sec:AbstractStratification} and \ref{sec:AbstractStratificationMoving} we will consider the force, torque and change in particle dynamics due to viscosity stratification.
	
	\begin{table}
		\begin{center}
			\caption{Forces and torques (in Cartesian basis) on an axi- and fore-aft symmetric particle in body-fixed coordinates (where component 3 or the $z$ axis aligns with the particle orientation) for different flows within equation \eqref{eq:BasisFlows} for a linearly stratified fluid with viscosity given by equation \eqref{eq:LinearViscosity}, are $\eta_0\mathbf{f}^\text{Stokes}+ \beta\mathbf{F}_\text{strat}^\text{body}\cdot\mathbf{d}^\text{body}{+\mathcal{O}(\beta^2)}$ and $\eta_0\mathbf{q}^\text{Stokes}+\beta\mathbf{Q}_\text{strat}^\text{body}\cdot\mathbf{d}^\text{body}{+\mathcal{O}(\beta^2)}$. The first part arises from fluid's uniform viscosity, $\eta_0$ and the second from its viscosity gradient with magnitude $\beta=||\nabla \eta||_2$. The unit of $f_i$ is $l$ and that of $q_i$, $F_{ij}^{\Gamma_k}$ and $Q_{ij}^{U_k}$ is $l^3$, where $l$ is the chosen length scale.
				\label{tab:NewtonianCoeffsStratifiedForces}}
			\begin{tabular}{ c c c c c}
				\hline
				Imposed flow & $\mathbf{f}^\text{Stokes}$ & $\mathbf{q}^\text{Stokes}$& $\mathbf{F}_\text{strat}^\text{body}$ & $\mathbf{Q}_\text{strat}^\text{body}$ \\
				\hline \hline
				$\hat{\mathbf{u}}^{(1)}$ & $f_1\hat{\mathbf{u}}^{(1)}$
				& $\mathbf{0}$& $\mathbf{0}$&$\begin{matrix}\frac{Q_{23}^{U_1}+Q_{32}^{U_1}}{2}\hat{\boldsymbol{\Gamma}}^{(5)}\\+\frac{Q_{23}^{U_1}-Q_{32}^{U_1}}{2}\hat{\boldsymbol{\Gamma}}^{(8)}\end{matrix}$\\
				\hline
				$\hat{\mathbf{u}}^{(2)}$ &$f_1\hat{\mathbf{u}}^{(2)}$
				& $\mathbf{0}$
				& $\mathbf{0}$&$\begin{matrix}-\frac{Q_{23}^{U_1}+Q_{32}^{U_1}}{2}\hat{\boldsymbol{\Gamma}}^{(4)}\\+\frac{Q_{23}^{U_1}-Q_{32}^{U_1}}{2}\hat{\boldsymbol{\Gamma}}^{(7)}\end{matrix}$\\
				\hline
				$\hat{\mathbf{u}}^{(3)}$ & $f_3\hat{\mathbf{u}}^{(3)}$	& $\mathbf{0}$&$\mathbf{0}$&$Q_{12}^{U_3}\hat{\boldsymbol{\Gamma}}^{(6)}$\\
				\hline
				$\mathbf{x}\cdot\hat{\boldsymbol{\Gamma}}^{(1)}$ & $\mathbf{0}$
				&$\mathbf{0}$& $F_{11}^{\Gamma_1}\hat{\boldsymbol{\Gamma}}^{(1)}$& $\mathbf{0}$\\
				\hline
				$\mathbf{x}\cdot\hat{\boldsymbol{\Gamma}}^{(2)}$ & $\mathbf{0}$
				& $\mathbf{0}$& $ \begin{bmatrix}
					F_{11}^{\Gamma_2}&0&0\\0&F_{11}^{\Gamma_2}&0 \\0&0&F_{33}^{\Gamma_2}
				\end{bmatrix}$& $\mathbf{0}$\\
				\hline
				$\mathbf{x}\cdot\hat{\boldsymbol{\Gamma}}^{(3)}$ & $\mathbf{0}$
				&$\mathbf{0}$& $F_{11}^{\Gamma_1}\hat{\boldsymbol{\Gamma}}^{(3)}$& $\mathbf{0}$ \\
				\hline
				$\mathbf{x}\cdot\hat{\boldsymbol{\Gamma}}^{(4)}$ & $\mathbf{0}$
				& $q_4\hat{\mathbf{u}}^{(2)}$		& $ \begin{matrix}\frac{F_{31}^{\Gamma_4}+F_{13}^{\Gamma_4}}{2}\hat{\boldsymbol{\Gamma}}^{(4)}\\+\frac{F_{31}^{\Gamma_4}-F_{13}^{\Gamma_4}}{2}\hat{\boldsymbol{\Gamma}}^{(7)}
				\end{matrix}$& $\mathbf{0}$ \\
				\hline
				$\mathbf{x}\cdot\hat{\boldsymbol{\Gamma}}^{(5)}$ & $\mathbf{0}$
				& $-q_4\hat{\mathbf{u}}^{(1)}$&	$ \begin{matrix}\frac{F_{13}^{\Gamma_4}+F_{31}^{\Gamma_4}}{2}\hat{\boldsymbol{\Gamma}}^{(5)}\\+\frac{F_{13}^{\Gamma_4}-F_{31}^{\Gamma_4}}{2}\hat{\boldsymbol{\Gamma}}^{(8)}\end{matrix}$& $\mathbf{0}$ \\
				\hline
				$\mathbf{x}\cdot\hat{\boldsymbol{\Gamma}}^{(6)}$ & $\mathbf{0}$
				& $q_3\hat{\mathbf{u}}^{(3)}$ & $F_{12}^{\Gamma_6}\hat{\boldsymbol{\Gamma}}^{(6)}$& $\mathbf{0}$ \\
				\hline
				$\mathbf{x}\cdot\hat{\boldsymbol{\Gamma}}^{(7)}$ & $\mathbf{0}$
				& $q_1\hat{\mathbf{u}}^{(2)}$& $ \begin{matrix}\frac{F_{31}^{\Gamma_7}+F_{23}^{\Gamma_8}}{2}\hat{\boldsymbol{\Gamma}}^{(7)}\\+\frac{F_{31}^{\Gamma_7}-F_{23}^{\Gamma_8}}{2}\hat{\boldsymbol{\Gamma}}^{(4)}\end{matrix}$& $\mathbf{0}$\\
				\hline
				$\mathbf{x}\cdot\hat{\boldsymbol{\Gamma}}^{(8)}$ & $\mathbf{0}$
				& $q_1\hat{\mathbf{u}}^{(1)}$& $ \begin{matrix}\frac{F_{31}^{\Gamma_7}+F_{23}^{\Gamma_8}}{2}\hat{\boldsymbol{\Gamma}}^{(8)}\\-\frac{F_{31}^{\Gamma_7}-F_{23}^{\Gamma_8}}{2}\hat{\boldsymbol{\Gamma}}^{(5)}\end{matrix}$& $\mathbf{0}$	\\
				\hline
			\end{tabular}
		\end{center}
	\end{table}
	\subsection{Constant viscosity fluid}\label{sec:Unstratified}
	
	In a fluid with spatially constant viscosity, $\eta_0$, a force, $\mathbf{g}$ and torque, $\mathbf{q}_{\boldsymbol{\Gamma}}$ leads to the following particle translation and rotation velocities,
	\begin{eqnarray}
		&\mathbf{u}_{\text{particle}}=\frac{1}{\eta_0}\Big[\mathbf{p}(\mathbf{p}\cdot{\mathbf{g}})\frac{f_1-f_3}{f_1f_3}+\frac{1}{f_1} {\mathbf{g}}\Big], \label{eq:NewtSedimentvel}\\
		&\boldsymbol{\omega}_{\text{particle}}=\frac{1}{\eta_0}\Big[\mathbf{p}(\mathbf{p}\cdot{\mathbf{q}_{\boldsymbol{\Gamma}}})\frac{q_1-q_3}{q_1q_3}+\frac{1}{q_1} {\mathbf{q}_{\boldsymbol{\Gamma}}}\Big]. \label{eq:NewtSedimentOmega}
	\end{eqnarray}
	The significance of particle shape-dependent factors $f_1$, $f_3$, $q_1$ and $q_3$ can be ascertained from table \ref{tab:NewtonianCoeffsStratifiedForces}. In particular, $f_1$ and $f_3$ respectively are the non-zero components of the hydrodynamic forces on the particle in a unit uniform flow of uniform viscosity fluid, aligned perpendicular and parallel to the axis. The factors $q_1$ and $q_3$ respectively are the non-zero components of hydrodynamic torques experienced by a fixed particle in the rotational flows $\mathbf{x}^\text{body}\cdot\boldsymbol{\Gamma}^{(7)}$ and $\mathbf{x}^\text{body}\cdot\boldsymbol{\Gamma}^{(6)}$. A particle fixed in an imposed uniform flow with velocity, $\mathbf{u}_\text{flow}$, experiences a force,
	\begin{equation}
		\mathbf{f}^\text{Stokes}_{\mathbf{u}_\text{flow}}=\eta_0[\mathbf{p}(\mathbf{p}\cdot\mathbf{u}_\text{flow})(f_3-f_1)+f_1 \mathbf{u}_\text{flow}],\label{eq:NewtForceU}
	\end{equation}
	and no torque. In a linear flow with an imposed velocity gradient,
	$	\boldsymbol{\Gamma} = \mathbf{E}+\boldsymbol{\Omega}$, with a symmetric (straining) part, $\mathbf{E}$, and an anti-symmetric (rotational) part, $\boldsymbol{\Omega}$ (such that intrinsic rotation of the imposed linear flow, $\boldsymbol{\omega}_\infty=-0.5\boldsymbol{\epsilon}:\boldsymbol{\Omega}$), a fixed particle experiences a torque,
	\begin{equation}
		\mathbf{q}_{\boldsymbol{\Gamma}}=\eta_0[\mathbf{p}(\mathbf{p}\cdot\boldsymbol{\omega}_\infty)(q_3-q_1)+q_1\boldsymbol{\omega}_\infty-q_4(\mathbf{E}\cdot\mathbf{p})\times\mathbf{p}],\label{eq:Torqueeta0}
	\end{equation}
	and no force. Therefore, from equation \eqref{eq:NewtSedimentOmega}, a torque-free particle in a linear flow rotates with an angular velocity,
	\begin{equation}
		\boldsymbol{\omega}^\text{Jeffery}_\text{particle}=\boldsymbol{\omega}_\infty-\frac{q_4}{q_1}(\mathbf{E}\cdot\mathbf{p})\times\mathbf{p},\label{eq:LabAngularVelocity}
	\end{equation}
	that leads to the time rate of change of the particle orientation vector given by,
	\begin{equation}
		\dot{\mathbf{p}}^\text{Jeffery}=	\boldsymbol{\omega}^\text{Jeffery}_\text{particle} \times \mathbf{p} = \boldsymbol{\omega}_\infty\times \mathbf{p} + \frac{q_4}{q_1}(\mathbf{E}\cdot \mathbf{p})\cdot (\mathbf{I}-\mathbf{p}\mathbf{p}).\label{eq:RotRateNewtonian}
	\end{equation}
	Here, $q_4$ is the non-zero component of the torque on a fixed particle in the straining flow given by $\mathbf{x}^\text{body}\cdot\boldsymbol{\Gamma}^{(4)}$ or $-\mathbf{x}^\text{body}\cdot\boldsymbol{\Gamma}^{(5)}$.
	The orientation trajectories obtained through the integration of the above equation are known as Jeffery orbits \citep{jeffery1922motion}.		
	
	\subsection{Fixed particle in stratified fluid}\label{sec:AbstractStratification}
	In a coordinate-free form, the {$\mathcal{O}(\beta)$} stratification-induced forces and torques on a fixed particle are, using values from table \ref{tab:NewtonianCoeffsStratifiedForces},
	\begin{equation}
		\mathbf{f}^\text{Stratified}_{\boldsymbol{\Gamma}}= \beta\mathbf{R}\cdot\mathbf{F}_\text{strat}^\text{body}\cdot\mathbf{R}^\text{T}\cdot\mathbf{d},\hspace{0.2in}	\mathbf{q}^\text{Stratified}_{\mathbf{u}_\text{flow}}= \beta\mathbf{R}\cdot \mathbf{Q}_\text{strat}^\text{body}\cdot\mathbf{R}^\text{T}\cdot\mathbf{d}.
	\end{equation}
	can be expressed as,
	\begin{equation}
		\mathbf{q}^\text{Stratified}_{\mathbf{u}_\text{flow}}= \beta[{Q_{32}^{U_1}} (\mathbf{p} \mathbf{p} \cdot(\mathbf{u}_\text{flow}\times\mathbf{d}))+{Q_{12}^{U_3}} (\mathbf{p} \cdot\mathbf{u}_\text{flow})\mathbf{d}\times \mathbf{p}+ {Q_{23}^{U_1}}(\mathbf{p}\cdot \mathbf{d})\mathbf{p}\times \mathbf{u}_\text{flow}],\label{eq:StratTorqueFixed}
	\end{equation}
	in a uniform flow with velocity, $\mathbf{u}_\text{flow}$, and,
	\begin{align}\begin{split}
			&\mathbf{f}^\text{Stratified}_{\boldsymbol{\Gamma}}=\beta\mathbf{R}\cdot	\mathbf{F}^\text{strat}_\text{linear,}\cdot\mathbf{R}^\text{T}\cdot \mathbf{d}
			=\\&\beta\Big(F_{11}^{\Gamma_1}\mathbf{E}+(F_{33}^{\Gamma_2}+F_{11}^{\Gamma_2}+F_{11}^{\Gamma_1}/2-F^{\Gamma_4}_{13}-F^{\Gamma_4}_{31})(\mathbf{p}\cdot\mathbf{E}\cdot\mathbf{p})\mathbf{pp}\\&+(F_{11}^{\Gamma_1}/2-F_{11}^{\Gamma_2})(\mathbf{p}\cdot\mathbf{E}\cdot\mathbf{p})\mathbf{I}+(F^{\Gamma_4}_{13}-F_{11}^{\Gamma_1})\mathbf{E}\cdot\mathbf{pp}+(F^{\Gamma_4}_{31}-F_{11}^{\Gamma_1})\mathbf{pp}\cdot\mathbf{E}\Big)\cdot \mathbf{d}\\&
			-\beta\Big(F_{12}^{\Gamma_6}\boldsymbol{\omega}_\text{flow}\times\mathbf{d} -(F_{12}^{\Gamma_6}-F^{\Gamma_7}_{31})\mathbf{pp}\cdot(\boldsymbol{\omega}_\text{flow}\times\mathbf{d})-(F_{12}^{\Gamma_6}-F^{\Gamma_8}_{23})(\boldsymbol{\omega}_\text{flow}\times\mathbf{p})\mathbf{p}\cdot\mathbf{d}\Big),\label{eq:StratForceLinear}
	\end{split}\end{align}
	in a linear imposed flow with $\mathbf{\Omega}$ and $\mathbf{E}$ as the vorticity and strain rate tensor. Here, $\boldsymbol{\omega}_\text{flow}=-0.5\boldsymbol{\epsilon}:\mathbf{\Omega}$ is the imposed fluid rotation in the perspective of the particle. The physical meaning of the coefficients $Q_{ij}^{U_k}$ ($i,j,k\in[1,3]$) and $F_{ij}^{\Gamma_k}$ ($i,j\in[1,3],k\in[1,8]$), can be inferred from table \ref{tab:NewtonianCoeffsStratifiedForces}. These are {$\beta$ normalized} stratification-induced torques, $Q_{ij}^{U_k}$, and forces, $F_{ij}^{\Gamma_k}$, along direction ${i}$ in a fluid with viscosity increasing linearly along $j$, on a fixed particle with axis of symmetry along direction 3, either in a uniform flow (for $Q_{ij}^{U_k}$) with unit velocity along direction $k$ or a linear flow with gradient $\hat{\boldsymbol{\Gamma}}^{(k)}$ (for $F_{ij}^{\Gamma_k}$).
	
	\subsection{Particle motion in viscosity-stratified fluid}\label{sec:AbstractStratificationMoving}
	A particle translating with a velocity $\mathbf{u}_\text{particle}$ in a quiescent fluid experiences a stratification-induced torque,
	\begin{equation}
		\mathbf{q}^\text{Stratified}_{\mathbf{u}_\text{particle}}=-\beta[{Q_{32}^{U_1}} (\mathbf{p} \mathbf{p} \cdot(\mathbf{u}_\text{particle}\times\mathbf{d}))+{Q_{12}^{U_3}} (\mathbf{p} \cdot\mathbf{u}_\text{particle})\mathbf{d}\times \mathbf{p}+ {Q_{23}^{U_1}}(\mathbf{p}\cdot \mathbf{d})\mathbf{p}\times \mathbf{u}_\text{particle}],\label{eq:StratTorquePart}
	\end{equation}
	in addition to the force, $-\eta_0[\mathbf{p}(\mathbf{p}\cdot\mathbf{u}_\text{particle})(f_3-f_1)+f_1 \mathbf{u}_\text{particle}]$ ($\mathbf{u}_\text{flow}$ replaced with $-\mathbf{u}_\text{particle}$ in equation \eqref{eq:NewtForceU}) from the uniform viscosity component of the fluid. While a particle rotating with an angular velocity $\boldsymbol{\omega}_\text{particle}$ in a quiescent fluid experiences a torque $-\eta_0[\mathbf{p}(\mathbf{p}\cdot\boldsymbol{\omega}_\text{particle})(q_3-q_1)+q_1\boldsymbol{\omega}_\text{particle}]$ ($\mathbf{E}=0$ and $\boldsymbol{\omega}_\infty=-\boldsymbol{\omega}_\text{particle}$ in equation \eqref{eq:Torqueeta0}) due to the constant part of viscosity. The linear stratification leads to a force,
	\begin{align}\begin{split}
			\mathbf{f}^\text{Stratified}_{\boldsymbol{\omega}_\text{particle}}=&\beta\Big(F_{12}^{\Gamma_6}\boldsymbol{\omega}_\text{particle}\times\mathbf{d} -(F_{12}^{\Gamma_6}-F^{\Gamma_7}_{31})\mathbf{pp}\cdot(\boldsymbol{\omega}_\text{particle}\times\mathbf{d})-(F_{12}^{\Gamma_6}-F^{\Gamma_8}_{23})(\boldsymbol{\omega}_\text{particle}\times\mathbf{p})\mathbf{p}\cdot\mathbf{d}\Big),\label{eq:StratForceLinearOmega}
	\end{split}\end{align}
	as a relative linear flow of the fluid with (anti-symmetric) velocity gradient $\boldsymbol{\Gamma} = -\boldsymbol{\epsilon}\cdot \boldsymbol{\omega}_\text{particle}$ is experienced by the particle.
	
	The stratification-induced torque on a translating particle and force on a rotating particle lead to a coupling between two types of motion not observed in a uniform viscosity fluid in the absence of inertia \citep{oppenheimer2016motion,datt2019active,anand2024sedimentation,gong2024active}. A particle sediments in a constant viscosity fluid along a linear path that depends on its shape and initial orientation. However, its orientation will change in the presence of viscosity stratification, leading to a different translation path. In a linear flow, while rotating as per Jeffery orbits, a particle translates at a velocity equal to the fluid's velocity at its centroid. However, a stratification-induced force will lead to a relative translation between the particle and fluid. For example, unlike a uniform viscosity scenario, it can migrate across the streamlines of a simple shear flow or be forced to move relative to the center of a uniaxial extensional flow.

	\subsubsection{Sedimenting particle}\label{sec:SedimentingParticle}
	First consider a particle settling under the action of a gravitational force, $\mathbf{g}$ in a quiescent fluid. The particle's translation and angular velocity are obtained via the following coupled equations,
	\begin{eqnarray}
		&	\mathbf{u}_{\text{particle}}=\frac{1}{\eta_0}\Big[\mathbf{p}(\mathbf{p}\cdot{(\mathbf{g}+\mathbf{f}^\text{Stratified}_{\boldsymbol{\omega}_\text{particle}})})\frac{f_1-f_3}{f_1f_3}+\frac{1}{f_1} {(\mathbf{g}+\mathbf{f}^\text{Stratified}_{\boldsymbol{\omega}_\text{particle}})}\Big],\label{eq:Sedimentuvel}\\	&\boldsymbol{\omega}_\text{particle}=\frac{1}{\eta_0}\Big[\mathbf{p}(\mathbf{p}\cdot{	\mathbf{q}^\text{Stratified}_{\mathbf{u}_\text{particle}} })\frac{q_1-q_3}{q_1q_3}+\frac{1}{q_1} {	\mathbf{q}^\text{Stratified}_{\mathbf{u}_\text{particle}} }\Big]. \label{eq:Sedimentomega}
	\end{eqnarray}
	where $\mathbf{q}^\text{Stratified}_{\mathbf{u}_\text{particle}}$ (a function of ${\mathbf{u}_\text{particle}}$) and $\mathbf{f}^\text{Stratified}_{\boldsymbol{\omega}_\text{particle}}$ (a function of $\boldsymbol{\omega}_\text{particle}$) are given by equations \eqref{eq:StratTorquePart} and \eqref{eq:StratForceLinearOmega}.
	Upon substitution of relevant variables, the governing equation for angular velocity is,
	\begin{eqnarray}
			&\boldsymbol{\omega}_{\text{particle}} = \frac{\beta }{\eta_0^2}[-t_1 (\mathbf{p} \mathbf{p} \cdot({\mathbf{g}}\times\mathbf{d}))-t_2 (\mathbf{p} \cdot{\mathbf{g}})\mathbf{d}\times \mathbf{p}-t_3(\mathbf{p}\cdot \mathbf{d})\mathbf{p}\times {\mathbf{g}}]+\mathcal{O}(\beta^2),\label{eq:angvelg}
	\end{eqnarray}
	where,
	\begin{eqnarray}
		t_1=\frac{Q_{32}^{U_1}}{f_1 q_3}, \quad	t_2=\frac{Q_{12}^{U_3}}{f_3 q_1} ,\quad	t_3=\frac{Q_{23}^{U_1}}{f_1 q_1}.\label{eq:t1t2t3}
	\end{eqnarray}
	{The $\mathcal{O}(\beta)$ effect of stratification is to rotate the particle's centerline, $\mathbf{p}$. We note that $\mathbf{f}^\text{Stratified}_{\boldsymbol{\omega}_\text{particle}}$ is $\mathcal{O}(\beta^2)$, since, from equation \eqref{eq:StratForceLinearOmega},  $\mathbf{f}^\text{Stratified}_{\boldsymbol{\omega}_\text{particle}}\sim\beta\mathcal{O}(\boldsymbol{\omega}_\text{particle})$ and $\boldsymbol{\omega}_\text{particle}$ is itself $\mathcal{O}(\beta)$, which can be discerned from equations \eqref{eq:StratForceLinearOmega} and \eqref{eq:Sedimentomega}. Therefore, up to $\mathcal{O}(\beta)$, the settling velocity of the particle is simply the Newtonian velocity in equation \eqref{eq:NewtSedimentvel} for the $\mathbf{p}$ altered by the stratification induced torque. At $\mathcal{O}(\beta^2)$, a modification in the particle's translational velocity due to $\mathbf{f}^\text{Stratified}_{\boldsymbol{\omega}_\text{particle}} $ is expected.} The time rate of change of the particle orientation vector, $\dot{\mathbf{p}}=\boldsymbol{\omega} \times \mathbf{p}$, is governed by,
	\begin{align}\begin{split}
				\dot{\mathbf{p}}=-\frac{\beta }{\eta_0^2}\cdot\Big[ t_3(\mathbf{p}\cdot \mathbf{d}){\mathbf{g}}-t_2 (\mathbf{p} \cdot{\mathbf{g}})\mathbf{d}\Big]\cdot(\mathbf{I}-\mathbf{pp})+\mathcal{O}(\beta^2).\label{eq:rotationratefull}
			\end{split}
	\end{align}	
\subsubsection{Freely moving particles in linear flows}\label{eq:LinearFlowParticle}
The coupling between the rotation and translation of a freely suspended particle in a linear flow with velocity gradient, $\boldsymbol{\Gamma}=\mathbf{E}+\boldsymbol{\Omega}$ leads to the following translation and angular velocity of the particle,
\begin{eqnarray}	&\mathbf{u}_{\text{particle}}=\mathbf{x}_p\cdot \boldsymbol{\Gamma}+\mathbf{u}^\text{Stratified},\quad\mathbf{u}^\text{Stratified}=\frac{1}{\eta_0}\Big[\mathbf{p}(\mathbf{p}\cdot{\mathbf{f}^\text{Stratified}_{\boldsymbol{\Gamma}}})\frac{f_1-f_3}{f_1f_3}+\frac{1}{f_1} {\mathbf{f}^\text{Stratified}_{\boldsymbol{\Gamma}}}\Big],\label{eq:uStratifiedLinFlow}\\
	&\begin{split}\boldsymbol{\omega}_\text{particle}=&\boldsymbol{\omega}_\infty-\frac{q_4}{q_1}(\mathbf{E}\cdot\mathbf{p})\times\mathbf{p}+ \frac{1}{\eta_0}\Big[\mathbf{p}(\mathbf{p}\cdot{ (	\mathbf{q}^\text{Stratified}_{\mathbf{u}_\text{flow-particle}})})\frac{q_1-q_3}{q_1q_3}+\frac{1}{q_1} {(	\mathbf{q}^\text{Stratified}_{\mathbf{u}_\text{flow-particle}}) }\Big].
\end{split}\end{eqnarray}
where, $\mathbf{f}^\text{Stratified}_{\boldsymbol{\Gamma}}$ (given in equation \eqref{eq:StratForceLinear} with $\mathbf{E}$ as the symmetric part of imposed velocity gradient, and $\boldsymbol{\omega}_\text{flow}=\boldsymbol{\omega}_\infty-\boldsymbol{\omega}_\text{particle}$) is the stratification-induced force on a fixed particle in a linear flow created by the imposed strain and relative rotation of the fluid in the perspective of the particle. The expressions for stratification-induced torques due to relative translation motion between particle and fluid, $\mathbf{q}^\text{Stratified}_{\mathbf{u}_\text{flow-particle}}$ is $\mathbf{q}^\text{Stratified}_{\mathbf{u}_\text{particle}}$ from equation \eqref{eq:StratTorquePart} with $\mathbf{u}_{\text{particle}}$ replaced with $\mathbf{u}^\text{Stratified}$. This relative velocity of the particle induced by stratification is governed by,
\begin{align}\begin{split}
			&\mathbf{u}^\text{Stratified}=\frac{\beta}{f_1\eta_0}\Bigg(F_{11}^{\Gamma_1}\mathbf{E}+(F_{11}^{\Gamma_1}/2+F_{11}^{\Gamma_2})(\mathbf{p}\cdot\mathbf{E}\cdot\mathbf{p})\mathbf{I}\\&+\Big[\frac{f_1}{f_3}F_{33}^{\Gamma_2}-F_{11}^{\Gamma_2}+F_{11}^{\Gamma_1}/2-F^{\Gamma_4}_{13}-\frac{f_1}{f_3}F^{\Gamma_4}_{31}+\frac{q_4}{q_1}\Big(\frac{f_1}{f_3}F_{31}^{\Gamma_7}-F_{23}^{\Gamma_8}\Big)\Big](\mathbf{p}\cdot\mathbf{E}\cdot\mathbf{p})\mathbf{pp}\\&+\Big(F^{\Gamma_4}_{13}-F_{11}^{\Gamma_1}+\frac{q_4}{q_1}F^{\Gamma_8}_{23}\Big)\mathbf{E}\cdot\mathbf{pp}+\Big(\frac{f_1}{f_3}F^{\Gamma_4}_{31}-F_{11}^{\Gamma_1}-\frac{q_4}{q_1}\frac{f_1}{f_3}F^{\Gamma_7}_{31}\Big)\mathbf{pp}\cdot\mathbf{E}\Bigg)\cdot\mathbf{d}+\mathcal{O}(\beta^2).\label{eq:StratificationVelocity}
\end{split}\end{align}
Since $\mathbf{q}^\text{Stratified}_{\mathbf{u}_\text{flow-particle}}\sim\beta\mathcal{O}(\mathbf{u}^\text{Stratified})\sim\mathcal{O}(\beta^2)$, up to $\mathcal{O}(\beta)$ the particle's centerline follows the Newtonian rotation rate equation \eqref{eq:RotRateNewtonian}.

In the next section, the force and torque components listed in table \ref{tab:NewtonianCoeffsStratifiedForces} are analytically obtained for spheroids using spheroidal harmonics. These coefficients appear in the particle dynamics equations \eqref{eq:rotationratefull} and \eqref{eq:StratificationVelocity}. Free settling in gravity using equations \eqref{eq:NewtSedimentvel} and \eqref{eq:rotationratefull} is explored in section \ref{sec:SpheroidsSedimenting} and the motion of particles freely suspended in linear flows calculated using equations \eqref{eq:RotRateNewtonian} and \eqref{eq:StratificationVelocity} is shown in section \ref{sec:SpheroidsLinear}.

\section{Fixed spheroids in viscosity stratified fluids}\label{sec:SpheroidResultsFix}
	As discussed in section \ref{sec:Formulation}, we only need to know the expressions for the Stokes flow fields around the particle to obtain the {$\mathcal{O}{(\beta)}$} stratification-induced force and torque from equation \eqref{eq:StratificationForceTorque}. We use the spheroidal harmonics formulation of \cite{dabade2015effects,dabade2016effect} to obtain these expressions in Mathematica. The stratification-induced force and torque expressions on spheroids are unwieldy. Hence, while we discuss the qualitative nature of the aspect ratio dependence of these quantities in this section, we provide the explicit formulae in appendix \ref{sec:ExpressionsExtra}.

	We have validated these expressions in three separate ways. First, in the limit $\kappa \rightarrow 1$ these values are compared with those obtained from simpler expressions of flow around a sphere, demonstrated in appendix \ref{sec:AppendixSphereCase}. {Second, for prolate spheroids we solve the stratified mass and momentum equations \eqref{eq:StratificationProblemMod3} at $\mathcal{O}(\beta)$,
	using the finite-difference based numerical method of \cite{sharma2023finite}. This solver is written in prolate spheroidal coordinates with the particle surface as the inner boundary and a nearly spherical surface in the far-field as the outer boundary. For the current study, we compute the flow around a fixed particle with either uniform or linear flow imposed at the outer boundary. The numerically evaluated stratification induced stress field, ${\boldsymbol{\sigma}}^\text{Stratified}=\beta ({\boldsymbol{\sigma}}^\text{Stratified})^{(1)}+\mathcal{O}(\beta^2)$ at $\mathcal{O}(\beta)$ is directly used to obtain the force and torque to compare with the analytical values obtained via a generalized reciprocal theorem that circumvents the stress field calculation.} Last, in the limit of large aspect ratio, values from the resistive force theory of \cite{kamal2023resistive} are compared with ours.

\subsection{Uniform flow}\label{sec:SpheroidFixUniform}
As mentioned in section \ref{sec:AbstractStratification}, $Q_{ij}^{U_k}$ (for $i,j,k\in[1,3]$) refers to the {$\beta$ normalized} stratification-induced torque along direction ${i}$ in a fluid with viscosity increasing linearly along $j$, on a fixed particle with axis of symmetry along direction 3, in a uniform flow with unit velocity along direction $k$. Out of the 27 coefficients in $Q_{ij}^{U_k}$ there are only three unique non-zero values. Therefore, the stratification-induced torque on a spheroid fixed in uniform flow of linearly stratified fluid for a general case can be described by the torque induced on the particle in three distinct scenarios which all require the imposed flow and the stratification direction to be mutually perpendicular. In the first case, the particle centerline, denoted by unit vector $\mathbf{p}$, lies normal to the flow-stratification plane and the stratification-induced torque, denoted by $Q_{32}^{U_1}$, lies along $\mathbf{p}$. In the remaining two scenarios, where either the flow direction (torque labeled as $Q_{12}^{U_3}$) or the stratification direction (torque denoted by $Q_{23}^{U_1}$) is along $\mathbf{p}$, the stratification-induced torque is normal to $\mathbf{p}$. In section \ref{sec:SpheroidFixUniformValid} we provide a validation of the expressions for $Q_{32}^{U_1}$, $Q_{23}^{U_1}$ and $Q_{12}^{U_3}$ before discussing the qualitative trends in the variation of these quantities with particle aspect ratio, $\kappa$, and providing a mechanistic explanation of their origins in section \ref{sec:SpheroidFixUniformMechanics}.

\subsubsection{Validation}\label{sec:SpheroidFixUniformValid}
\begin{figure}
	\centering
	\subfloat{\includegraphics[width=0.6\textwidth]{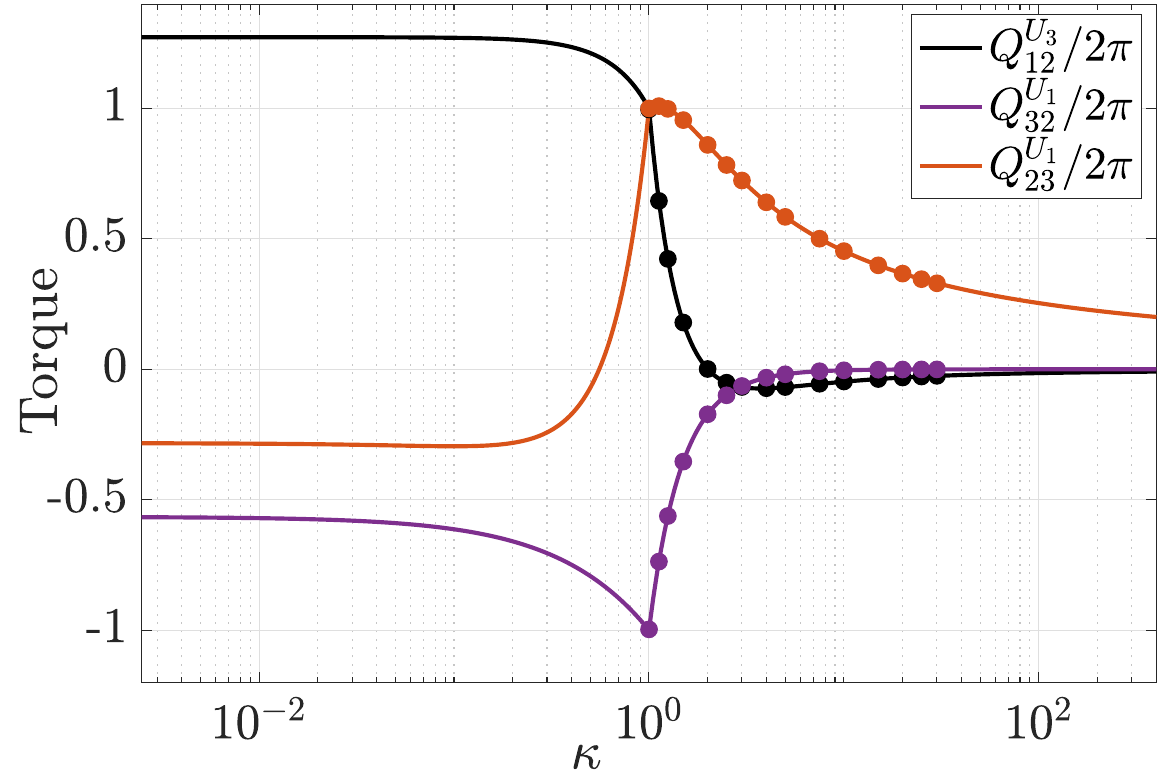}}
	\caption {Viscosity-stratification-induced torque in uniform flow past a spheroid with aspect ratio $\kappa$ and major axis equal to 1. The markers for $1\le\kappa\le30$ are the values obtained from the numerical code. \label{fig:UniformFlowTorques}}
\end{figure}

The analytical expressions for stratification-induced torques $Q_{32}^{U_1}$, $Q_{23}^{U_1}$ and $Q_{12}^{U_3}$ for a spheroid fixed in uniform flow ($\hat{\mathbf{u}}^{(i)}, i\in[1,3]$ from equation \eqref{eq:BasisFlows}) are shown in equation \eqref{eq:Q32Q12} within appendix \ref{sec:SpheroidFixUniformAppend} and are plotted in figure \ref{fig:UniformFlowTorques} as solid curves. The corresponding numerically obtained values for a prolate spheroid with aspect ratio, $1\le\kappa\le30$ are shown as circular markers in the same figure where a close match between the values obtained from the two different techniques can be observed. Our formulas show that in the limit of a sphere of radius, $l$, i.e. $\kappa\rightarrow1$,
\begin{equation} \lim_{\kappa\rightarrow1}(Q_{23}^{U_1})=\lim_{\kappa\rightarrow1}(Q_{12}^{U_3})=-\lim_{\kappa\rightarrow1}(Q_{32}^{U_1})=2\pi l^3,
\end{equation}
which are the same results obtained for a sphere in appendix \ref{sec:AppendixSphereCase} (equation \eqref{eq:SphereUniformTorque}) without using the spheroidal harmonics.

For a slender fiber, i.e., a prolate spheroid with $\kappa\rightarrow\infty$,
\begin{eqnarray}\begin{split} &\lim_{\kappa\rightarrow\infty}(Q_{32}^{U_1})=-\frac{32\pi}{3}l^3\frac{\frac{\kappa}{\sqrt{\kappa^2-1}}-1}{2\log(2\kappa)+1}+\mathcal{O}(\kappa^{-2}),\\
		&\lim_{\kappa\rightarrow\infty}(Q_{23}^{U_1})=\frac{8\pi}{3}l^3\frac{1}{\log(2\kappa)-\frac{1}{4\log(2\kappa)}}+\mathcal{O}\Big(\frac{\kappa^{-2}}{\log(\kappa)}\Big), \\
		&\lim_{\kappa\rightarrow\infty}(Q_{12}^{U_3})=-\frac{8\pi}{3}l^3\frac{1}{(2\log(2\kappa)-1)^2}+\mathcal{O}\Big(\frac{\kappa^{-2}}{\log(\kappa)}\Big).\end{split}\label{eq:FiberTorques}
\end{eqnarray}
\cite{kamal2023resistive} evaluated the torque equivalent to $Q_{23}^{U_1}$ for a slender fiber using resistive force theory. According to their equation (3.21), a slender prolate spheroid of length $2l$ fixed in a fluid with viscosity gradient $\eta_0/(2l)$ parallel to a fiber undergoing a uniform flow with velocity $U_2$ perpendicular to the fiber experiences a torque $4\pi/[3(\log(2\kappa)+1)]l^2 \eta_0 U_2+\mathcal{O}(\log(\kappa)^{-2})$ along the axis normal to the flow-stratification plane. Their torque value re-scaled for $U_2=1$ and viscosity gradient $\beta$ is $(\beta 8\pi/3)l^3/ (\log(2\kappa)+1)+\mathcal{O}(\log(\kappa)^{-2})$. The equivalent torque (equation \eqref{eq:FiberTorques}), from our study in a fluid is $\beta \lim\limits_{\kappa\rightarrow\infty}Q_{23}^{U_1}=(8\pi/3) l^3\beta(1/[\log(2\kappa)-(1/4(\log(2\kappa)))])+\mathcal{O}(\kappa^{-2}\log(\kappa)^{-1})$. Hence, in the large $\kappa$ limit our expression matches with that of \cite{kamal2023resistive}.

\subsubsection{Mechanistic origin of stratification-induced torque}\label{sec:SpheroidFixUniformMechanics}
The decomposition of torque provided in equations \eqref{eq:ForceTorqueA} and \eqref{eq:ForceTorqueB} shows that the stratification-induced torque arises from two distinct fluid stresses (see equation \eqref{eq:TotalStressDecomposed}) in a viscosity stratified fluid flow around a particle. The {$\beta$ normalized} stress component $(\eta'/\eta_0)\boldsymbol{\sigma}^\text{Stokes}$ depicts the stress on a particle in a uniform viscosity fluid, $\boldsymbol{\sigma}^\text{Stokes}$, but incremented by the local variation in fluid viscosity around the particle. It leads to the stratification-induced torques ${Q_{32}^{U_1}}_\text{A}$, ${Q_{23}^{U_3}}_\text{A}$ and ${Q_{12}^{U_3}}_\text{A}$. The remaining part of the stratification-induced torques, ${Q_{32}^{U_1}}_\text{B}$, ${Q_{23}^{U_3}}_\text{B}$ and ${Q_{12}^{U_3}}_\text{B}$ arise due to the fluid stress resulting from the changes in stratification-induced velocity ($\mathbf{u}^\text{Stratified}$) and pressure ($p^\text{Stratified}$) {at $\mathcal{O}(\beta)$}.

\begin{figure}
	\centering
	\subfloat{\includegraphics[width=0.475\textwidth]{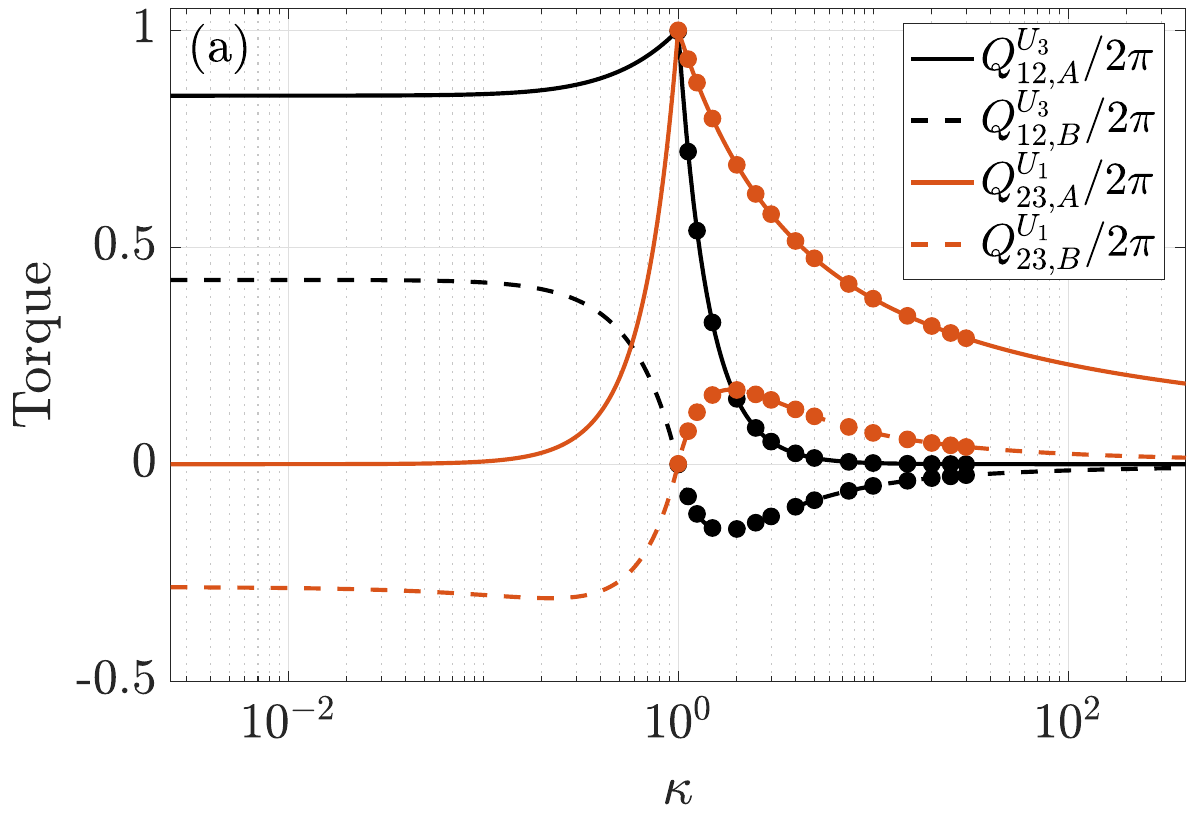} \label{fig:TorquesSpheroidDecompA}}
	\subfloat{\includegraphics[width=0.475\textwidth]{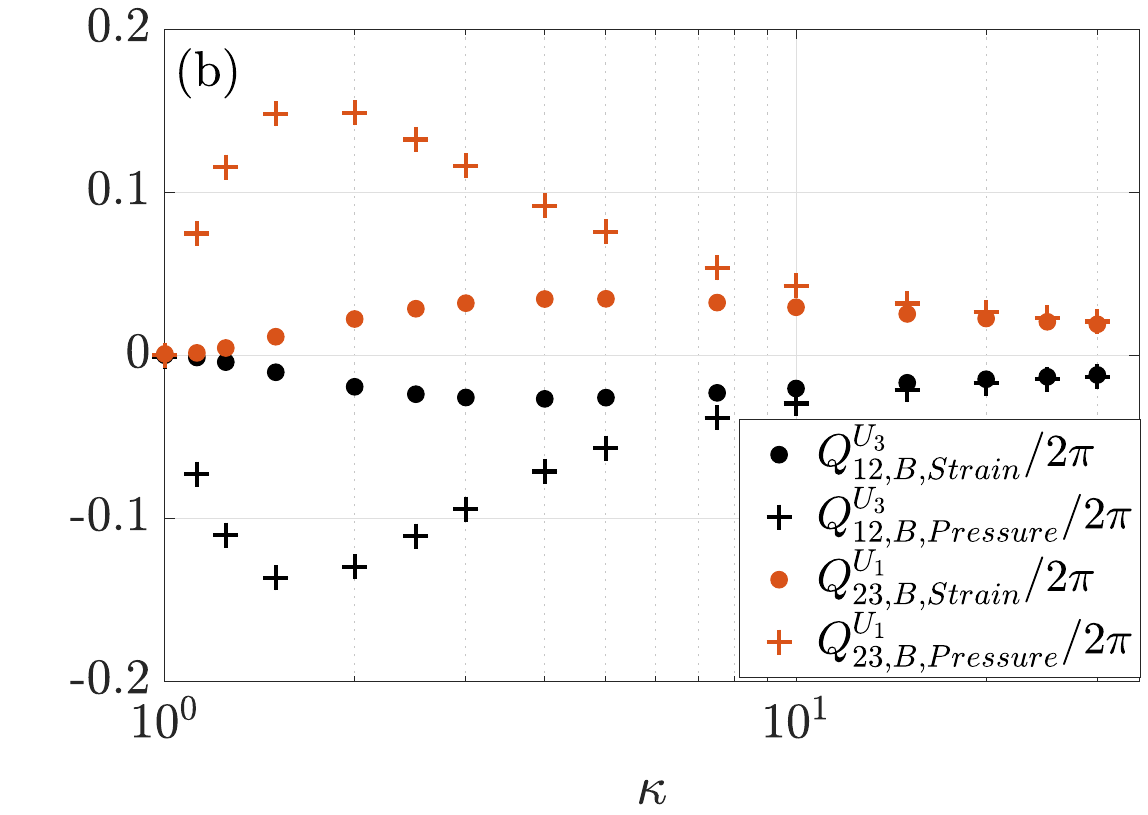} \label{fig:TorquesSpheroidDecompB}}
	\caption {Decomposition of stratification-induced torques in uniform flow on spheroids vs. aspect ratio: (a) Decomposition into $\mathbf{q}^\text{Stratified}_\text{A}$ and $\mathbf{q}^\text{Stratified}_\text{B}=\mathbf{q}^\text{Stratified}-\mathbf{q}^\text{Stratified}_\text{A}$ given by equation \eqref{eq:DecompAB} (the markers show values from the numerical code), and (b) Decomposition of $\mathbf{q}^\text{Stratified}_\text{B}$ into that from stratification-induced pressure and strain obtained from the numerical code. \label{fig:TorquesSpheroidDecomp}}
\end{figure}

First consider the case when the imposed uniform flow and viscosity stratification are perpendicular to one another and also to the particle centerline $\mathbf{p}$. Then the stratification-induced torque, ${Q_{32}^{U_1}}$, acts along $\mathbf{p}$. We find,
\begin{equation}
	{Q_{32}^{U_1}}_\text{B}=0 \rightarrow Q_{32}^{U_1}={Q_{32}^{U_1}}_\text{A},
\end{equation}
for all $\kappa$. Therefore, ${Q_{32}^{U_1}}$ is only due the Stokes stress acting in a variable viscosity environment. The fluid stress, $\boldsymbol{\sigma}^\text{Stratified}$, created by the stratification-induced velocity and pressure does not contribute to this torque. However, both $(\eta'/\eta_0)\boldsymbol{\sigma}^\text{Stokes}$ and $\boldsymbol{\sigma}^\text{Stratified}$ contribute towards ${Q_{12}^{U_3}}$ and ${Q_{23}^{U_1}}$.
These are the stratification-induced torques induced normal to $\mathbf{p}$ when either the stratification or the flow direction is along $\mathbf{p}$. The expressions for decomposed torques ${Q_{23}^{U_1}}_\text{A}$, ${Q_{12}^{U_3}}_\text{A}$, ${Q_{23}^{U_1}}_\text{B}$, and ${Q_{12}^{U_3}}_\text{B}$ are shown in equation \eqref{eq:DecompAB} and plotted in figure \ref{fig:TorquesSpheroidDecompA} as a function of particle aspect ratio, $\kappa$. The torques ${Q_{23}^{U_1}}_\text{B}$ and ${Q_{12}^{U_3}}_\text{B}$ are zero for $\kappa=1$, i.e., the stratification-induced velocity and pressure do not play a role in the stratification-induced torque on a sphere. However, for $\kappa\ne 1$, ${Q_{23}^{U_1}}_\text{B}$ and ${Q_{12}^{U_3}}_\text{B}$ are present in addition to ${Q_{23}^{U_1}}_\text{A}$ and ${Q_{12}^{U_3}}_\text{A}$.

From figure \ref{fig:UniformFlowTorques}, we can observe that the net stratification-induced torques, ${Q_{23}^{U_1}}$ and ${Q_{12}^{U_3}}$ change signs at $\kappa\approx0.55$ and $\kappa\approx2.0$ respectively. As seen in figure \ref{fig:TorquesSpheroidDecompA}, the change in signs at $\kappa\approx0.55$ and $\kappa\approx2.0$ is due to the competition between the torque arising from $(\eta'/\eta_0)\boldsymbol{\sigma}^\text{Stokes}$ (${Q_{23}^{U_1}}_\text{A}$ and ${Q_{12}^{U_3}}_\text{A}$) and that from $\boldsymbol{\sigma}^\text{Stratified}$ (${Q_{23}^{U_1}}_\text{B}$ and ${Q_{12}^{U_3}}_\text{B}$). In the upcoming discussion of section \ref{sec:SpheroidsSedimenting}, this sign change will play an important role in the particle's rotational dynamics.

The torques ${Q_{23}^{U_1}}_\text{B}$ and ${Q_{12}^{U_3}}_\text{B}$ can be further decomposed into those arising from the stratification-induced pressure, ${p}^\text{Stratified}$, and those from viscous force per unit area $2(\eta_0)\boldsymbol{e}^\text{Stratified}$, where $\boldsymbol{e}^\text{Stratified}=\nabla\mathbf{u}^\text{Stratified}+(\nabla\mathbf{u}^\text{Stratified})^T$ {at $\mathcal{O}(\beta)$}. This decomposition can be accessed for the prolate spheroid through the numerical calculation and is shown in figure \ref{fig:TorquesSpheroidDecompB} for ${Q_{23}^{U_1}}_\text{B}$ and ${Q_{12}^{U_3}}_\text{B}$, marked with subscripts ``Pressure" and ``Strain." Here we observe that the torque due to the stratification-induced pressure and viscous force per unit area, act in the same direction, but the former dominates {for prolate spheroids with $\kappa \lessapprox 10$, beyond which both are of almost equal magnitude}. Furthermore, $Q_{12,B,\text{Pressure}}^{U_3}$ and $Q_{23,B,\text{Pressure}}^{U_1}$ (as well as $Q_{12,A,\text{Strain}}^{U_3}$ and $Q_{23,B,\text{Strain}}^{U_1}$) are equal and opposite.

Similarly, the torques ${Q_{23}^{U_1}}_\text{A}$ and ${Q_{12}^{U_3}}_\text{A}$ that arise from the Stokes stress acting in a variable-viscosity environment can be decomposed into that arising from $(\eta'/\eta_0)p^\text{Stokes}$ and $2\eta'\mathbf{e}^\text{Stokes}$ (not shown).
Since the pressure acts normal to the surface, and for a sphere the line of action of the pressure force acts through the particle surface, pressure does not contribute to the torque. Therefore, for a sphere the entire stratification-induced torque arises from $2\eta'\mathbf{e}^\text{Stokes}$, the Stokes strain rate acting in a variable-viscosity environment.

The stratification-induced pressure distribution, $p^\text{Stratified}$, is the dominant contributor to the torques ${Q_{23}^{U_1}}_\text{B}$ and ${Q_{12}^{U_3}}_\text{B}$ {for prolate spheroids with $\kappa \lessapprox 10$}. Using patterns of $\eta'p^\text{Stokes}$ and $p^\text{Stratified}$ at the surface of prolate spheroids we can understand the mechanisms in various torque distributions observed in figure \ref{fig:TorquesSpheroidDecomp}.

\begin{figure}
	\centering
	\subfloat{\includegraphics[width=0.3\textwidth]{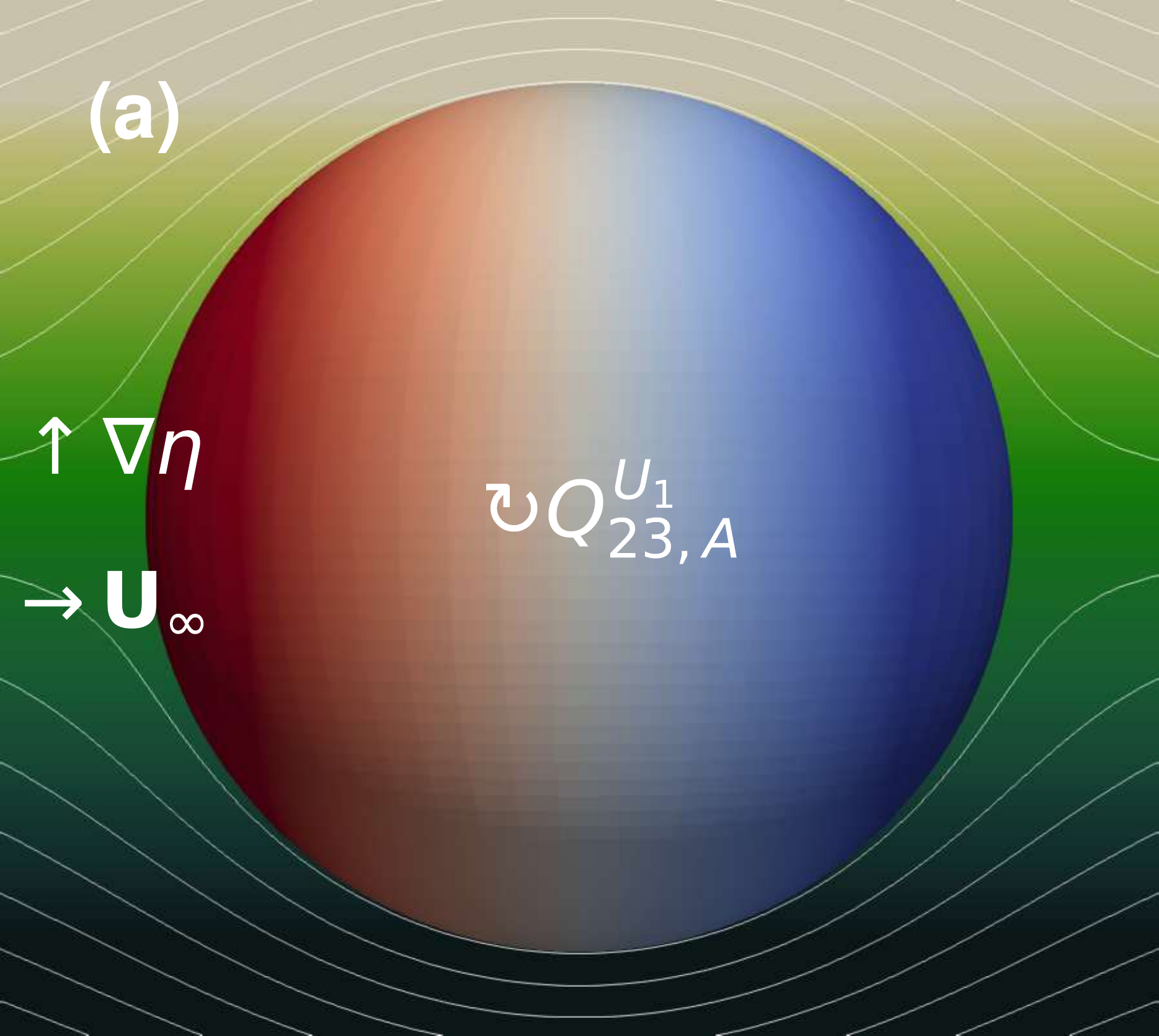} \label{StokesPressureAR1Q23U1}}
	\subfloat{\includegraphics[width=0.3\textwidth]{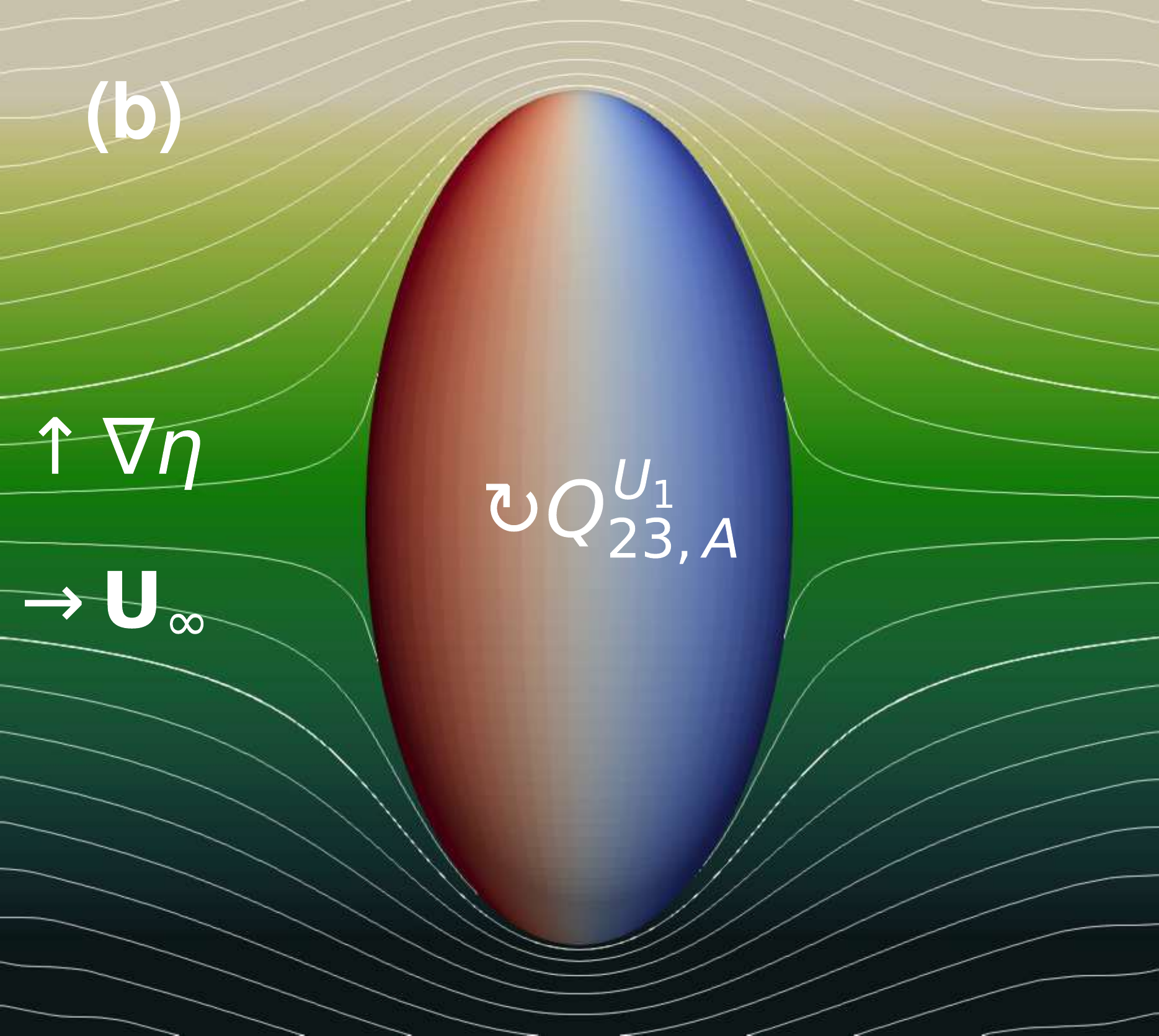} \label{StokesPressureAR2Q23U1}}
	\subfloat{\includegraphics[width=0.3\textwidth]{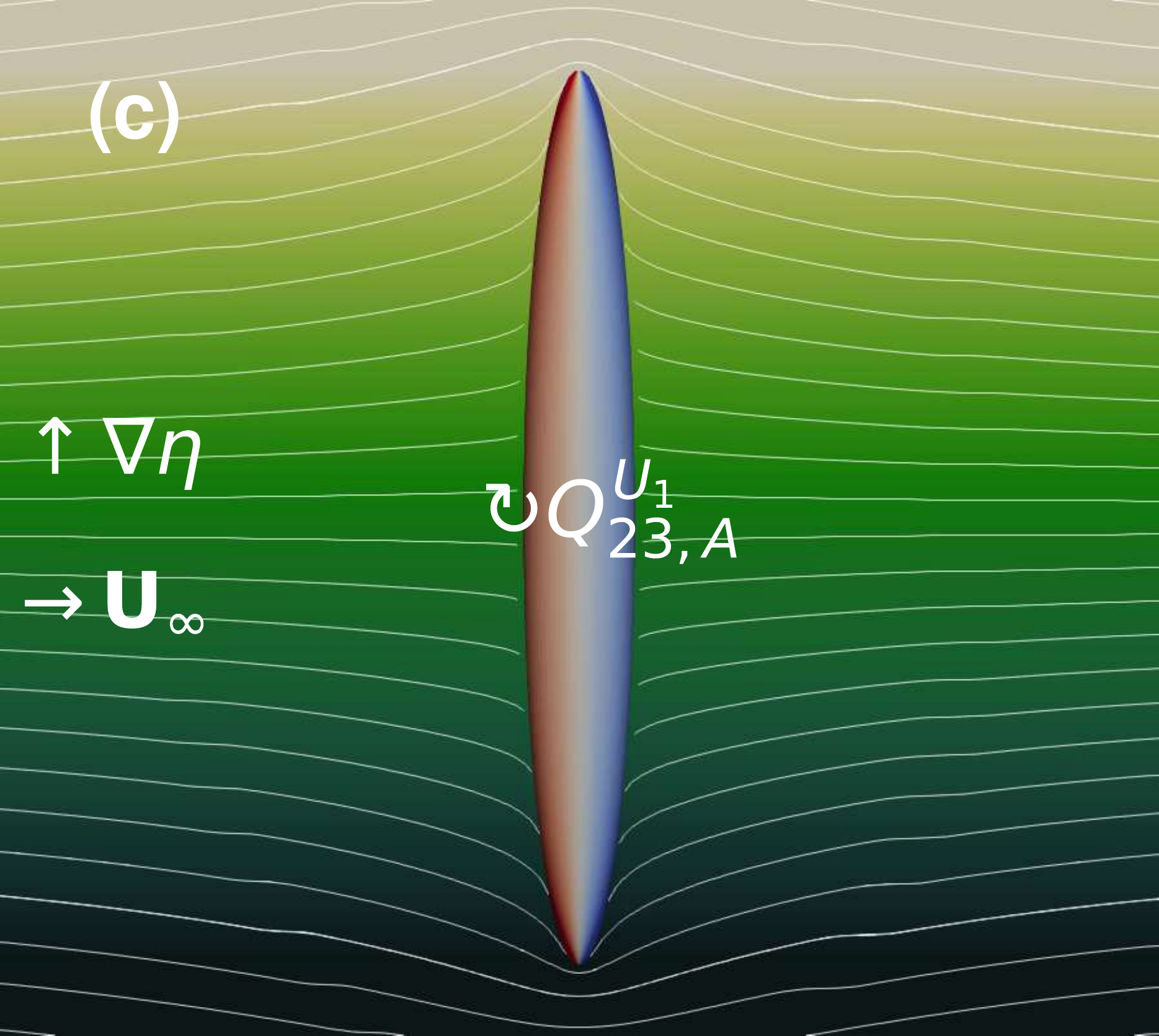} \label{StokesPressureAR8Q23U1}}
	\subfloat{\includegraphics[width=0.1\textwidth]{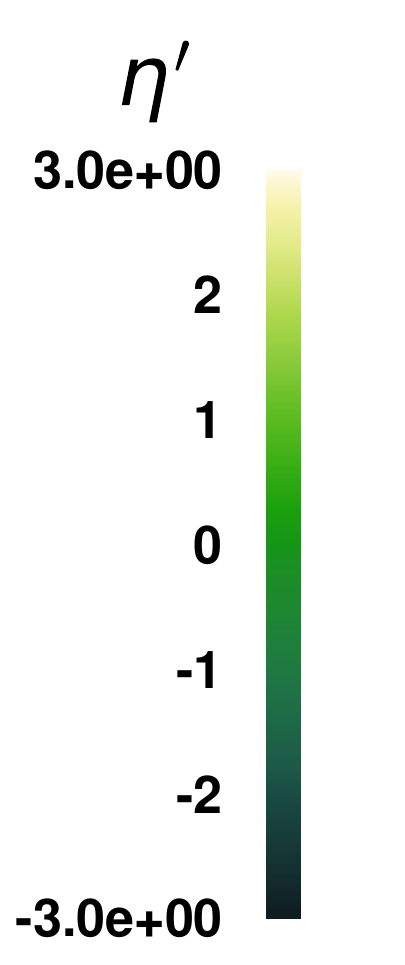}}\\
	\subfloat{\includegraphics[width=0.3\textwidth]{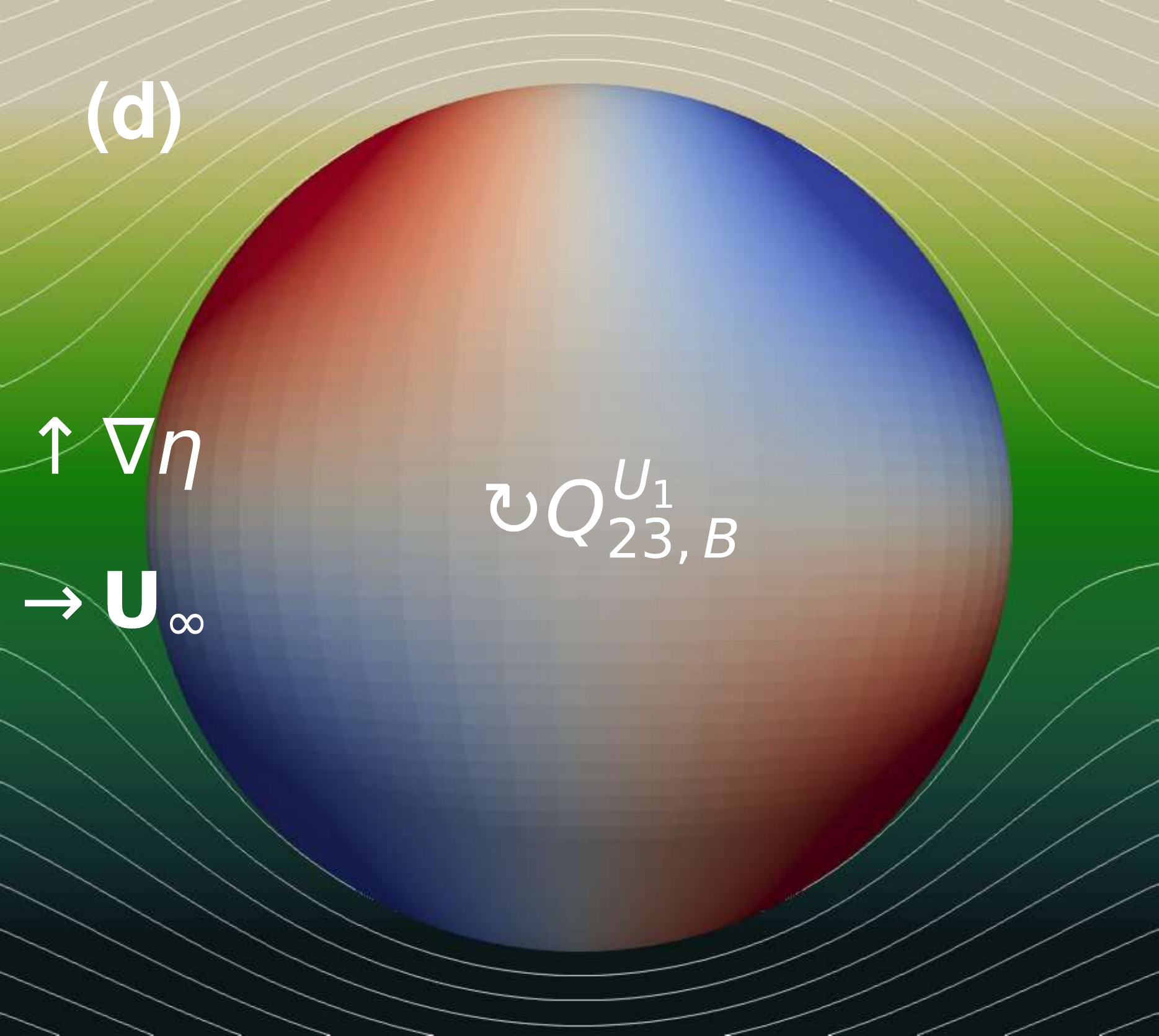} \label{StratPressureAR1Q23U1} }
	\subfloat{\includegraphics[width=0.3\textwidth]{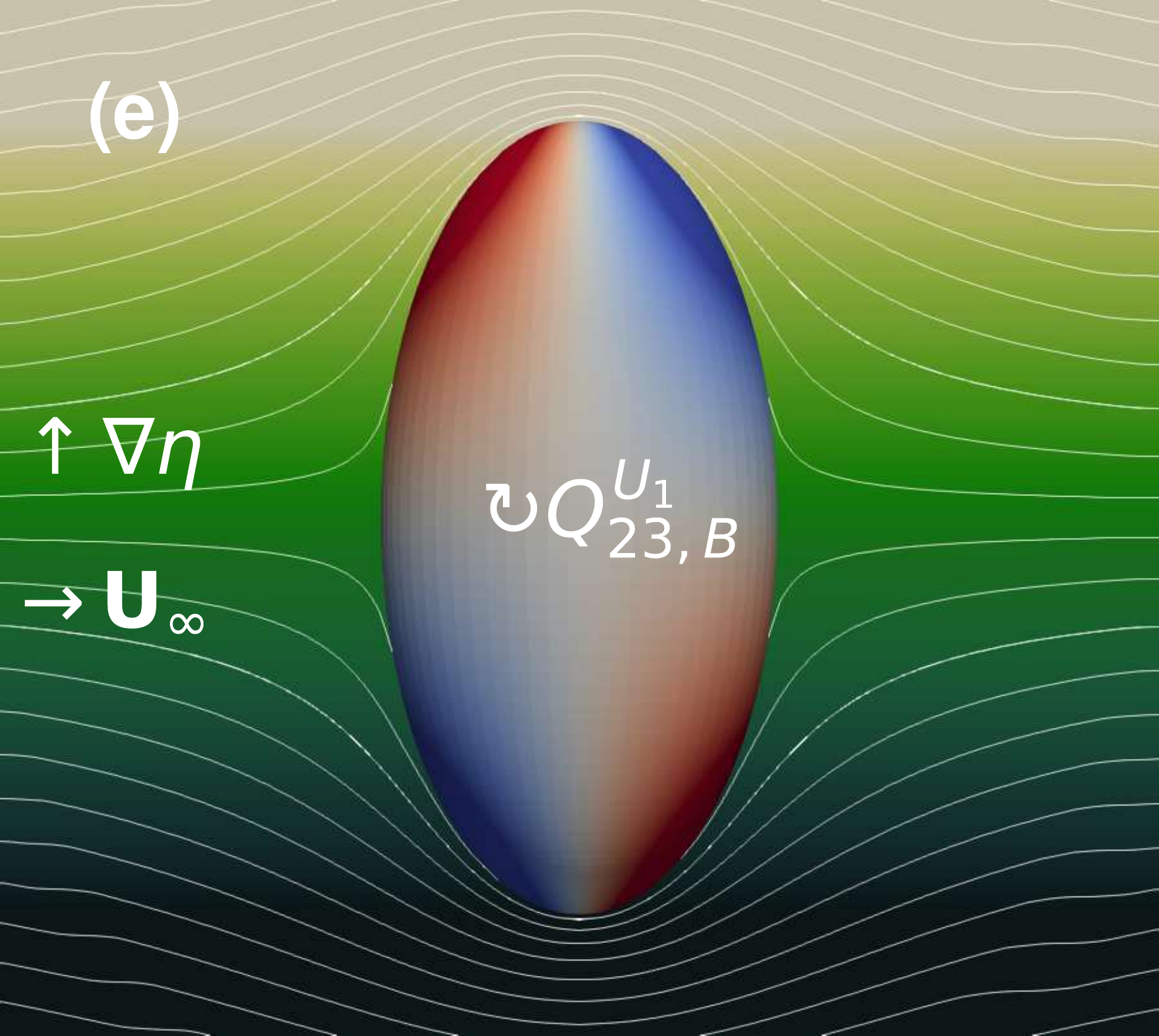} \label{StratPressureAR2Q23U1} }
	\subfloat{\includegraphics[width=0.3\textwidth]{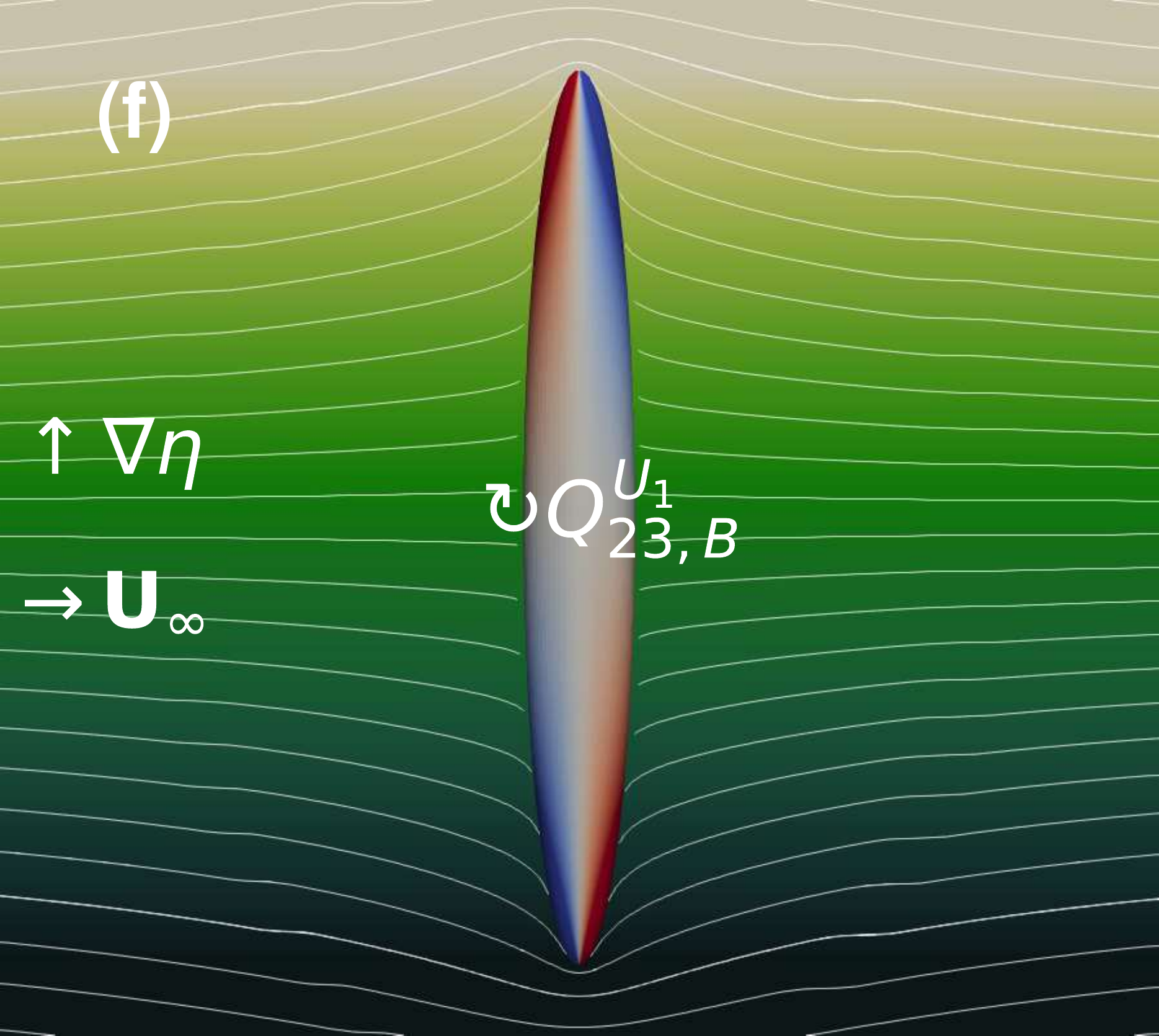} \label{StratPressureAR8Q23U1}}
	\subfloat{\includegraphics[width=0.1\textwidth]{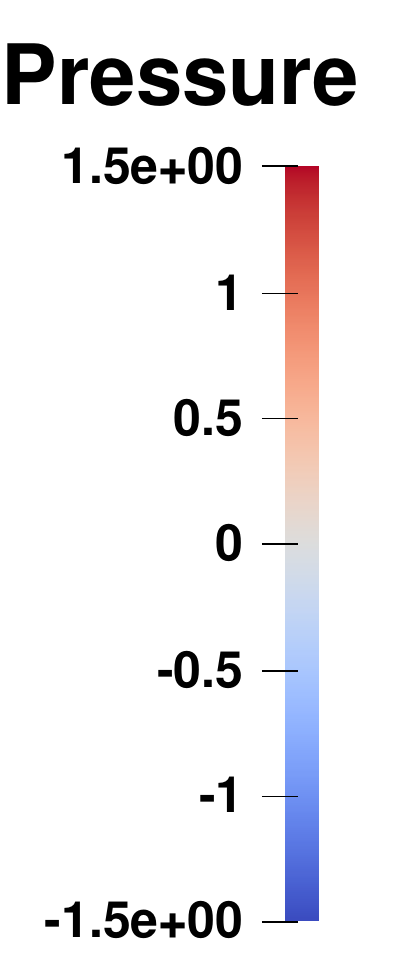}}
	\caption {(a) to (c) Stokes pressure $p^\text{Stokes}$ in uniform flow on a particle surface leading to $Q_{23, A}^{U_1}$ via $\eta'/\eta_0p^\text{Stokes}$ on (left) $\kappa=1$, (middle) $\kappa=2$ and (right) $\kappa=8$ prolate spheroid. (d) to (e) Stratification-induced pressure $p^\text{Stratified}/\beta$ leading to $Q_{23, B}^{U_1}$ for the same $\kappa$ as (a) to (c) respectively. Flow ($\mathbf{u}_\infty$) and stratification ($\nabla\eta$) are respectively perpendicular and parallel to the particle axis of symmetry. Background contours show viscosity variation, $\eta'$ with light (dark) green representing $\eta'>0$ ($\eta'<0$). Streamlines are of the Stokes flow, $\mathbf{u}^\text{Stokes}$. Red (blue) indicates a positive (negative) surface pressure ($p^\text{Stokes}$ or $p^\text{Stratified}$). \label{fig:StokesPresQ23U1}}
\end{figure}
\begin{figure}
	\centering
	\subfloat{\includegraphics[width=0.3\textwidth]{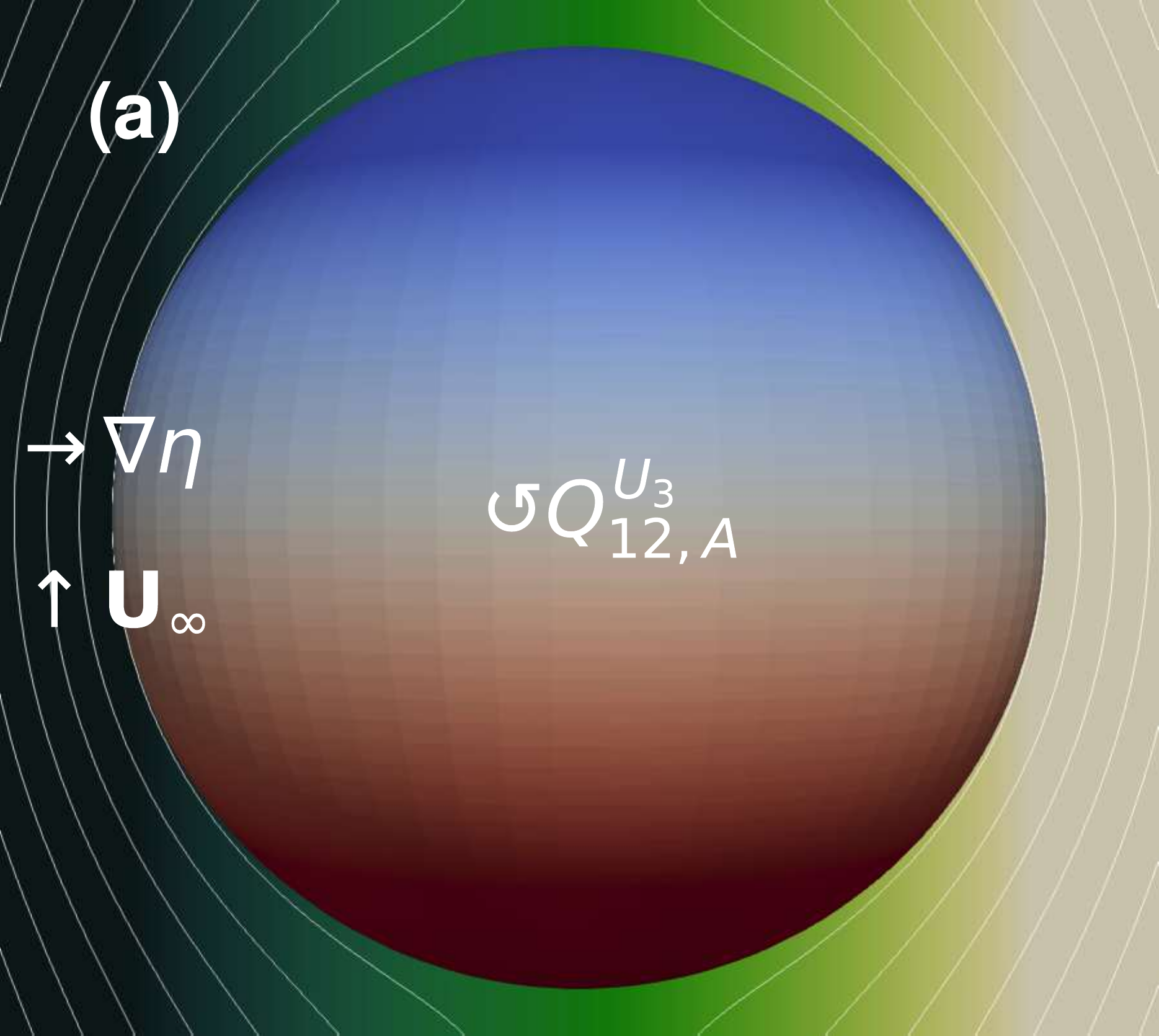} \label{StokesPressureAR1Q12U3}}
	\subfloat{\includegraphics[width=0.3\textwidth]{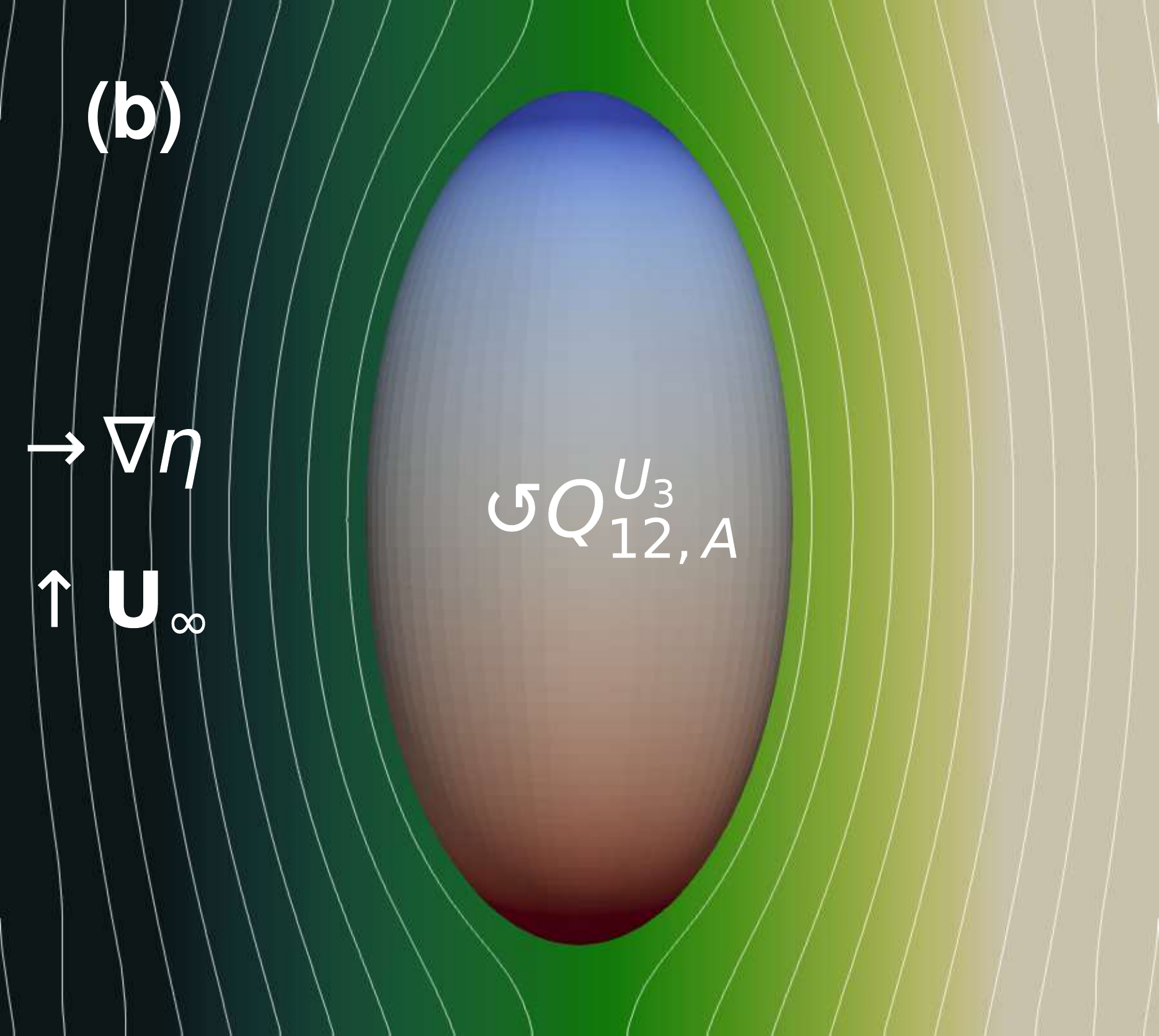}\label{StokesPressureAR2Q12U3} }
	\subfloat{\includegraphics[width=0.3\textwidth]{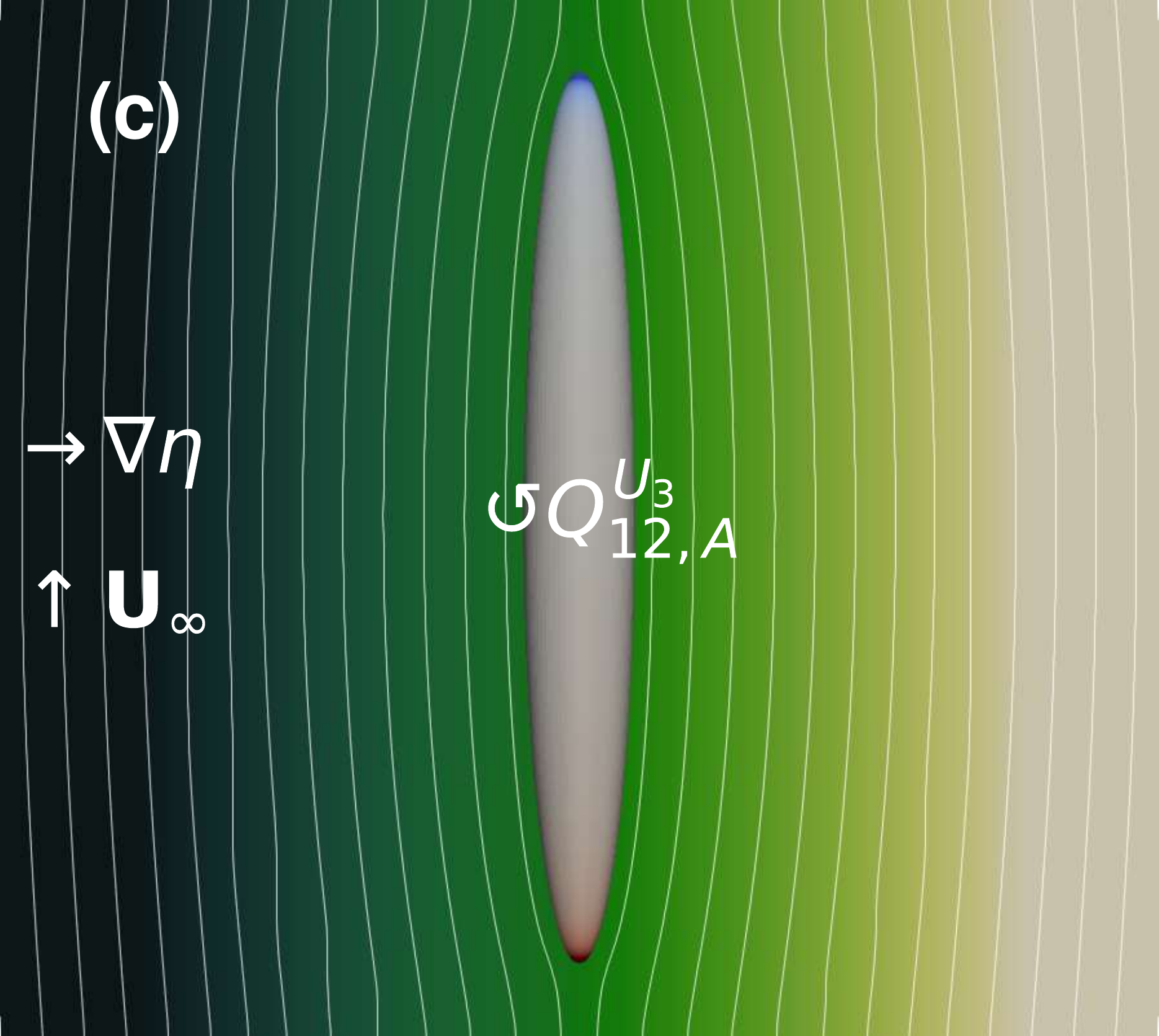}\label{StokesPressureAR8Q12U3} }
	\subfloat{\includegraphics[width=0.1\textwidth]{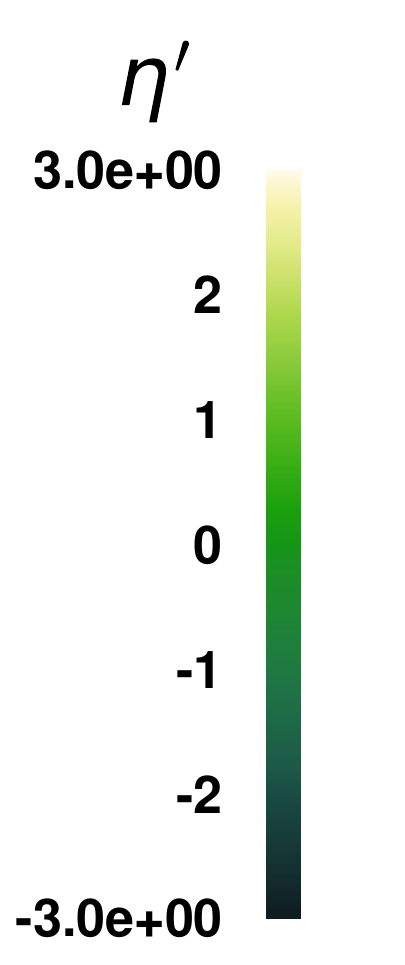}}\\
	\subfloat{\includegraphics[width=0.3\textwidth]{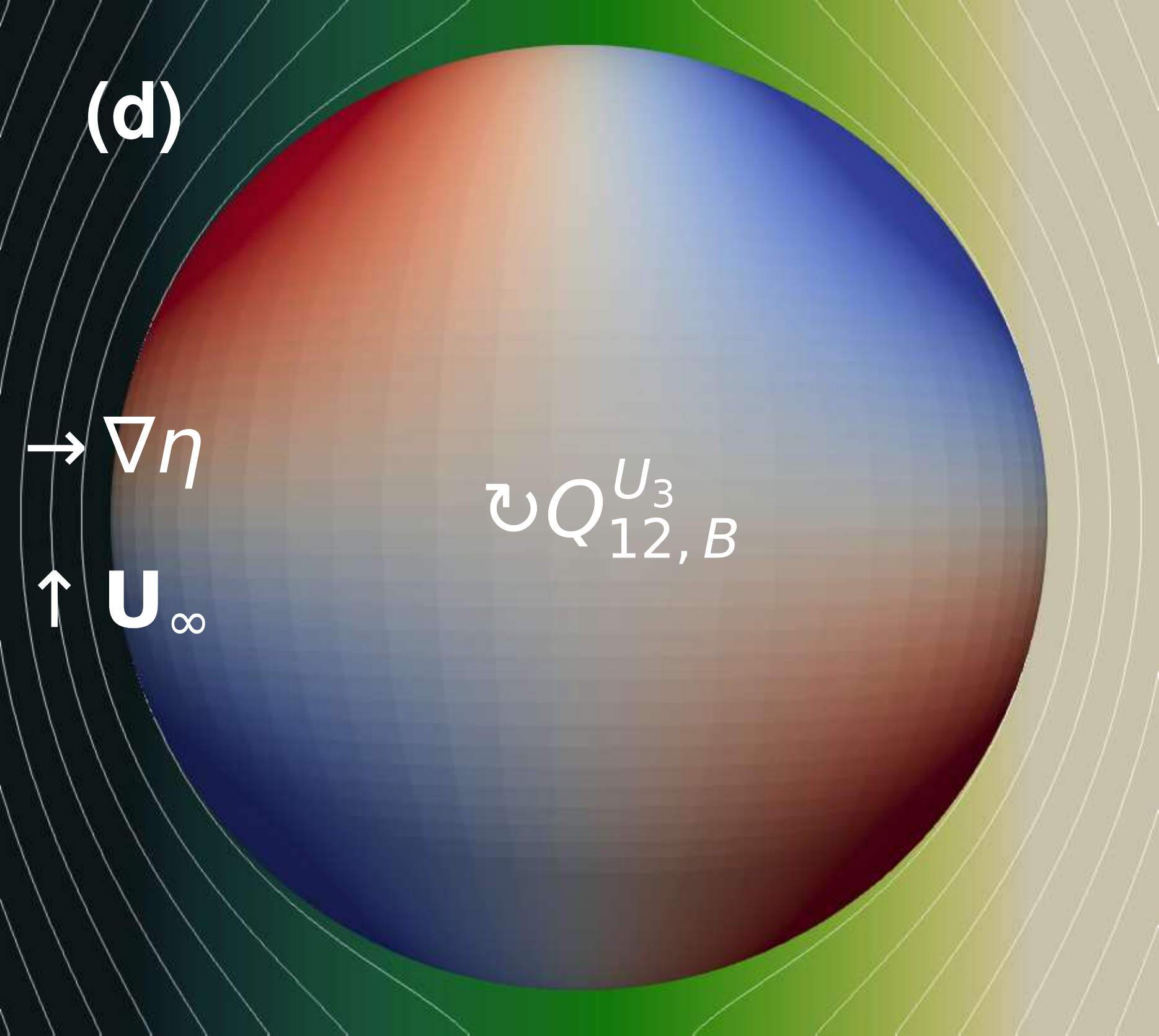} }
	\subfloat{\includegraphics[width=0.3\textwidth]{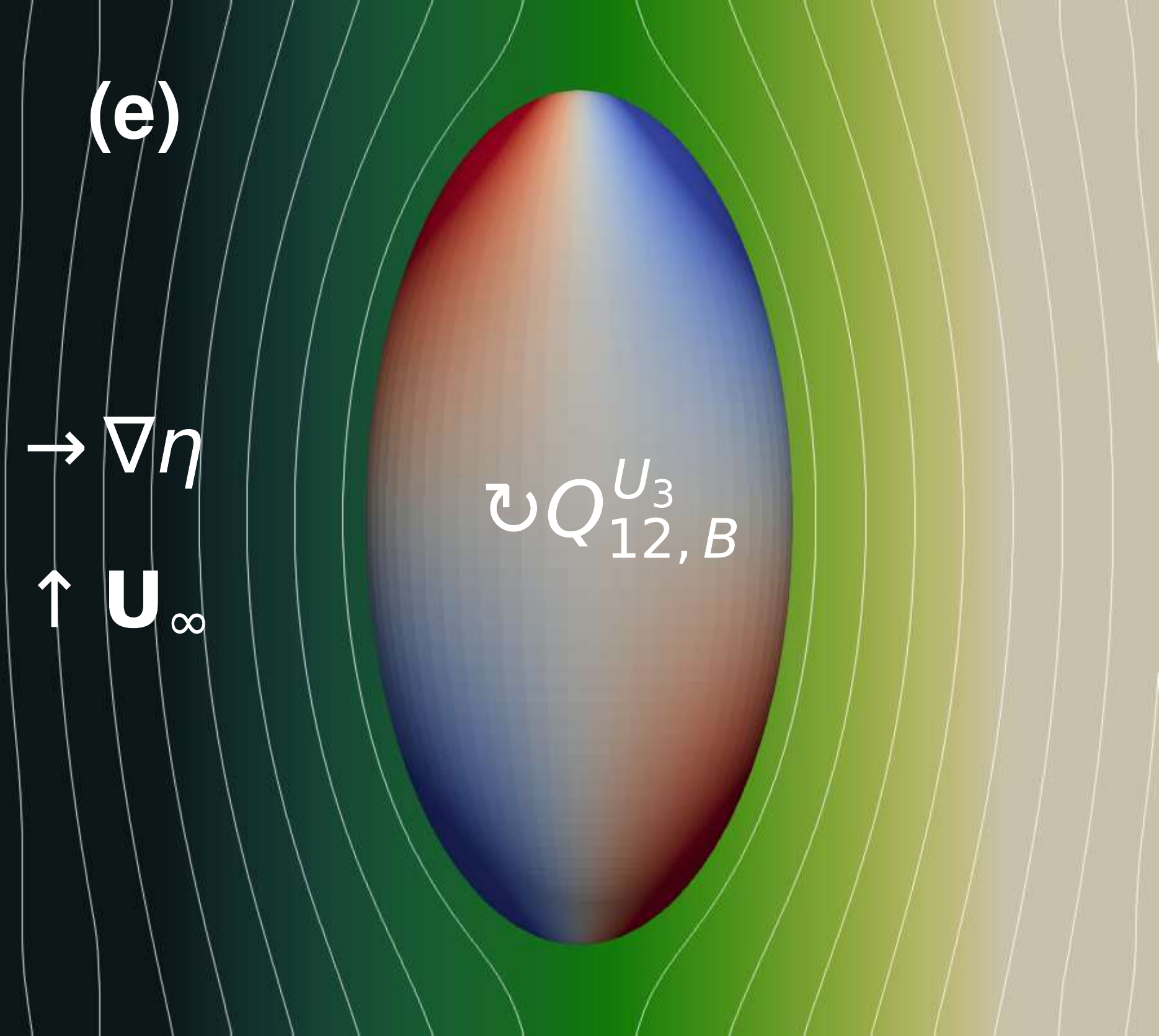} }
	\subfloat{\includegraphics[width=0.3\textwidth]{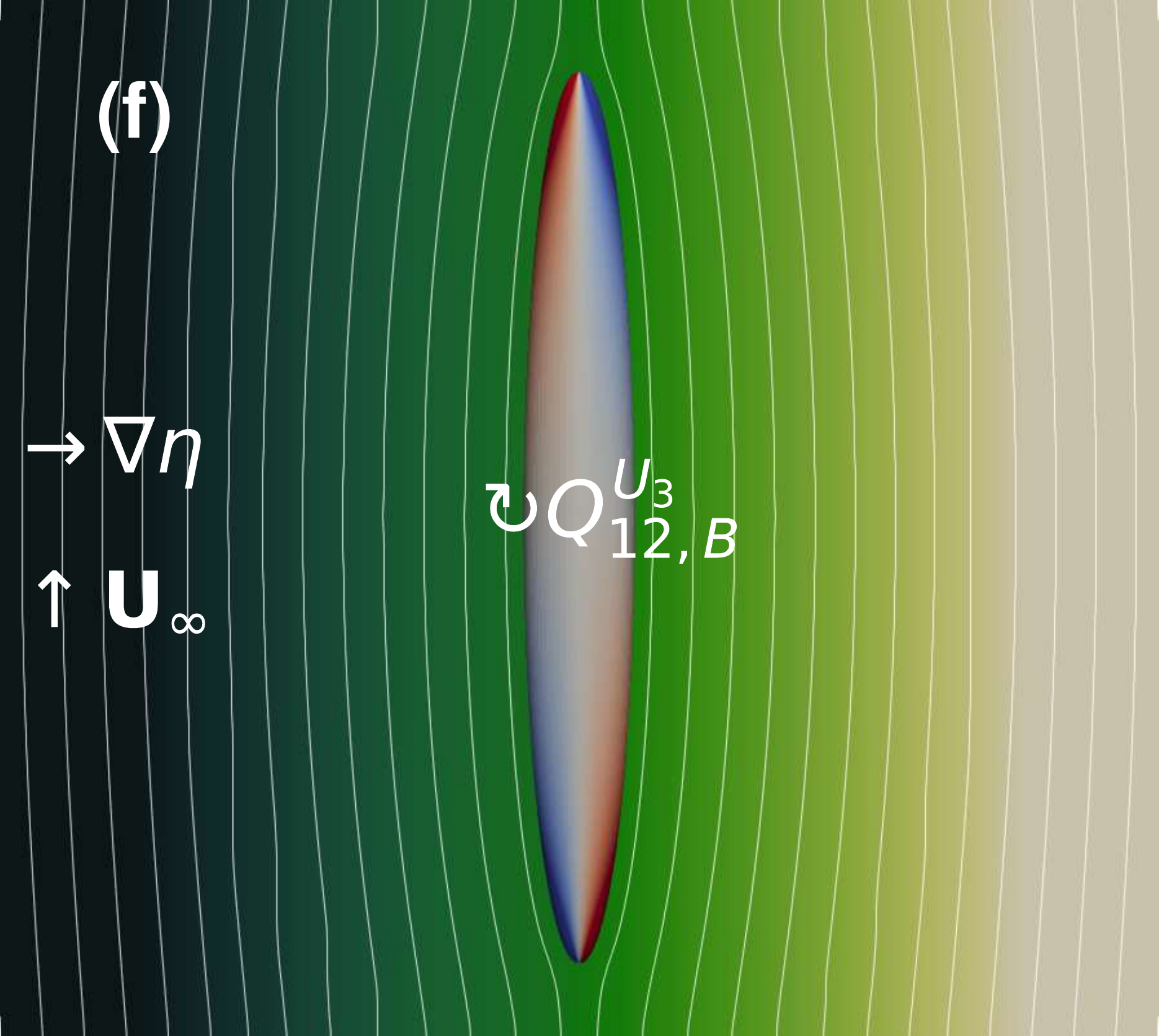} }
	\subfloat{\includegraphics[width=0.1\textwidth]{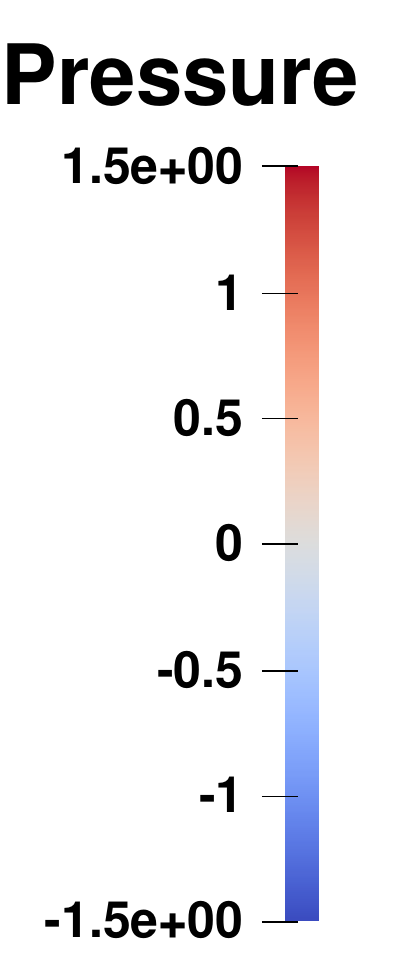}}
	\caption {Same legend as figure \ref{fig:StokesPresQ23U1} but for torques $Q_{12, A}^{U_3}$ and $Q_{12, B}^{U_3}$, i.e., flow ($\mathbf{u}_\infty$) and stratification ($\nabla\eta$) are respectively parallel and perpendicular to the particle axis of symmetry. \label{fig:StokesPresQ12U3}}
\end{figure}
First consider the pressure distribution responsible for torque $Q_{23,B,\text{Pressure}}^{U_1}$ (and $Q_{23,A,\text{Pressure}}^{U_1}$), i.e., for the scenario when imposed flow and stratification are perpendicular and parallel to $\mathbf{p}$ respectively. Figure \ref{fig:StokesPresQ23U1} shows the pressures $p^\text{Stokes}$ and $p^\text{Stratified}$ on $\kappa=1$, 2 and 8 prolate spheroids along with the background viscosity variation, $\eta'$. Here, the variable component of viscosity, $\eta'=0$ at the center of the particle. For every $\kappa$, the Stokes pressure, $p^\text{Stokes}$ is positive on the particle surface facing the flow. On this flow-facing/ upstream side of the particle, the stratification-induced pressure, $p^\text{Stratified}$ is positive when $\eta'>0$ (second quadrant) and negative when $\eta'<0$ (third quadrant). On the downstream side, $p^\text{Stokes}$, is negative and the sign of $p^\text{Stratified}$ is also reversed from the upstream side. Overall, the $p^\text{Stratified}$ distribution remains qualitatively similar as the particle aspect ratio is altered (figure \ref{fig:StokesPresQ23U1}). The force due to pressure acts along the particle's center for a sphere leading to a zero moment arm for each surface element or a torque per unit area of zero, i.e. $\mathbf{x}\times \mathbf{n}=0$ on a sphere since $\mathbf{x}=l\mathbf{n}$ for the sphere with radius, $l$. As $\kappa$ is changed from 1, the surface normal no longer points towards the particle center and there is a non-zero torque per unit area due to pressure. Upon increasing $\kappa$ from 1, the magnitude of $\mathbf{x}\times \mathbf{n}$ near the particle ends increases and the peak locations of $|p^\text{Stratified}|$ move towards the particle ends (figures \ref{StratPressureAR1Q23U1}, \ref{StratPressureAR2Q23U1} and \ref{StratPressureAR8Q23U1}) leading to an initial increase in $Q_{23,B,\text{Pressure}}^{U_1}$ with $\kappa$ in figure \ref{fig:TorquesSpheroidDecompB}. However, the decreasing surface area and reduction in $|p^\text{Stratified}|$, upon increasing $\kappa$ for a fixed major axis length, causes the final decrease of $Q_{23,B,\text{Pressure}}^{U_1}$ with $\kappa$. Therefore, a maximum in $Q_{23,B,\text{Pressure}}^{U_1}$ is observed at $\kappa\lessapprox2$ in figure \ref{fig:TorquesSpheroidDecompB}. The pressure distribution is such that the torque $Q_{23,B,\text{Pressure}}^{U_1}$ is along $\nabla \eta \times \mathbf{u}_\infty$ or clockwise in the view shown in figures \ref{fig:StokesPresQ23U1} and is depicted as positive for all $\kappa>1$ in figure \ref{fig:TorquesSpheroidDecompB}.

While figure \ref{fig:StokesPresQ23U1} depicted the case where the viscosity gradient direction $\mathbf{d}$ and the particle's centerline $\mathbf{p}$ are aligned, and both perpendicular to $\mathbf{u}_\infty$, figure \ref{fig:StokesPresQ12U3} shows the case where $\mathbf{u}_\infty$ and $\mathbf{p}$ are aligned, and both are perpendicular to $\mathbf{d}$. In other words the distinction between figures \ref{fig:StokesPresQ12U3} and \ref{fig:StokesPresQ23U1} is that the viscosity stratification, $\nabla \eta$, and flow, $\mathbf{u}_\infty$, directions are swapped.
The distribution of $p^\text{Stratified}$ shown in the bottom panels of figure \ref{fig:StokesPresQ12U3} that leads to $Q_{12,B,\text{Pressure}}^{U_3}$ shown in figure \ref{fig:TorquesSpheroidDecompB} corresponds to the scenario when flow and stratification are qualitatively similar to the $p^\text{Stratified}$ distributions shown in the bottom panel of figure \ref{fig:StokesPresQ23U1} that are discussed above. Therefore, while $Q_{23,B,\text{Pressure}}^{U_1}$ along the direction $\nabla \eta \times \mathbf{u}_\infty$, $Q_{12,B,\text{Pressure}}^{U_3}$ is along $\mathbf{u}_\infty\times\nabla \eta$, leading to an opposite sign of $Q_{23,B,\text{Pressure}}^{U_1}$ and $Q_{12,B,\text{Pressure}}^{U_3}$ for each $\kappa$ shown in figure \ref{fig:TorquesSpheroidDecompB}. The pressure distributions $p^\text{Stokes}$ shown in the top panels of figures \ref{fig:StokesPresQ23U1} and \ref{fig:StokesPresQ12U3} are such that the torque due to $(\eta'/\eta_0)p^\text{Stokes}$ is positive in both cases, which leads to a positive contribution to $Q_{23,A}^{U_1}$ and $Q_{12,A}^{U_3}$ in figure \ref{fig:TorquesSpheroidDecompA}.

\subsection{Linear flows}\label{sec:LinearFlows}
As stated in section \ref{sec:AbstractStratification}, $F_{ij}^{\Gamma_k}$ (for $i,j\in[1,3],k\in[1,8]$) denotes the stratification-induced force along direction ${i}$ in a fluid with viscosity increasing linearly along direction $j$ with $\beta=1$, on a particle with axis of symmetry along direction 3, in linear flows with gradient $\hat{\boldsymbol{\Gamma}}^{k}$. The flows corresponding to various $\hat{\boldsymbol{\Gamma}}^{k}, k\in[1,8]$ labeled in equation \eqref{eq:BasisFlows} are defined relative to the particle centerline directed along a unit vector $\mathbf{p}$. These are: planar extensional flow normal to $\mathbf{p}$ ($k=1$), uniaxial extensional flow with extensional axis along $\mathbf{p}$ ($k=2$), planar straining flow in the plane perpendicular ($k=3$) and parallel ($k=4$ and 5) to $\mathbf{p}$ and purely rotational flows with vorticity directed along ($k=6$) and perpendicular ($k=7$ and 8) to $\mathbf{p}$. Sections \ref{sec:LinearFlowsValidation} and \ref{sec:LinearFlowsMechanics} below discuss the validation and mechanistic origin of $F_{ij}^{U_k}$ (for $i,j\in[1,3],k\in[1,8]$).
\subsubsection{Validation}\label{sec:LinearFlowsValidation}
\begin{figure}
	\centering
	\subfloat{\includegraphics[width=0.47\textwidth]{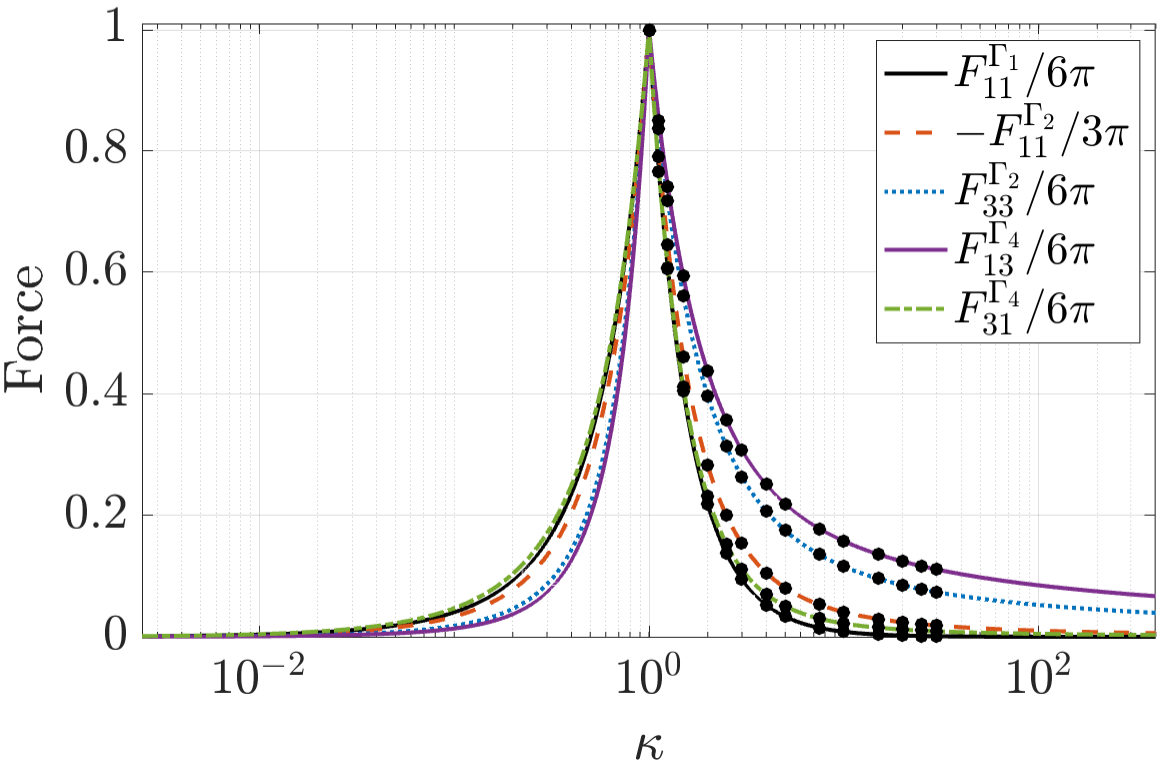}\label{fig:LinearFlowForcesSpheroidA}}
	\subfloat{\includegraphics[width=0.47\textwidth]{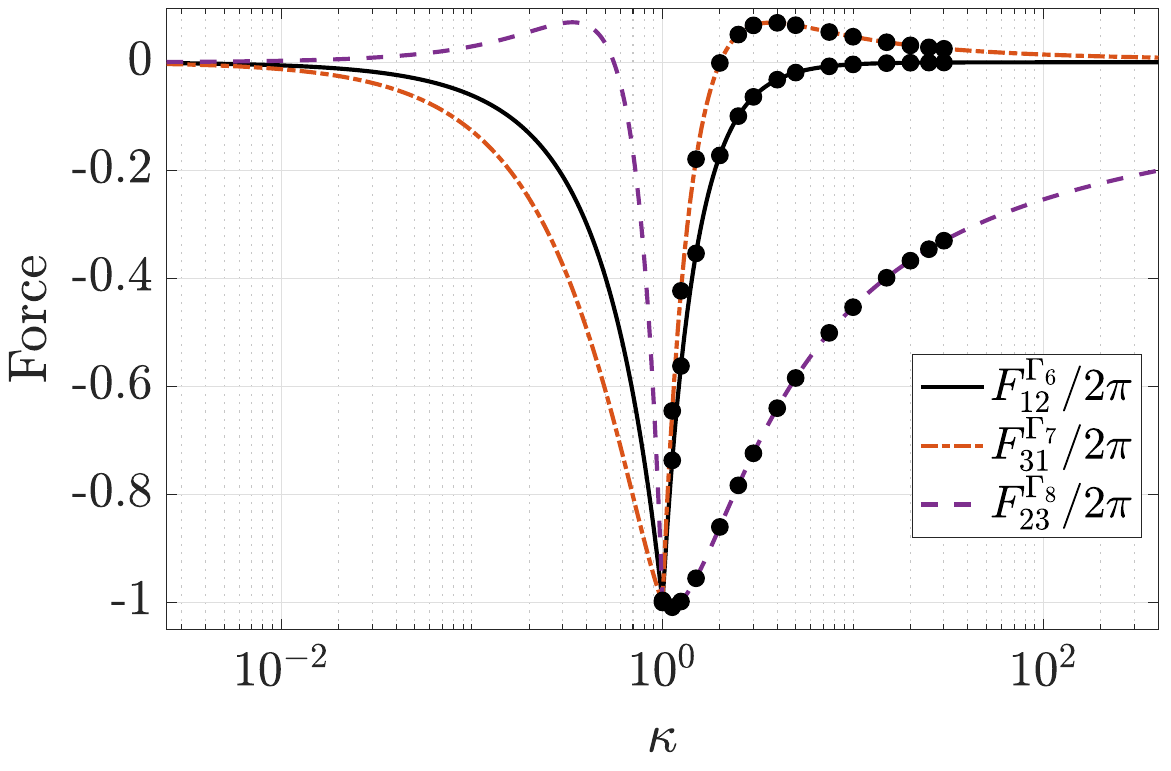}\label{fig:LinearFlowForcesSpheroidB}}
	\caption {Variation of viscosity-stratification-induced forces on a spheroid aspect ratio, $\kappa$, and major axis of 1 in a linear flow. The markers for $1\le\kappa\le30$ are the values obtained from the numerical code. \label{fig:LinearFlowForcesSpheroid}}
\end{figure}

The analytical expressions for the non-zero $F_{ij}^{\Gamma_k},i,j\in[1,3],k\in[1,8]$ are shown in appendix \ref{sec:LinearFlowsAppend} (equation \ref{eq:StratLin}). Solid curves in figure \ref{fig:LinearFlowForcesSpheroid} show the $\kappa$ variation of these forces along with black symbols obtained from the finite difference based numerical solution of equation \eqref{eq:StratificationProblem}. Similar to the torques in the previous section (figure \ref{fig:UniformFlowTorques}) a close match between the symbols and solid curves of figure \ref{fig:LinearFlowForcesSpheroid} obtained from two different techniques serves as a point of validation.
Using resistive force theory for a slender fiber, \cite{kamal2023resistive} obtain the force equivalent to $F_{33}^{\Gamma_2}$ and $F_{23}^{\Gamma_8}$ (their equations 3.28 and 3.32) as $4\pi l^3 /(3\log(2\kappa)-2)$ and $-8\pi l^3 /(3\log(2\kappa)-1)$ with an error of $\mathcal{O}(\log(\kappa)^{-2})$. From our expressions in equation \ref{eq:StratLin}, in the limit of a slender prolate spheroid, i.e., $\kappa\rightarrow\infty$ with a major radius, $l$,
\begin{eqnarray}
	\begin{split}
		&\lim\limits_{\kappa\rightarrow\infty}(F_{11}^{\Gamma_2})=\lim\limits_{\kappa\rightarrow\infty}(F_{31}^{\Gamma_4})=\mathcal{O}((\log(\kappa))^{-2}),\lim\limits_{\kappa\rightarrow\infty}(F_{11}^{\Gamma_1})=\lim\limits_{\kappa\rightarrow\infty}(F_{31}^{\Gamma_7})=\mathcal{O}(\kappa^{-2}),\\&\lim\limits_{\kappa\rightarrow\infty}(F_{12}^{\Gamma_6})=\mathcal{O}(\kappa^{-2}(\log(\kappa))^{-1}), \lim\limits_{\kappa\rightarrow\infty}(F_{23}^{\Gamma_8})=-\frac{8\pi l^3}{3}\frac{1}{\log(2\kappa)-\frac{0.25}{\log(2\kappa)}}+\mathcal{O}(\kappa^{-2}),\\&\lim\limits_{\kappa\rightarrow\infty}(F_{13}^{\Gamma_4})=\frac{32\pi l^3}{3}\frac{\log(2\kappa)}{\log(2\kappa^2)^2+\log(4)\log(2\kappa^2)+\log(2)^2-1}+\mathcal{O}(\kappa^{-2}),\\
		&\lim\limits_{\kappa\rightarrow\infty}(F_{33}^{\Gamma_2})=\frac{4\pi l^3}{3}\frac{1}{\log(2\kappa)\frac{\log(2\kappa)-2}{\log(2\kappa)-1}+\frac{0.75}{\log(2\kappa)-1}}+\mathcal{O}(\kappa^{-2}).
\end{split} \end{eqnarray}
Hence, values from our expressions in the large $\kappa$ limit match those of \cite{kamal2023resistive}. In the limit of the sphere of radius $l$, $\kappa \rightarrow 1$,
\begin{align}
	\begin{split}
		&\lim\limits_{\kappa\rightarrow1}(F_{11}^{\Gamma_1})=-2\lim\limits_{\kappa\rightarrow1}(F_{11}^{\Gamma_2})=\lim\limits_{\kappa\rightarrow1}(F_{33}^{\Gamma_2})=\lim\limits_{\kappa\rightarrow1}(F_{13}^{\Gamma_4})=\lim\limits_{\kappa\rightarrow1}(F_{31}^{\Gamma_4})=6\pi l^3,\\ &\lim\limits_{\kappa\rightarrow1}(F_{12}^{\Gamma_6})=\lim\limits_{\kappa\rightarrow1}(F_{31}^{\Gamma_7})=\lim\limits_{\kappa\rightarrow1}(F_{23}^{\Gamma_8})=-2\pi l^3.
	\end{split}
\end{align}
Substituting these formulae into equation \eqref{eq:StratForceLinear}, we find the force on a fixed sphere in linear flow, with gradient $\boldsymbol{\Gamma}=\mathbf{E}+\boldsymbol{\Omega}$ is,
\begin{equation}
	\mathbf{f}^\text{Stratified}_{\boldsymbol{\Gamma}}=\mathbf{f}_\text{sphere}=2\pi \beta l^3(3\mathbf{E}-\boldsymbol{\Omega})\cdot\mathbf{d}.\label{eq:FSphere}
\end{equation}
This is the same expressions as obtained without spheroidal harmonics but directly from the calculation using the flow fields around a sphere in appendix \ref{sec:AppendixSphereCase} (equations \eqref{eq:SphereForceStratLin} and \eqref{eq:SphereForceStratLin}). This provides another validation of our use of spheroidal harmonics.

\subsubsection{Mechanistic origin of stratification-induced force}\label{sec:LinearFlowsMechanics}
For brevity, we only discuss the mechanisms on a sphere in this section. The signs of most of the stratification-induced-torques on a $\kappa\ne 1$ spheroid are the same as those for a sphere (figure \ref{fig:LinearFlowForcesSpheroid}) and can be explained in a qualitatively similar manner as those for a sphere considered here.

Following the decomposition introduced in equations \eqref{eq:ForceTorqueA} and \eqref{eq:ForceTorqueB}, the stratification-induced force, $\mathbf{f}_\text{sphere}$ from equation \eqref{eq:FSphere} can be split into two parts. The first arises from the Stokes stress acting in a variable-viscosity environment $(\eta'/\eta_0)\boldsymbol{\sigma}^\text{Stokes}$, and its force is, $\mathbf{f}_\text{A,sphere}=4\pi \beta l^3(\mathbf{E}+\boldsymbol{\Omega})\cdot\mathbf{d}$. The second is due to the stratification-induced stress, $\boldsymbol{\sigma}^\text{Stratified}$, and its force is $\mathbf{f}_\text{B,sphere}=\mathbf{f}_\text{sphere}-\mathbf{f}_\text{A,sphere}$. Within $\mathbf{f}_\text{A,sphere}$, the pressure, $(\eta'/\eta_0)p^\text{Stokes}$ contributes an amount $\mathbf{f}_\text{A,sphere}^\text{Pressure}=(8\pi/3)\beta l^3\mathbf{E}\cdot\mathbf{d}$ and the remaining $(4\pi/3)\beta l^3(\mathbf{E}+3\boldsymbol{\Omega})\cdot\mathbf{d}$ arises from the viscous stress, $2\eta'\mathbf{e}^\text{Stokes}$. A freely suspended sphere's rotation effectively negates the rotation part ($\boldsymbol{\Omega}$) of the imposed linear flow and the sign of the force arising from the straining part ($\mathbf{E}$) is same from all the decomposed components discussed above ($\boldsymbol{\sigma}^\text{Stratified}$, $2\eta'\mathbf{e}^\text{Stokes}$ and $(\eta'/\eta_0)p^\text{Stokes}$). Therefore, we use the contours of $p^\text{Stokes}$ along with the background viscosity variation to shed light on the origins of the force arising from $(\eta'/\eta_0)p^\text{Stokes}$, i.e., $\mathbf{f}_\text{A,sphere}^\text{Pressure}=(8\pi/3)\beta l^3\mathbf{E}\cdot\mathbf{d}$ in uniaxial extension and simple shear flows. The other components of the stratification-induced force in straining flows follow a similar mechanism.

\begin{figure}
	\centering
	\subfloat{\includegraphics[width=0.4\textwidth]{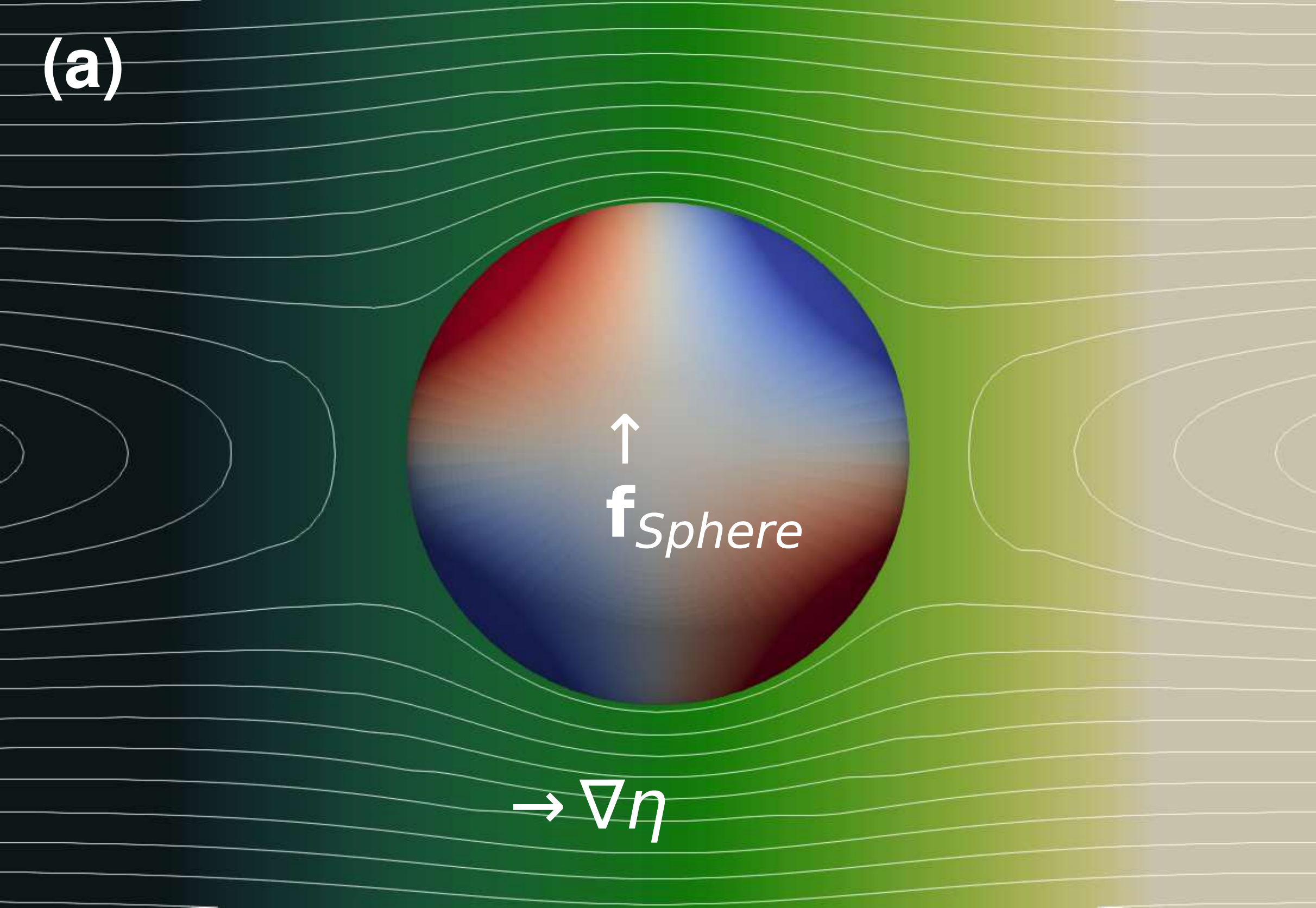}\label{ShearForceSphere}} \hfill
	\subfloat{\includegraphics[width=0.4\textwidth]{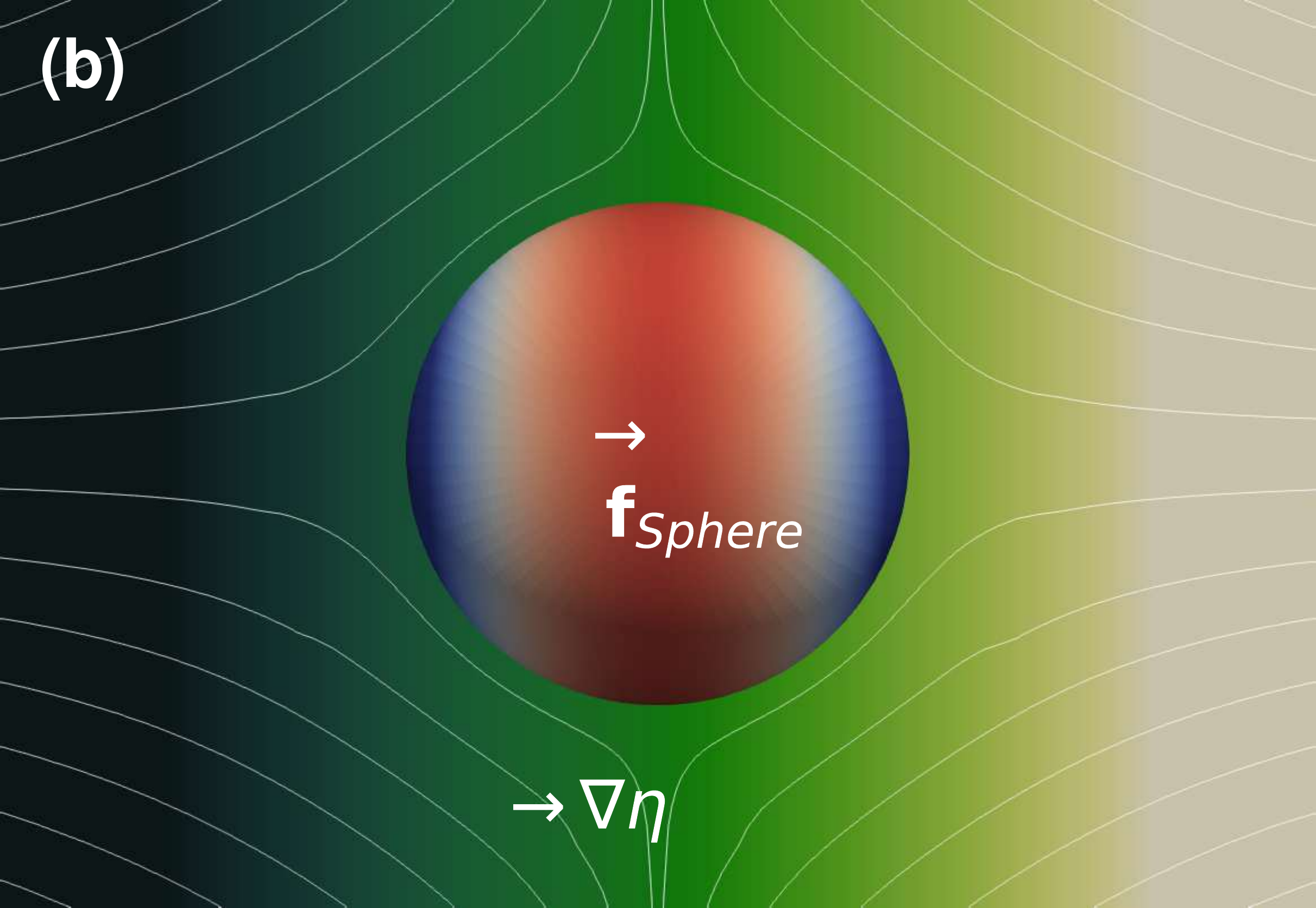}\label{ExtensionalForceSphere}}
	\subfloat{\includegraphics[width=0.17\textwidth]{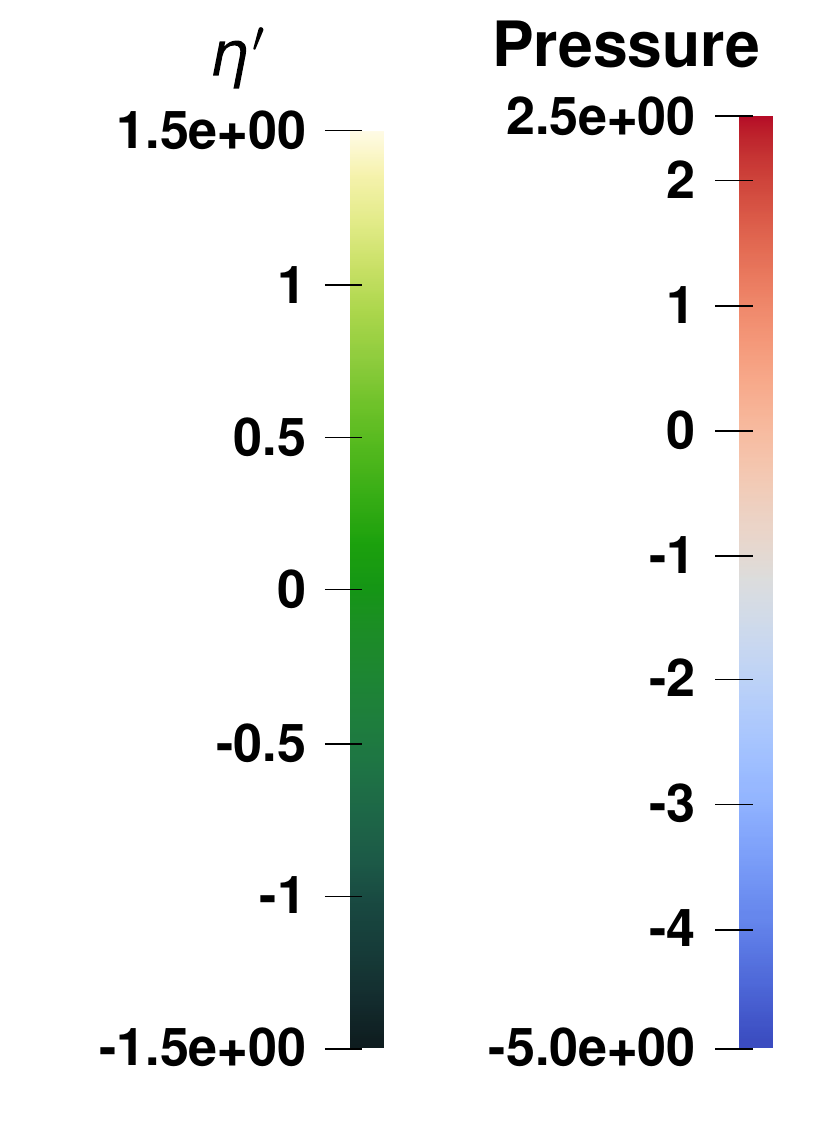}}
	\caption {Stokes pressure $p^\text{Stokes}$ on the particle surface in (a) simple shear flow and (b) uniaxial extensional flow with viscosity increasing along the horizontal direction. Red (blue) represents a positive (negative) $p^\text{Stokes}$ and light (dark) green represents higher (lower) $\eta'(=\eta-\eta_0)$. Streamlines are of the Stokes velocity field for the respective flows. In (a) the imposed shear flow is towards the right on the top and left on the bottom half of the figure. The extensional flow in (b) is axisymmetric about the horizontal/ extensional axis. It approaches the particle's center along the vertical (compression plane) and leaves it along the extensional axis. \label{fig:SchematicsLinearFlow}}
\end{figure}
Consider a simple shear flow with strain rate $E_{ij}=0.5(\delta_{i2}\delta_{j1}+\delta_{j2}\delta_{i1})$ such that 1 is the flow, 2 the velocity gradient and 3 the vorticity direction such that, $\mathbf{E}\cdot\mathbf{d}=0.5\begin{bmatrix}
	d_2&d_1&0
\end{bmatrix}^T$. A sphere in simple shear flow of uniform viscosity fluid is force-free. However, if the fluid's viscosity increases along the velocity gradient direction ($d_1=0$, $d_2=1$ and $d_3=0$), the particle will experience a force along the flow direction (perpendicular to the viscosity gradient). Alternatively, if the viscosity increment is along the flow direction ($d_1=1$, $d_2=d_3=0$), the stratification-induced force is in the velocity gradient direction (again perpendicular to viscosity gradient). The contribution of this force coming from $\eta'p^\text{Stokes}$ for a horizontal (flow direction) viscosity stratification can be understood through figure \ref{ShearForceSphere}. Relative to the mean, $p^\text{Stokes}$ does not create a force in a constant viscosity fluid as this pressure is both left-right and top-down anti-symmetric. The pressure $(\eta'/\eta_0)p^\text{Stokes}$ is also top-down anti-symmetric, but left-right symmetric. This leads to a net force upwards, i.e., across the streamlines of imposed flow. Similarly if viscosity increases upwards instead ($d_1=0$, $d_2=1$, $d_3=0$), $\eta'p^\text{Stokes}$, is left right anti-symmetric, but top down symmetric creating a force towards the flow direction. The pressure distribution is the same on the back half of the sphere (not shown). Therefore, if viscosity stratification is entirely along the vorticity direction (perpendicular to the plane of the picture) of the imposed simple shear flow the stratification-induced force is zero.

Stratification also leads to a force on an otherwise force-free sphere in uniaxial extensional flow (strain rate $E_{ij}=\delta_{i1}\delta_{j1}-0.5(\delta_{i2}\delta_{j2}+\delta_{i3}\delta_{j3})$). If the viscosity increases along the extensional axis this force is towards the higher viscosity region. This can be explained through the $p^\text{Stokes}$ contours on the particle surface along with background $\eta'$ shown in figure \ref{ExtensionalForceSphere}. The distribution of $p^\text{Stokes}$ on the particle surface is such that it is negative near the extensional axis and positive near the compression plane. Despite local compressive and extensional forces the net force is zero, so a sphere does not move in a constant viscosity fluid. However, if viscosity increases towards the right as in figure \ref{ExtensionalForceSphere}, such that the two horizontal ends that are pulling the surface of the sphere outwards are in different viscosity environments, the force due to pressure $(\eta'/\eta_0)p^\text{Stokes}$ is towards the right. Similarly, if viscosity increases upwards and considering that $p^\text{Stokes}$ pushes the particle inwards on top and bottom, $(\eta'/\eta_0)p^\text{Stokes}$ causes the particle to go downwards. Therefore, in a uniaxial extensional flow if the viscosity gradient lies along the compression (extensional) direction, the sphere will move towards the lower (higher) viscosity region. A sphere freely suspended in uniaxial extensional flow of uniform viscosity fluid has a saddle fixed point at the origin of the imposed flow. However, for the case of viscosity stratified fluid, the saddle point for the particle trajectory is shifted towards the lower viscosity fluid relative to the stagnation point of the imposed flow. The effect of stratification induced pressure (not shown) is similar to that of $(\eta'/\eta_0)p^\text{Stokes}$ described above through figure \ref{fig:SchematicsLinearFlow} for both shear and extensional flow.

In the next two sections, we will consider the motion of a freely suspended spheroid in settling due to gravity and a freely suspended neutrally buoyant particle in a linear flow field due to stratification-induced forces and torques. 

\section{Freely sedimenting spheroids in viscosity gradients}\label{sec:SpheroidsSedimenting}
\subsection{{Spheres ($\kappa=1$)}}\label{sec:SphereSediment}
The rotational and translational velocities of a sedimenting sphere in a linearly stratified fluid are coupled via equations \eqref{eq:Sedimentuvel} and \eqref{eq:angvelg} which in the limit $\kappa\rightarrow1$ leads to angular and translation velocities $\boldsymbol{\omega}_{\text{particle}}$ and $\mathbf{u}_{\text{particle}}$ given by
	\begin{equation}
		\boldsymbol{\omega}_{\text{particle}}=\beta\frac{\mathbf{g}\times \mathbf{d}}{24\pi l \eta_0^2}+\mathcal{O}(\beta^2),	\hspace{0.2in} \mathbf{u}_{\text{particle}}=\frac{\mathbf{g}}{6\pi l \eta_0}-\beta\frac{l^2}{3\eta_0}\boldsymbol{\omega}_{\text{particle}}\times \mathbf{d}+\mathcal{O}(\beta^2)=\frac{\mathbf{g}}{6\pi l \eta_0}+\mathcal{O}(\beta^2).\label{eq:SphereSettle}
	\end{equation}
	From the $\mathcal{O}(\beta)$ analysis conducted here we observe that the sedimenting velocity of the sphere in a linearly stratified fluid does not change from that in a uniform viscosity fluid. However, at $\mathcal{O}(\beta)$ the stratification induces a rotation to the particle. A fluid rotating relative to a sphere at an angular velocity $\boldsymbol{\omega}_\text{flow}$ experiences a stratification induced force proportional to $\beta \boldsymbol{\omega}_\text{flow}$ (equation \eqref{eq:StratForceLinear} and figure \ref{fig:LinearFlowForcesSpheroidB}). The relative rotation here arises at $\mathcal{O}(\beta)$, and part of the stratification induced velocity at $\mathcal{O}(\beta^2)$ lies along $\boldsymbol{\omega}_{\text{particle}}\times \mathbf{d}\propto \mathbf{d(d} \cdot\mathbf{g})$. In other words, beyond the formally valid $\mathcal{O}(\beta)$ effects from the calculation conducting here, we may expect the velocity of a sedimenting sphere to have a finite component proportional to $\beta^2 d_g \mathbf{d}$, where the parameter, 
	\begin{equation}
		d_g=\frac{1}{|| \mathbf{g}||_2}\mathbf{d}\cdot\mathbf{g},\label{eq:dg}
	\end{equation}
	measures the alignment between gravity and the viscosity variation direction ($d_g=$ 0, -1 and 1 imply that viscosity increases perpendicular, opposite and towards the gravity direction, respectively). This parameter is qualitatively important in discussion throughout section \ref{sec:SpheroidsSedimenting} where it plays a role in the $\mathcal{O}(\beta)$ effect on spheroids. For spheres, when $d_g\ne 0$ and $\mathbf{d}$ is not aligned with gravity $\mathbf{g}$, we may expect that a sedimenting sphere will fall in a curved path instead of straight line along gravity. Quantitative conclusions about this change in the sphere's trajectory can not be made from the $\mathcal{O}(\beta)$ calculation conducted in this paper. Two spheres in an inertia-less fluid with uniform velocity fall with no relative motion. However, another impact of the $\mathcal{O}(\beta)$ rotation rate induced by stratification is likely to be a change in the relative motion of the spheres.

\subsection{Spheroids with $\kappa\ne 1$}\label{sec:SpheroidSediment}
We discussed above that a {sedimenting sphere (a spheroid with $\kappa=1$) starts rotating due to viscosity stratification at $\mathcal{O}(\beta)$ and may undergo  a horizontal drift at $\mathcal{O}(\beta^2)$. However, a similar stratification induced rotation on a spheroid with $\kappa\ne 1$ leads to change in settling behavior at $\mathcal{O}(\beta)$,} as its settling velocity (even in a constant viscosity fluid) depends on its centerline orientation, $\mathbf{p}$. As a consequence, for a non-spherical spheroid, the rotational-translational coupling due to stratification leads to novel sedimenting behavior even at $\mathcal{O}(\beta)$ obtained from equation \eqref{eq:NewtSedimentvel} and \eqref{eq:rotationratefull}. We consider $\eta_0=1$ in the results presented below. The non-linear dynamics defined by the equation \eqref{eq:rotationratefull} for the particle rotation suggests a rich set of behaviors where, depending upon the values of $t_2$ and $t_3$ defined in equation \eqref{eq:t1t2t3} and the parameter $d_g$ (equation \eqref{eq:dg}), the orientation phase space has neutral orbits, spirals and fixed points (saddle, stable and unstable). Our complete analysis of these dynamics is given in appendix \ref{sec:Rotationatsmallbeta}. We summarize our results below.

\begin{figure}
	\centering
	\subfloat{\includegraphics[width=0.46\textwidth]{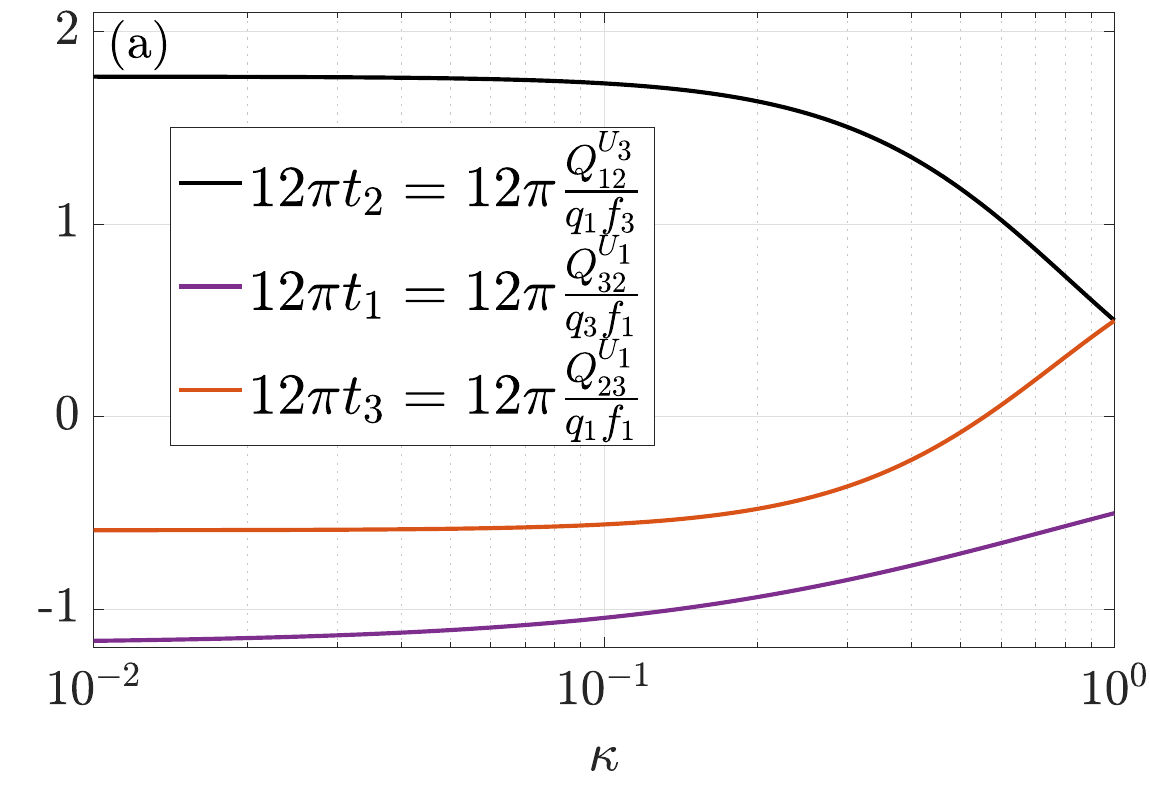}}
	\subfloat{\includegraphics[width=0.46\textwidth]{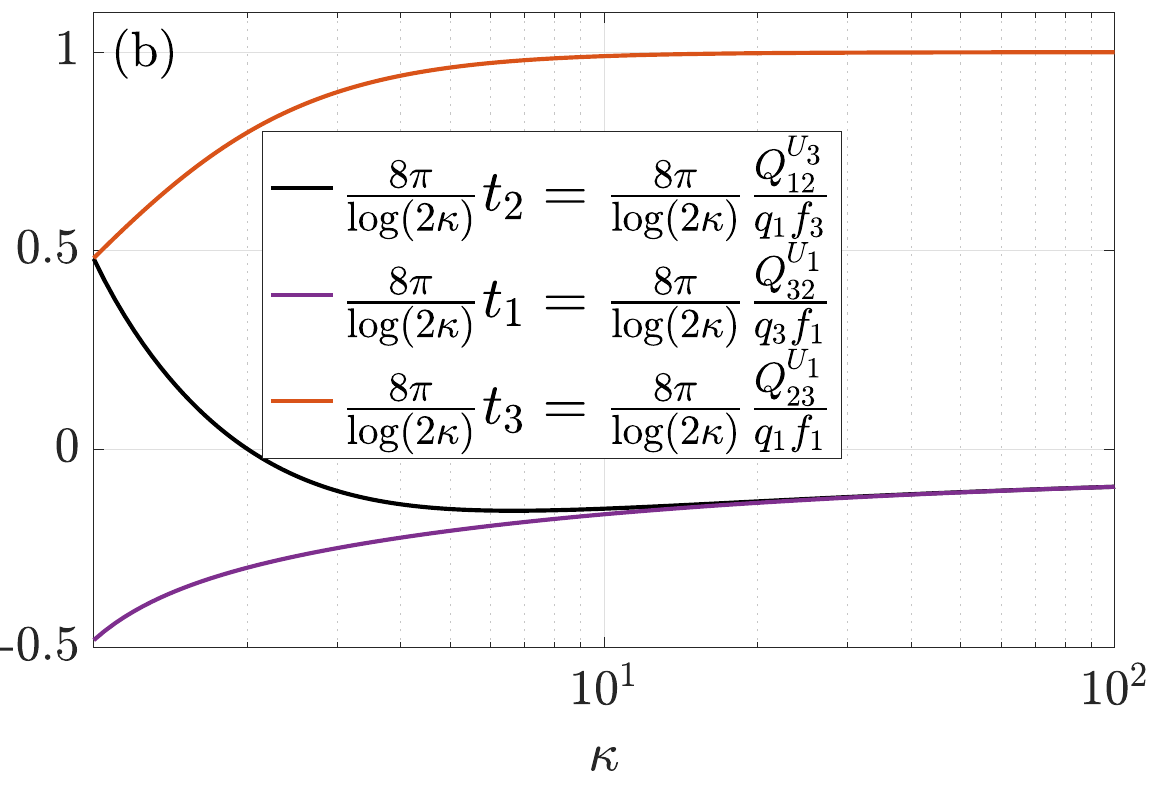}}
	\caption {Variation of shape-dependent parameters $t_i,i\in[1,3]$ (defined in equation \eqref{eq:t1t2t3}) with aspect ratio, $\kappa$, for the rotation of (a) oblate and (b) prolate spheroids during sedimentation in a stratified environment. The parameters $t_2$ and $t_3$ change the particle's orientation, $\mathbf{p}$, at $\mathcal{O}(\beta)$ (equation \eqref{eq:rotationratefull}) analyzed in section \ref{sec:SpheroidSediment}, whereas $t_1$ causes rotation of the particle about it centerline that alters $\mathbf{p}$ if higher orders in $\beta$ are included through full the rotation rate equation \eqref{eq:rotationratefull}. The focal length is chosen such that the major axis of the particle is $l=1$ for each particle type irrespective of $\kappa$. The terms are multiplied with $12\pi$ for oblate and $8\pi/\log(2\kappa)$ for prolate particles. \label{fig:SedimentingTerms}}
\end{figure}

The parameters $t_2$ and $t_3$ are only dependent on the particle aspect ratio $\kappa$, and this variation for a wide range of $\kappa$ is shown in figure \ref{fig:SedimentingTerms} (the expressions for $t_2$ and $t_3$ for a spheroid are provided in equations \eqref{eq:t1t2t3Prolate} and \eqref{eq:t1t2t3Oblate}). We observe that, particle orientation dynamics (and hence the translational dynamics, which is coupled with $\mathbf{p}$) depend only on $\kappa$ and $d_g$. Thus, in $\kappa-d_g$ phase space, we obtain qualitatively different orientation behaviors demarcated by the boundaries shown in figure \ref{fig:PhaseDiagram}.

To describe the particle dynamics in more physical terms than in appendix \ref{sec:Rotationatsmallbeta}, it is useful to define a coordinate system aligned with the gravity vector $\mathbf{g}$ and the stratification direction $\mathbf{d}$. In this coordinate system, two of the three orthogonal axes are within the gravity-stratification (GS) plane: one of the axes in this plane is along the unit vector along gravity, $\hat{\mathbf{g}}$, and along the other axis, $\hat{\mathbf{e}}$, the viscosity gradient is non-decreasing such that $\hat{\mathbf{g}}\cdot\hat{\mathbf{e}}=0$ and $\mathbf{d}\cdot\hat{\mathbf{e}}=||\mathbf{d}\cdot\hat{\mathbf{e}}||_2=\sqrt{1-d_g^2}>0$. The third orthogonal axis is normal to the gravity-stratification (GS) plane and is defined by the vector $\hat{\mathbf{g}}\times\hat{\mathbf{e}}$. Thus, the gravity and stratification directions can be expressed as
\begin{eqnarray}
	\mathbf{g}=||\mathbf{g}||\hat{\mathbf{g}},\hspace{0.05in}\mathbf{d}=d_g\hat{\mathbf{g}}+\sqrt{1-d_g^2}\hat{\mathbf{e}},\label{eq:CoordinateSystem1}
\end{eqnarray}
and the particle orientation vector is expressed in this newly defined coordinate system as
\begin{eqnarray} \mathbf{p}=p_g\hat{\mathbf{g}}+p_e\hat{\mathbf{e}}+\sqrt{1-p_g^2-p_e^2}\hat{\mathbf{g}}\times\hat{\mathbf{e}}.\label{eq:CoordinateSystem2}
\end{eqnarray}
The invariant objects (neutral orbits, spirals, limit cycles and stable/ unstable/ saddle fixed points) of the dynamical system either lie on the $\hat{\mathbf{g}}\times\hat{\mathbf{e}}$ axis or within the GS plane. The sign of the parameter $s=d_g(t_3-t_2)$ also plays a key role. The key points from appendix \ref{sec:Rotationatsmallbeta} (the different regions, $L_1$, $L_2$ and $R_i,i\in[1,6]$, are shown in figure \ref{fig:PhaseDiagram}) are:
\begin{itemize}
	\item Region $L_1$: When gravity and stratification are perpendicular, i.e., $d_g=0$, and $t_2t_3>0$, the particle's orientation follows a non-uniform neutral orbit with a time period $2\pi/(\sqrt{t_2t_3}||\mathbf{g}||_2\beta/\eta_0^2)$. This is similar to the Jeffery orbits observed for particles with Bretherton constant $|\lambda|<1$ in simple shear flows.
	\item Region $L_2$: If gravity and stratification are collinear, i.e. $d_g=\pm1$, the particle aligns with gravity if $s<0$ and perpendicular to gravity if $s>0$.
	\item Regions $R_1=GS_\text{spiral}^\perp$ and $R_2=GS_\text{spiral}^\parallel$: For $t_2t_3>0$ and $0<|d_g|<|d_g|_\text{spiral}=2\sqrt{t_2t_3}/t_2+t_3$, the GS plane is a limit cycle (stable if $s>0$ and unstable otherwise) and the $\hat{\mathbf{g}}\times\hat{\mathbf{e}}$ axis is a spiral (stable if $s<0$ and unstable otherwise). Therefore, the particle's axis of symmetry spirals towards the gravity-stratification (GS) plane if $s<0$ ($R_2$) or towards the $\hat{\mathbf{g}}\times\hat{\mathbf{e}}$ axis if $s>0$ ($R_1$).
	\item Regions $R_3=GS_1^\perp$ and $R_4=GS_3^\parallel$: When $t_2t_3>0$ and $|d_g|>|d_g|_\text{spiral}$, the $\hat{\mathbf{g}}\times\hat{\mathbf{e}}$ axis is a stable node when $s>0$ and unstable otherwise. Hence, for $s>0$, the particle aligns along the $\hat{\mathbf{g}}\times\hat{\mathbf{e}}$ axis irrespective of the initial condition ($R_3$). For $s<0$ ($R_4$), a stable fixed point occurs on the GS plane, attracting all particle orientation trajectories.
	\item Regions $R_5=GS_1^\parallel$ and $R_6=GS_2^\parallel$: The axis $\hat{\mathbf{g}}\times\hat{\mathbf{e}}$ is a saddle node for $t_2t_3<0$, but trajectories approach the node faster along its stable direction than they depart along the unstable direction when $s>0$ ($R_5$). If $s<0$, the unstable direction is sampled faster ($R_6$). In both scenarios the particle orients at a stable fixed point in the GS plane, but, in region $R_5$, a larger proportion of the trajectories approach $\hat{\mathbf{g}}\times\hat{\mathbf{e}}$ before departing it towards the GS plane.
\end{itemize}

Since $s/d_g=t_3-t_2>0$ for prolate spheroids and $s/d_g<0$ for oblate spheroids (figure \ref{fig:SedimentingTerms}), the orientation dynamics behavior observed for a positive $d_g$ for prolate spheroids is qualitatively replicated by a negative $d_g$ in oblate spheroids. As discussed above, the GS plane has fixed points when $|d_g|>|d_g|_\text{spiral}$. The fixed points come in pairs due to the fore-aft symmetry of the particle, so there are an even number (four) of fixed points on the GS plane. When the GS plane is a stable attractor, two of the fixed points are stable nodes and the other two are saddle nodes. In the case when the GS plane is an unstable attractor, two fixed points are saddle nodes and the other two unstable nodes. Figure \ref{fig:SteadyBranches} shows the contours of $p_g^{(0,2)}$, i.e., the magnitude of the projection of the least unstable fixed points (stable nodes for a stable attractor on GS and saddle nodes for an unstable attractor on GS), $\pm\mathbf{p}^\text{stable}_\text{GS}$, projected along the gravity direction, in $\kappa-d_g$ space, where,
\begin{equation}
	\pm p_g^{(0,2)}=\pm \mathbf{p}^\text{stable}_\text{GS}\cdot\hat{\mathbf{g}}.\label{eq:p_g}
\end{equation}
The 0 in the superscript refers to a fixed point, and 2 refers to the second of the three pairs of fixed points (listed in equation \eqref{eq:FixedPointsVortStrat}) of the corresponding dynamical system given by equation \eqref{eq:rotationratefull}.
\begin{figure}
	\centering
	\includegraphics[width=0.75\textwidth]{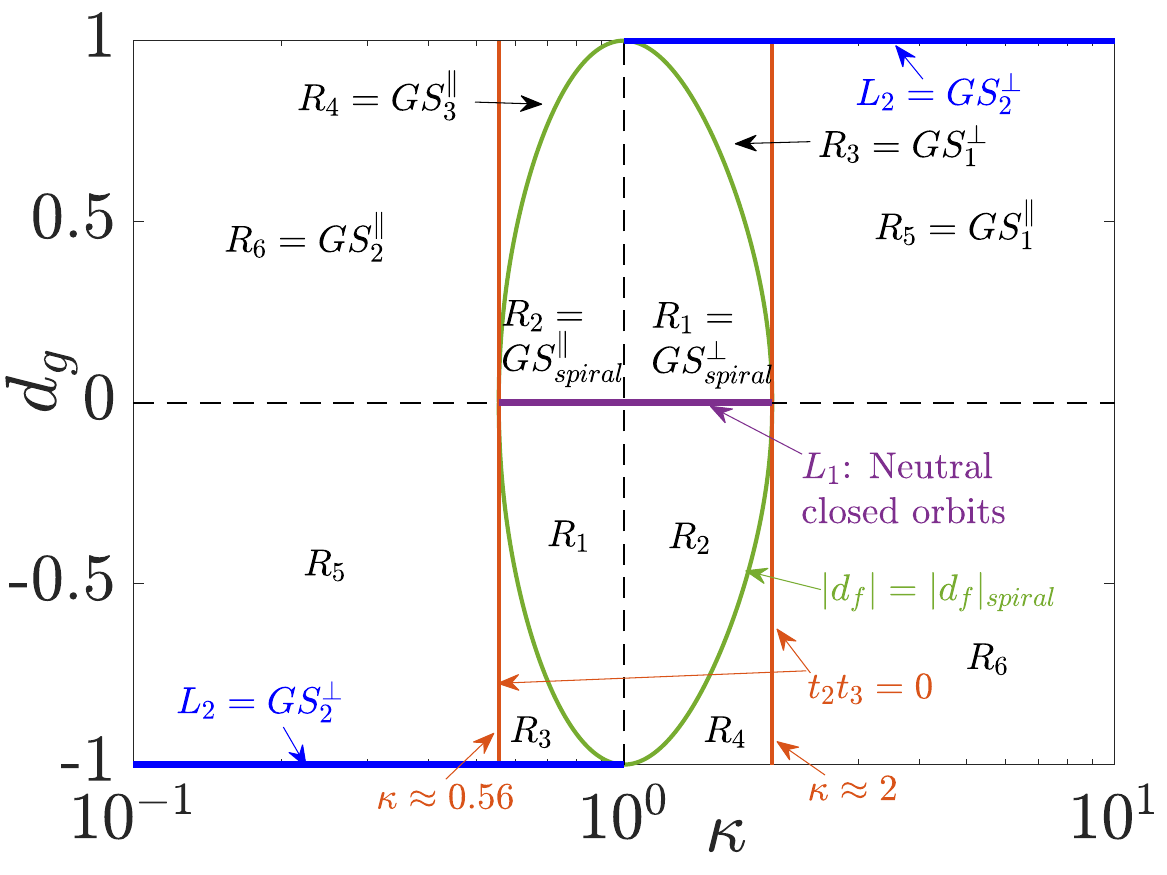}
	\caption {Phase diagram in $d_g-\kappa$ space of the orientation dynamics described by equation \eqref{eq:rotatateg} for spheroidal particles, i.e., $t_2$ and $t_3$ given by equations \eqref{eq:t1t2t3Prolate} and \eqref{eq:t1t2t3Oblate}. Inside the boundaries marked $t_2t_3=0$, $t_2t_3$ is negative, and it is positive otherwise. Here, GS refers to the gravity-stratification plane, and the superscripts $\parallel$ and $\perp$ refer to the final orientation of the particle's axis of symmetry lying parallel and perpendicular to the GS plane respectively. In the regions labeled with subscript ``spiral'' the orientation trajectory towards its equilibrium position/orbit occurs in a spiraling motion, instead of a monotonic drift towards these positions. \label{fig:PhaseDiagram}}
\end{figure}
\begin{figure}
	\centering
	\includegraphics[width=0.75\textwidth]{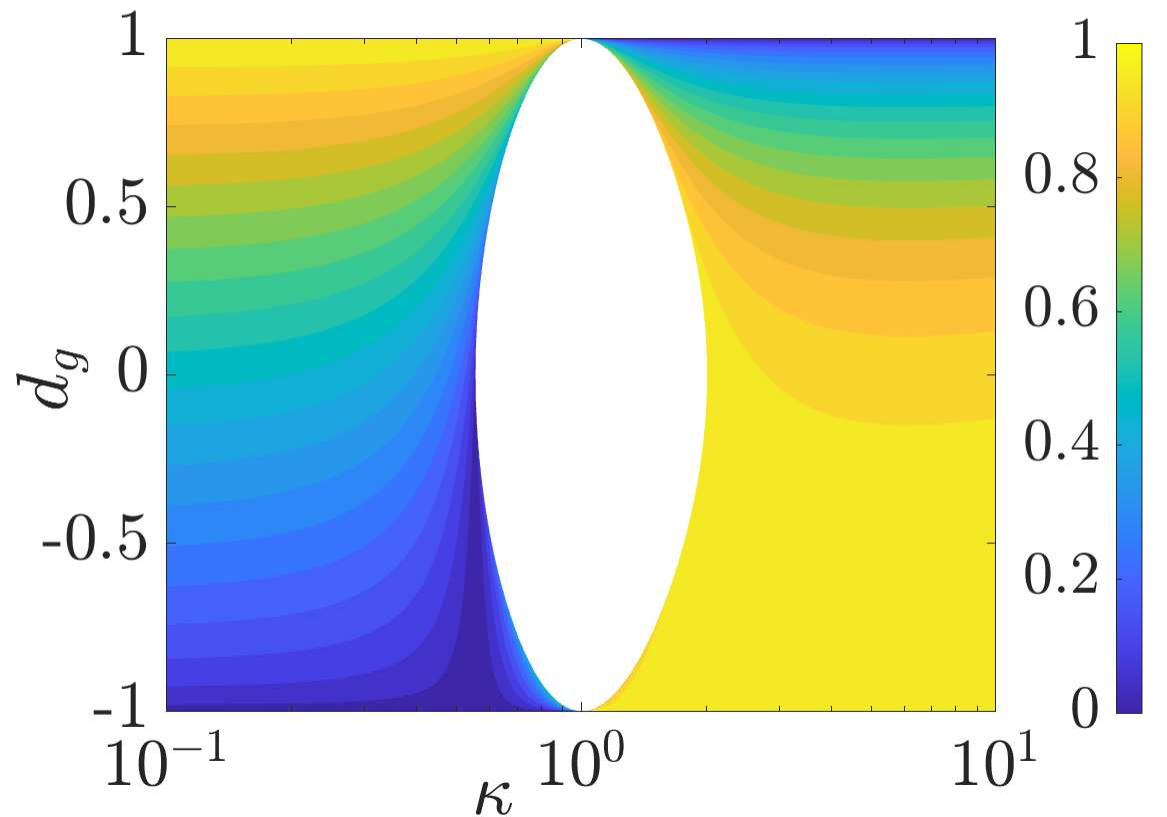}
	\caption {Contours of $p_g^{(0,2)}$ in $d_g-\kappa$ phase space for spheroids. As defined in equation \eqref{eq:FixedPointsVortStrat}, $p_g^{(0,2)}$ is the magnitude of the projection of the location of the stable fixed point (when it exists) within the gravity-stratification (GS) plane along the gravity direction $\hat{\mathbf{g}}$. Yellow ($p_g^{(0,2)}\approx 1$) indicates the stable fixed point location to be closer to $\hat{\mathbf{g}}$ and blue ($p_g^{(0,2)}\approx 0$) indicates a location closer to $\hat{\mathbf{e}}$, i.e., perpendicular to $\hat{\mathbf{g}}$. \label{fig:SteadyBranches}}
\end{figure}

\begin{figure}
	\centering
	\subfloat{\includegraphics[width=0.46\textwidth]{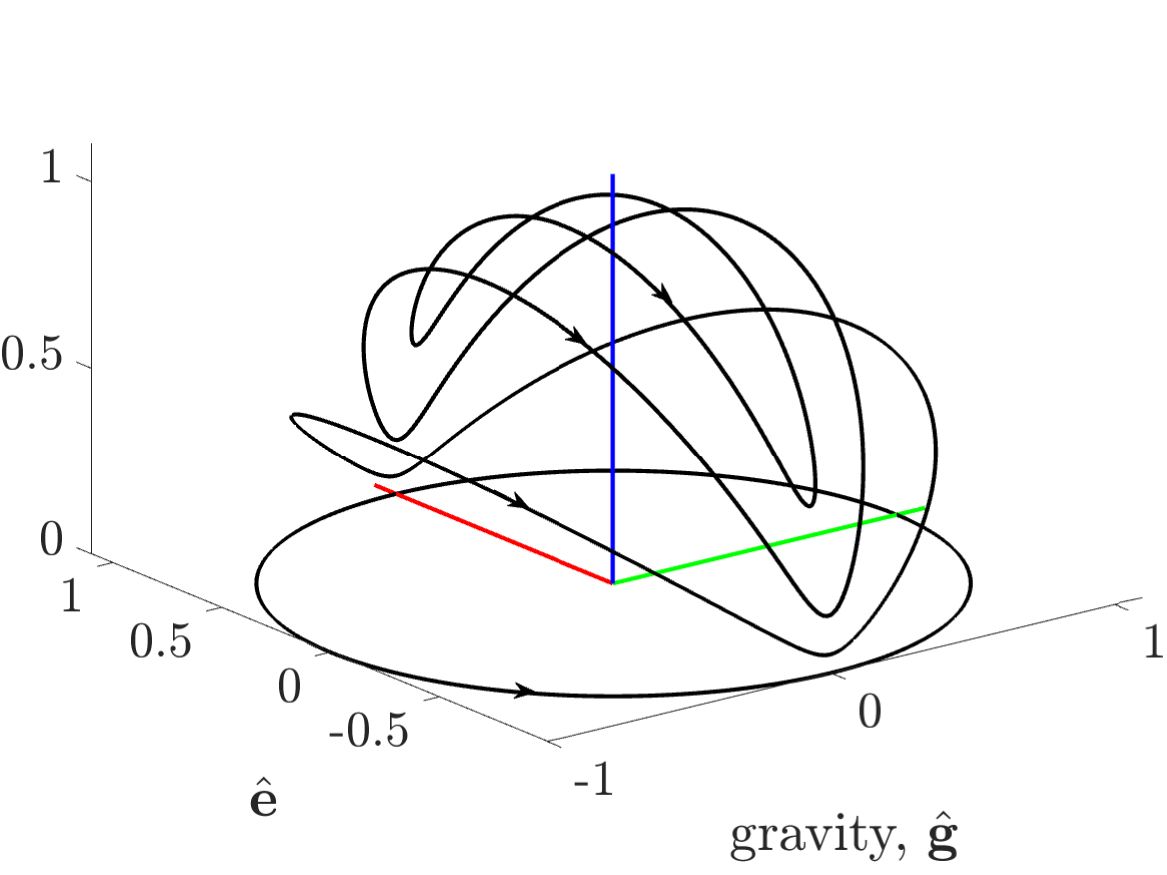}\label{fig:kapap56df0orient}}
	\subfloat{\includegraphics[width=0.46\textwidth]{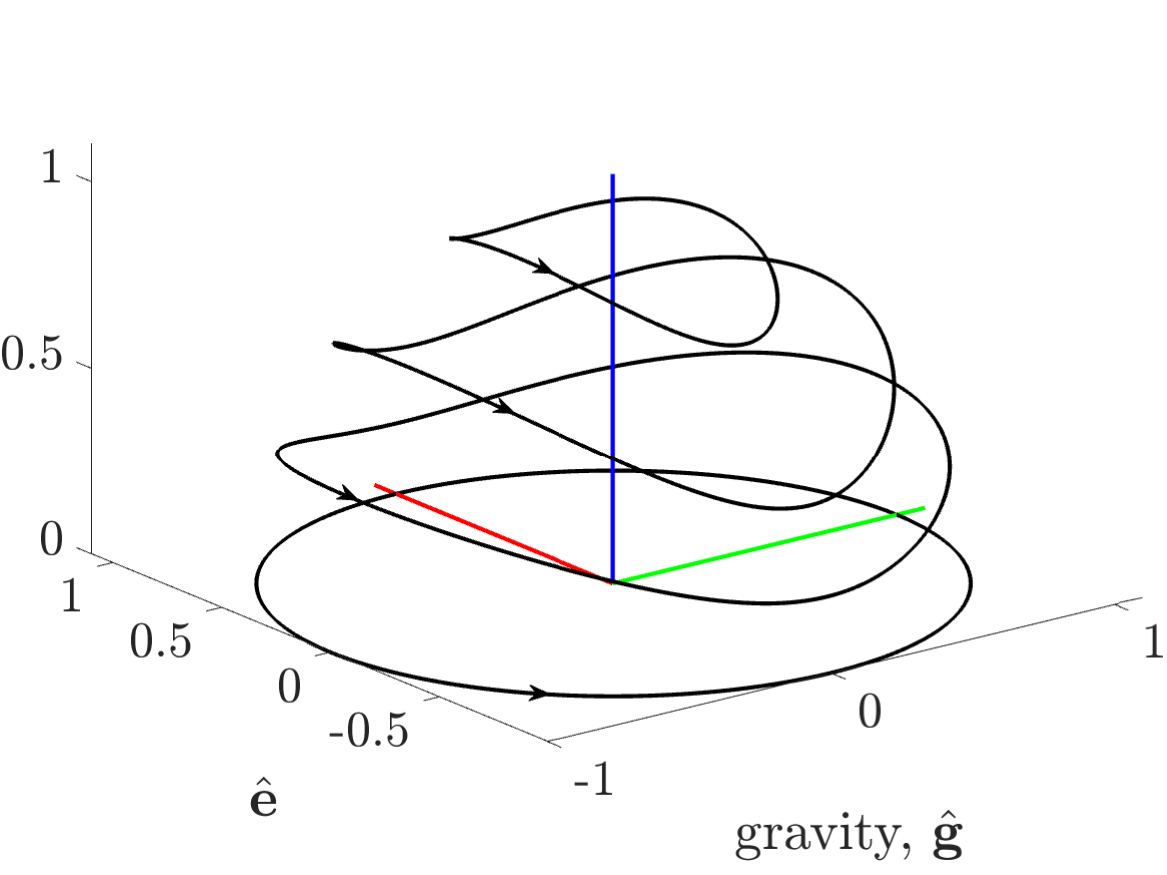}}\\
	\subfloat{\includegraphics[width=0.46\textwidth]{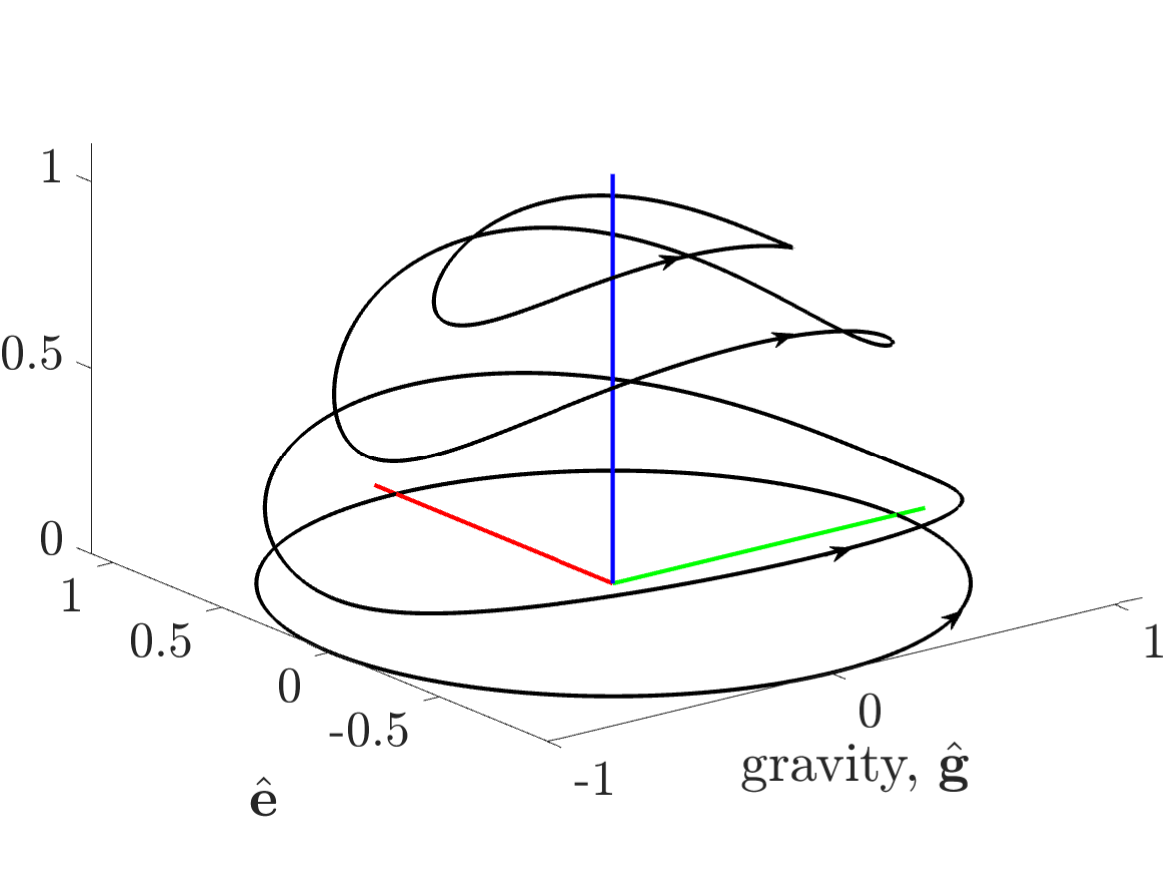}}
	\subfloat{\includegraphics[width=0.46\textwidth]{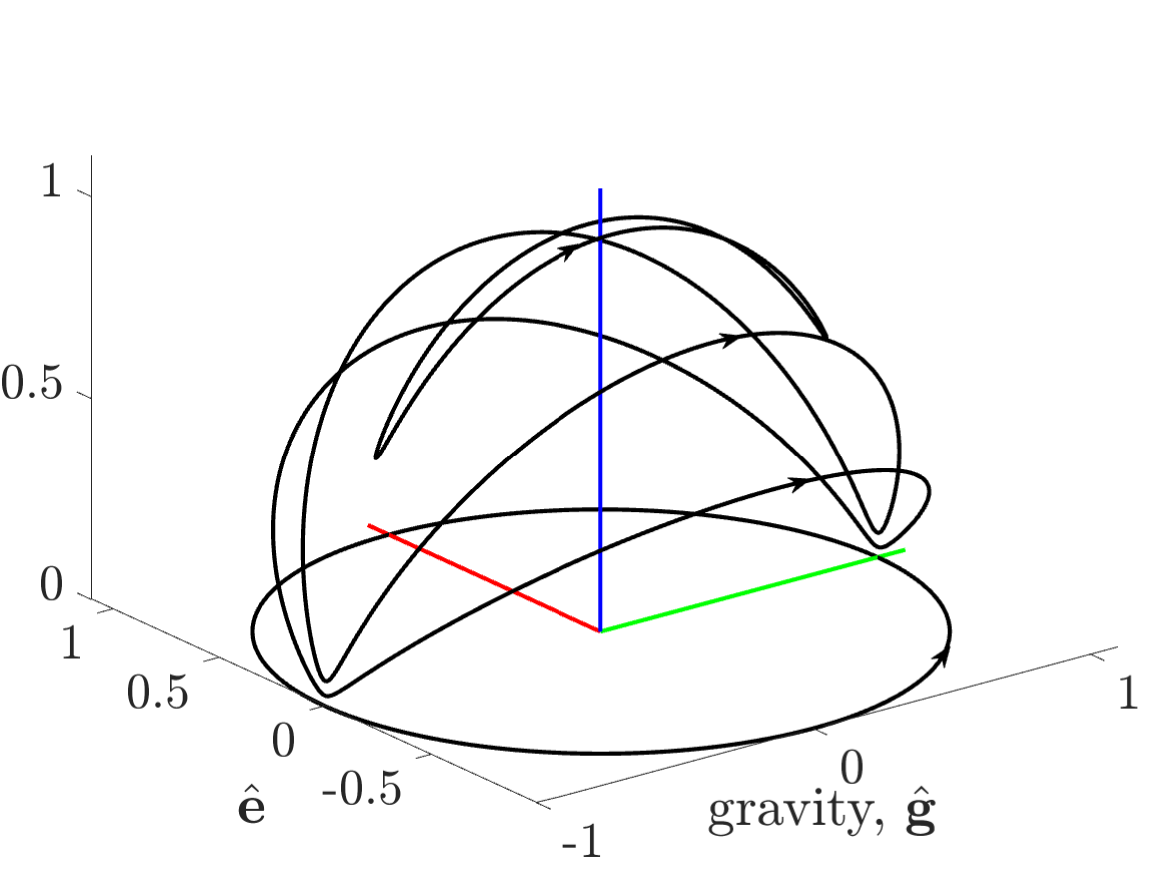}\label{fig:kapa2df0orient}}
	\caption {Orientation dynamics in the region $L_1$, i.e. $d_g=0$ and for (a) $\kappa=0.56$, (b) $\kappa=0.7$, (c) $\kappa=1.5$, and (d) $\kappa=2.0$. Each curve represent the trajectory of the orientation vector of the particle's axis. The magnitudes of $\eta_0$, $\beta$ and $\mathbf{g}$ change the rate of rotation along these orientation trajectories, but not their shape. {Green, red and blue axes represent the direction of vectors $\hat{\mathbf{g}}$, $\hat{\mathbf{e}}$ and $\hat{\mathbf{g}}\times \hat{\mathbf{e}}$ respectively. For $d_g=0$ considered here the red axis ($\hat{\mathbf{e}}$) correspond to the viscosity stratification direction.} \label{fig:L1SpheroidRotation}}
\end{figure}

\begin{figure}
	\centering
	\subfloat{\includegraphics[width=0.49\textwidth]{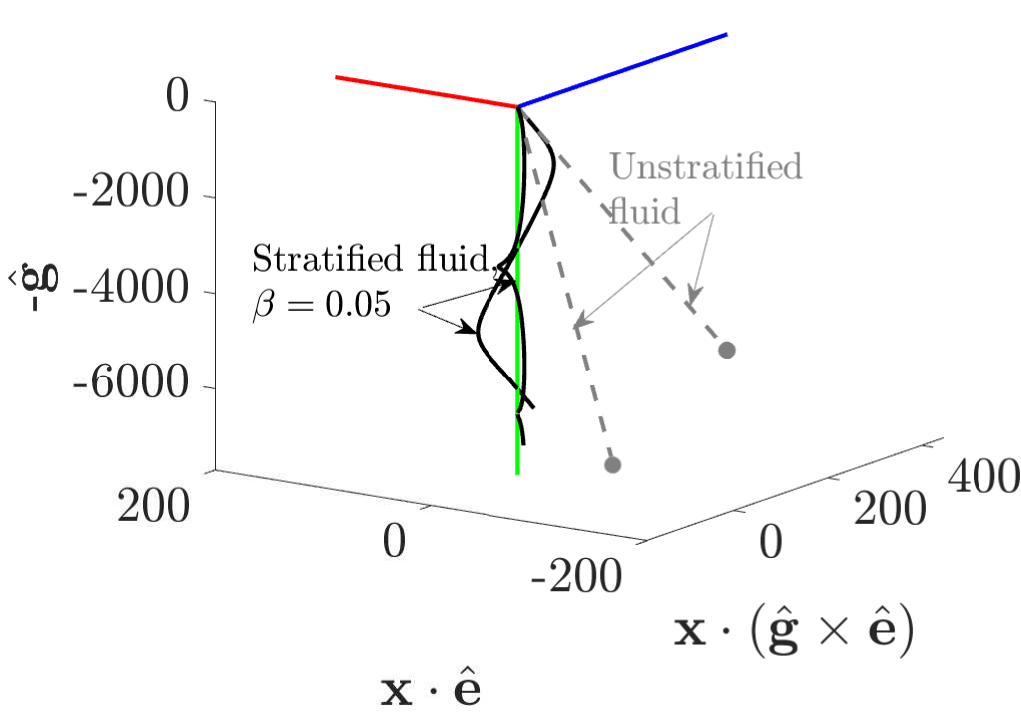}\label{fig:kapa2df0epsp05trans}}
	\subfloat{\includegraphics[width=0.49\textwidth]{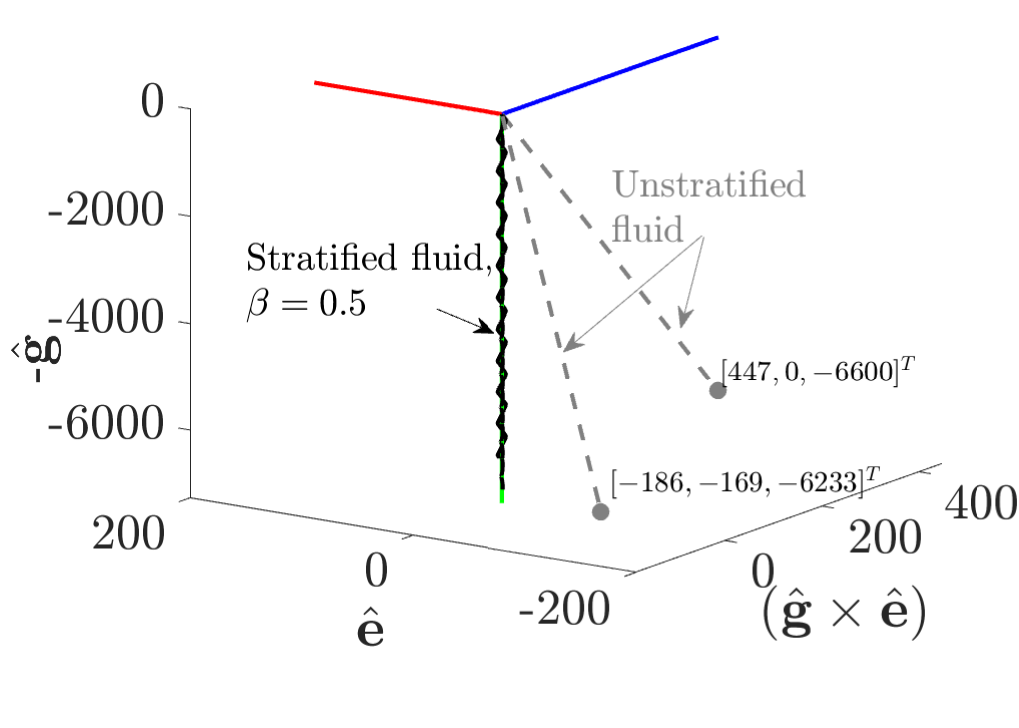}\label{fig:kapa2df0epsp5trans}}
	\caption{Translation dynamics of $\kappa=2.0$ particle with $d_g=0$, $||\mathbf{g}||=1$ and $\beta=0.05$ and 0.5 compared with that for unstratified fluid ($\beta=0$). Each curve represents the trajectory of the particle's centroid.  {Green, red and blue axes represent the direction of vectors $\hat{\mathbf{g}}$, $\hat{\mathbf{e}}$ and $\hat{\mathbf{g}}\times \hat{\mathbf{e}}$ respectively. For $d_g=0$ considered here the red axis ($\hat{\mathbf{e}}$) correspond to the viscosity stratification direction.} \label{fig:L1ProlateTranslation}}
\end{figure}

\begin{figure}
	\centering
	\subfloat{\includegraphics[width=0.49\textwidth]{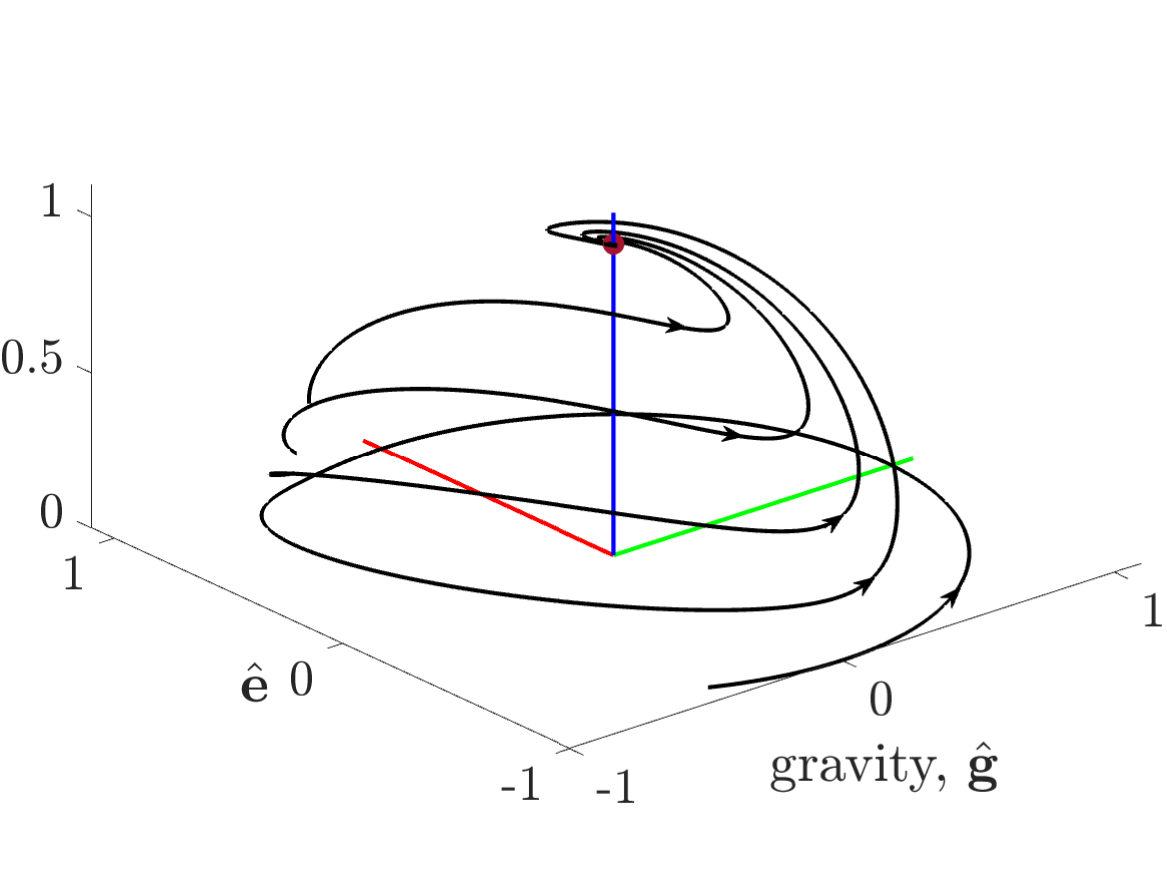}\label{fig:kapap65dfmp5}}
	\subfloat{\includegraphics[width=0.49\textwidth]{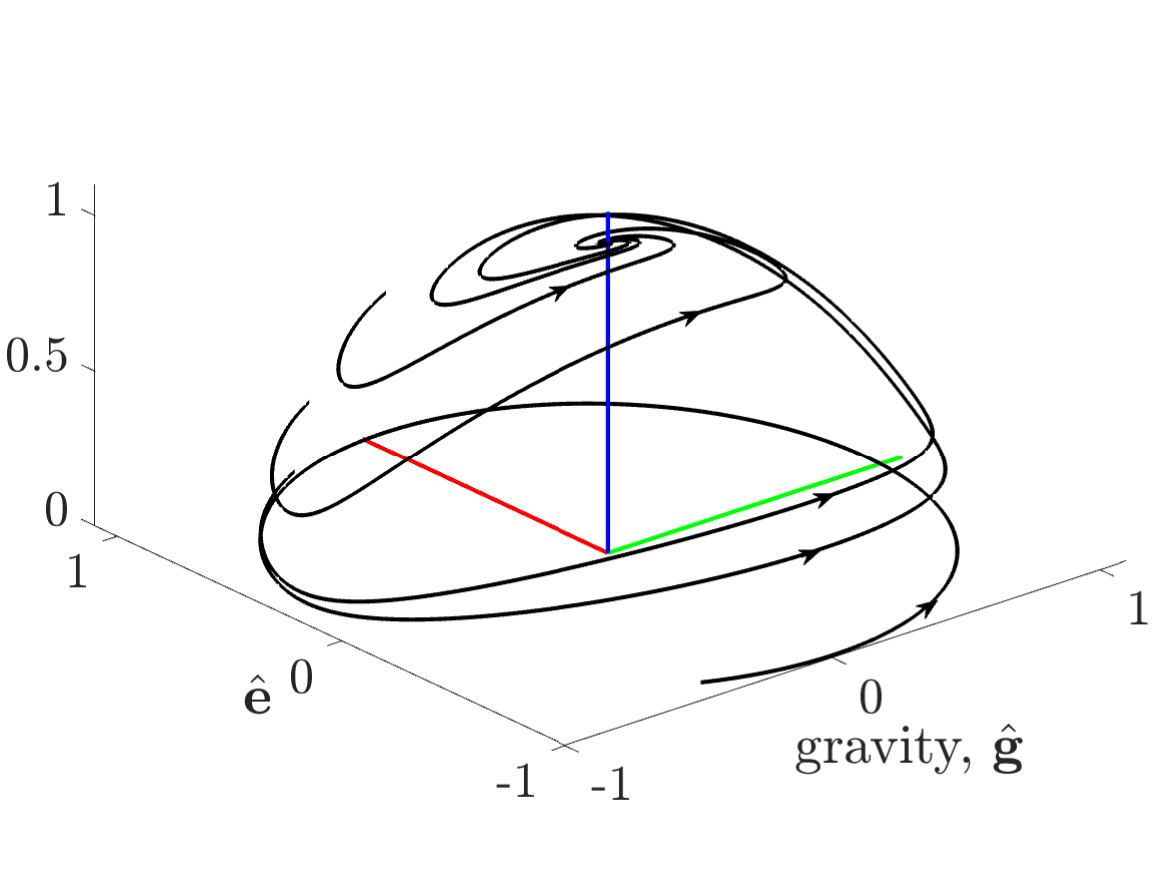}\label{fig:kapa1p5dfp5}}
	\caption{Orientation dynamics in $R_1$ exemplified with (a) oblate particle, $\kappa=0.65$ with $d_g=-0.5$ and (b) prolate particle, $\kappa=1.5$ with $d_g=0.5$. The magnitudes of $\eta_0$, $\beta$ and $\mathbf{g}$ change the rate of rotation along these orientation trajectories, but not their shape.  {Green, red and blue axes represent the direction of vectors $\hat{\mathbf{g}}$, $\hat{\mathbf{e}}$ and $\hat{\mathbf{g}}\times \hat{\mathbf{e}}$ respectively. For $d_g<0$ considered here, the stratification direction lies in the plane of red ($\hat{\mathbf{e}}$) and green ($\hat{\mathbf{g}}$) axes and  the viscosity increases in the positive $\hat{\mathbf{e}}$ and negative $\hat{\mathbf{g}}$ directions.} \label{fig:R1SpheroidRotation}}
\end{figure}

\begin{figure}
	\centering
	\subfloat{\includegraphics[width=0.49\textwidth]{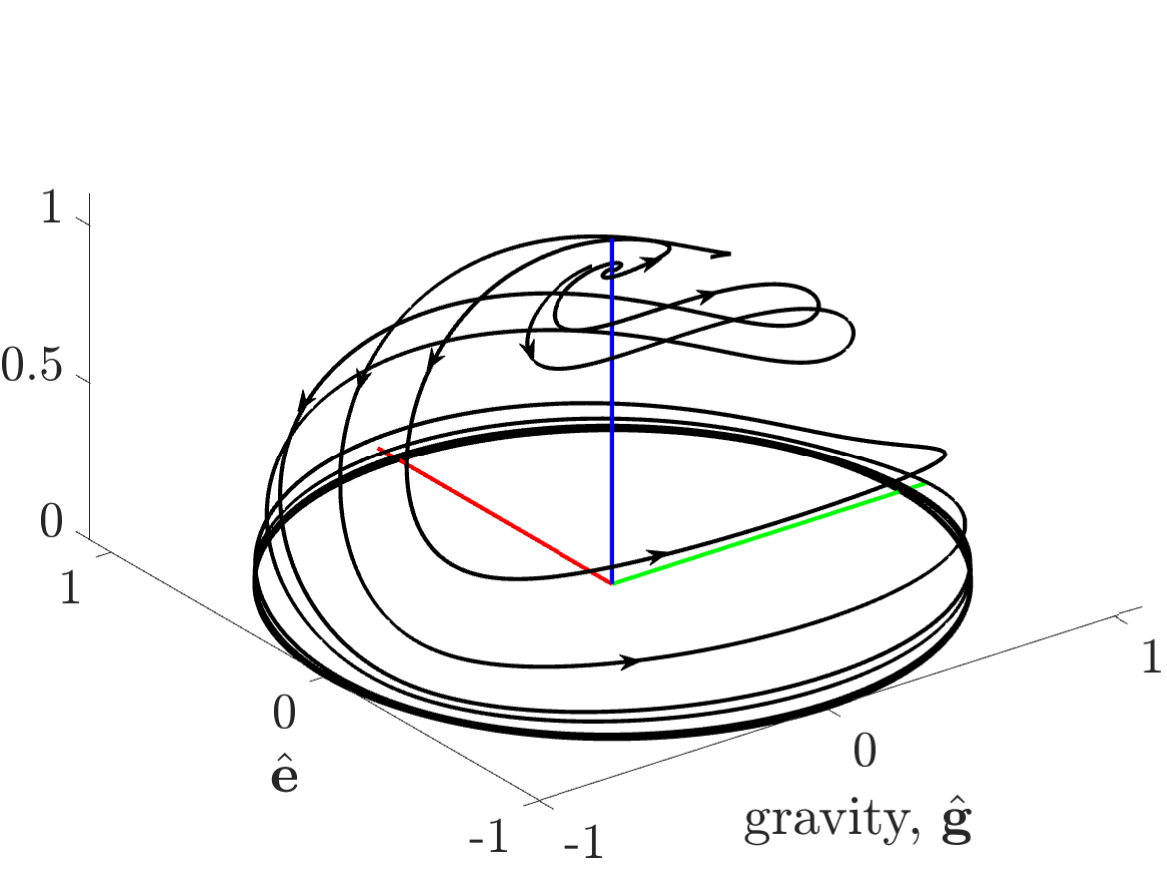}\label{fig:kapa1p5dfmp5}}
	\subfloat{\includegraphics[width=0.49\textwidth]{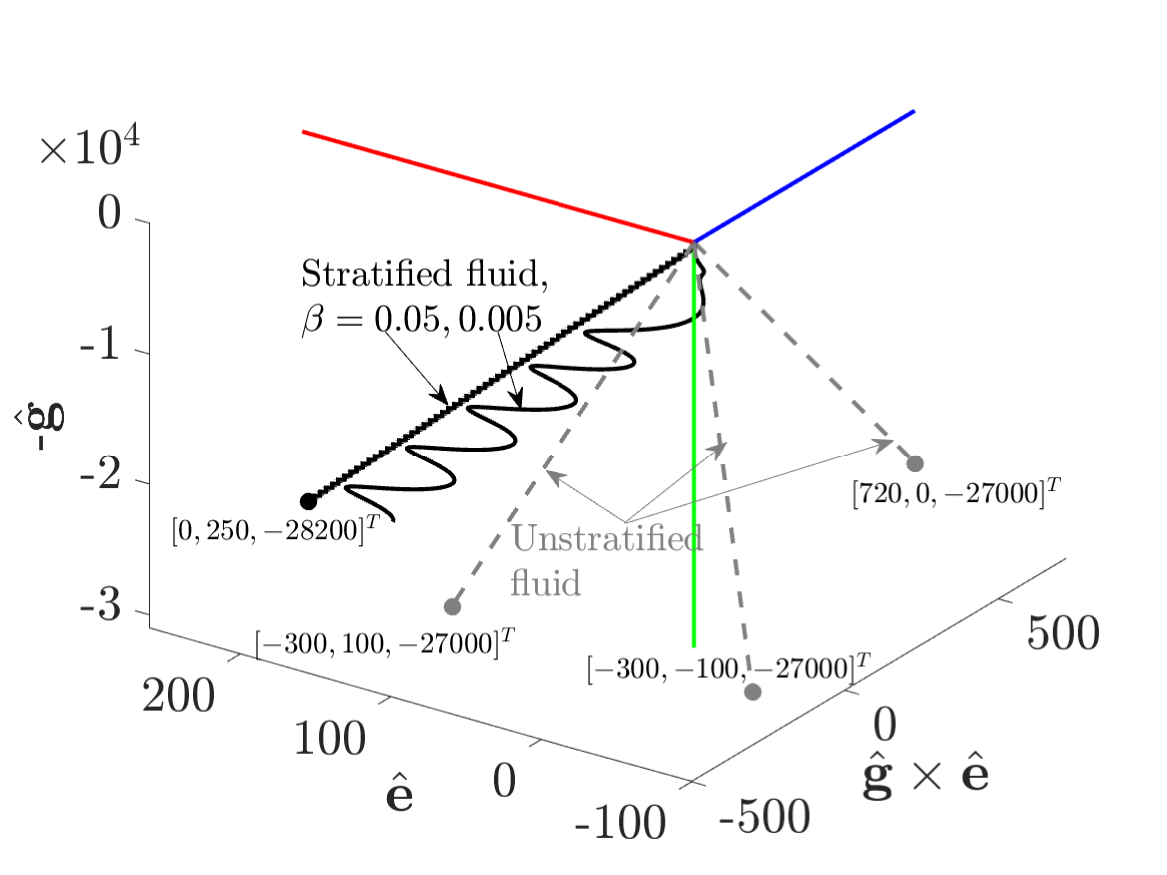}\label{fig:kapa1p5dfmp5Translationepsp05p005}}
	\caption{(a) Orientation dynamics (the magnitudes of $\eta_0$, $\beta$, and $\mathbf{g}$ change the rate of rotation along these orientation trajectories, but not their shape) and (b) Translation dynamics in $R_2$ for a prolate spheroid, $\kappa=1.5$ with $d_g=-0.5$. {Green, red and blue axes represent the direction of vectors $\hat{\mathbf{g}}$, $\hat{\mathbf{e}}$ and $\hat{\mathbf{g}}\times \hat{\mathbf{e}}$ respectively. For $d_g<0$ considered here, the stratification direction lies in the plane of red ($\hat{\mathbf{e}}$) and green ($\hat{\mathbf{g}}$) axes and  the viscosity increases in the positive $\hat{\mathbf{e}}$ and negative $\hat{\mathbf{g}}$ directions.} \label{fig:R2ProlateSpheroidRotationTranslation}}
\end{figure}

\begin{figure}
	\centering
	\subfloat{\includegraphics[width=0.49\textwidth]{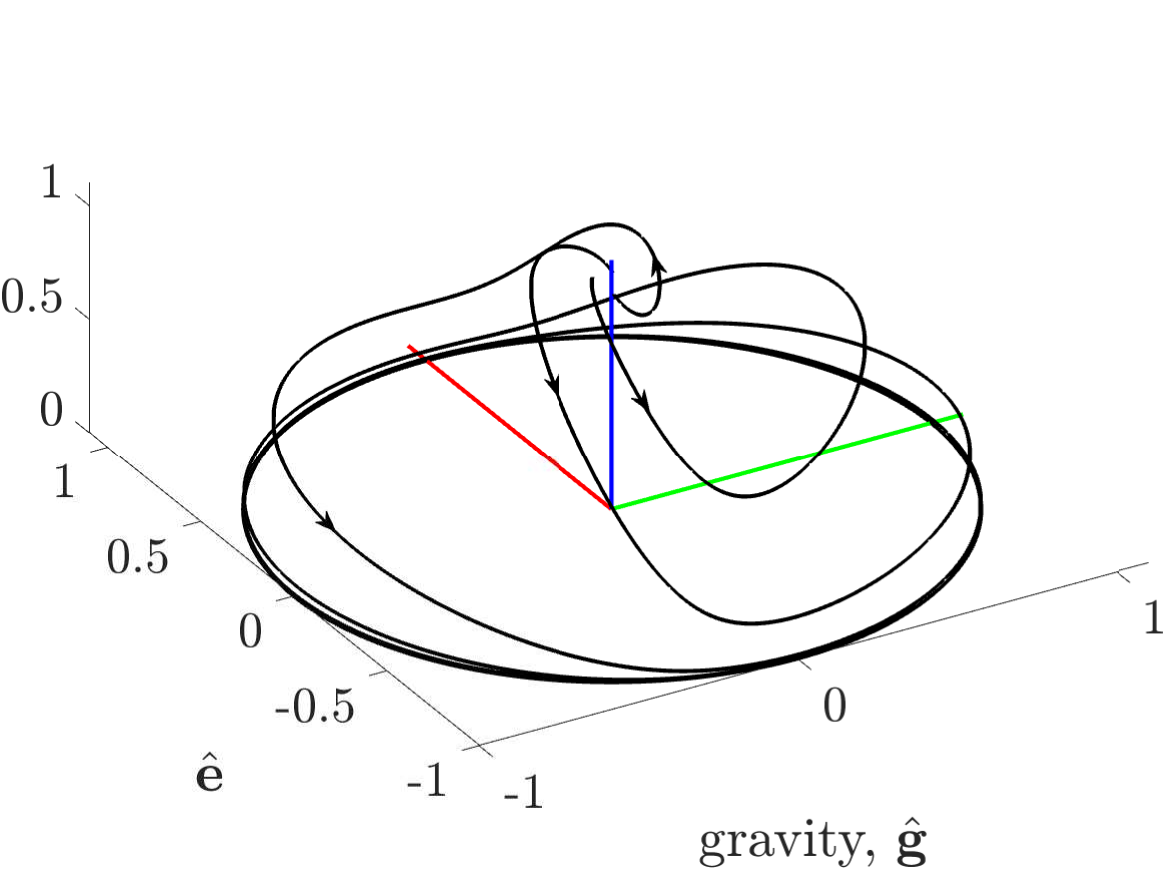}\label{fig:kapap65dfp5}}
	\subfloat{\includegraphics[width=0.49\textwidth]{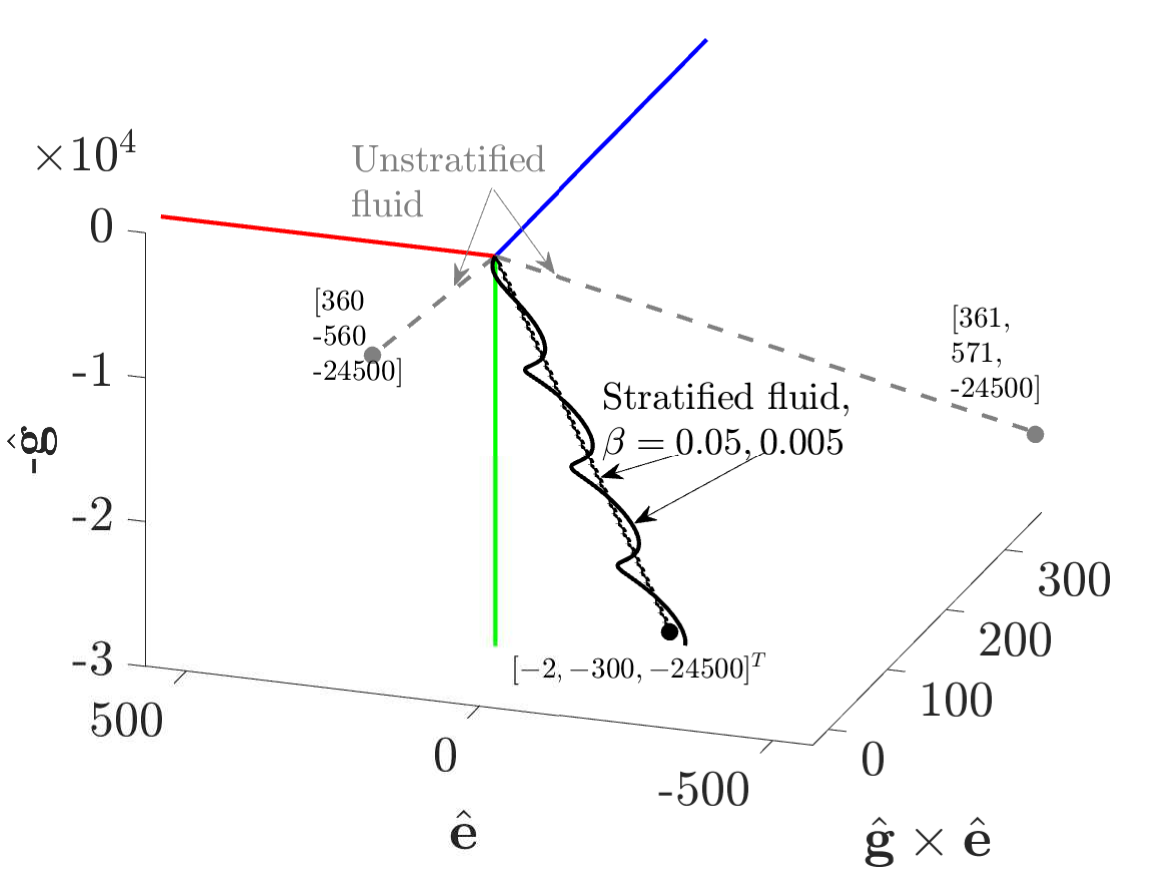}\label{fig:kapap65dfp5Translationepsp05p005}}
	\caption{(a) Orientation dynamics and (b) Translation dynamics in $R_2$ for an oblate particle, $\kappa=0.65$ with $d_g=0.5$. {Green, red and blue axes represent the direction of vectors $\hat{\mathbf{g}}$, $\hat{\mathbf{e}}$ and $\hat{\mathbf{g}}\times \hat{\mathbf{e}}$ respectively. For $d_g>0$ considered here, the stratification direction lies in the plane of red ($\hat{\mathbf{e}}$) and green ($\hat{\mathbf{g}}$) axes and the viscosity increases  in the positive $\hat{\mathbf{e}}$ and positive $\hat{\mathbf{g}}$ directions.}  \label{fig:R2OblateSpheroidRotationTranslation}}
\end{figure}

\begin{figure}
	\centering
	\subfloat{\includegraphics[width=0.49\textwidth]{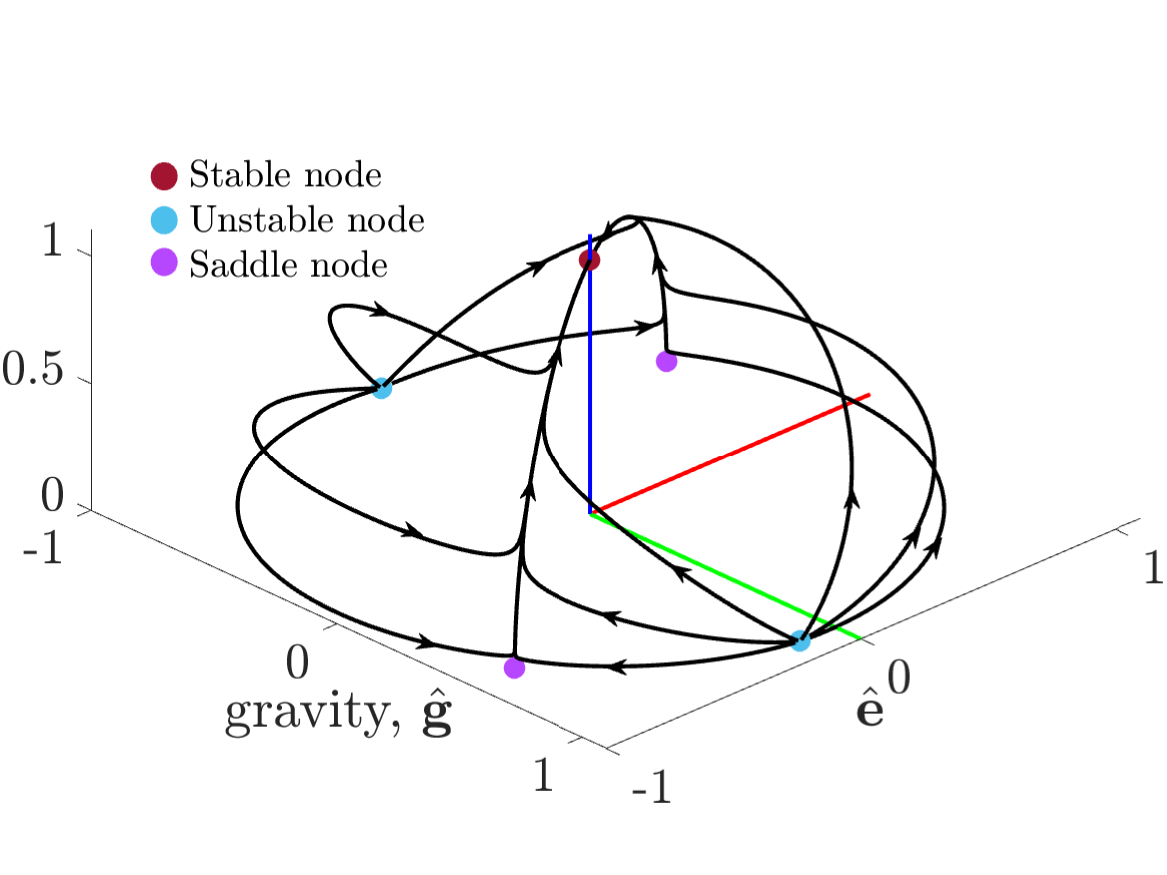}\label{fig:kapapp65dfmp9}}
	\subfloat{\includegraphics[width=0.49\textwidth]{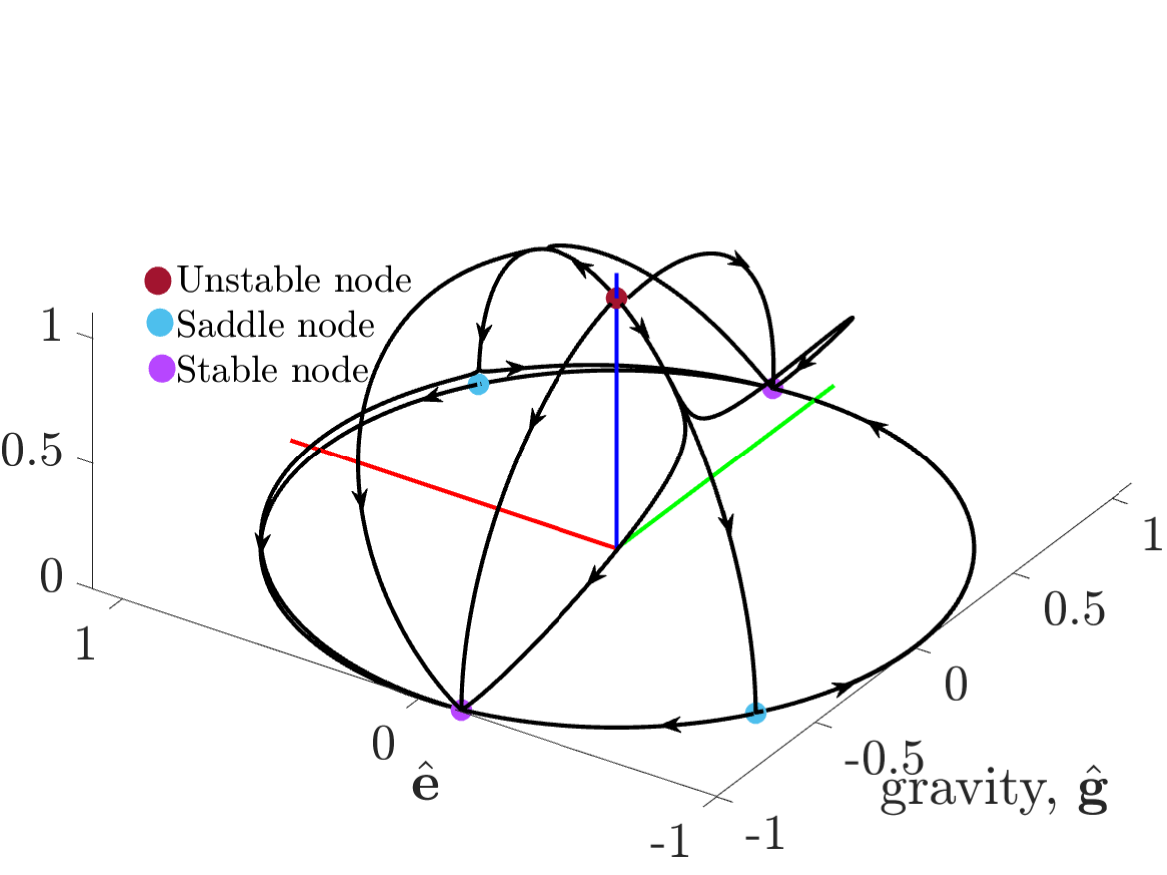}\label{fig:kapap1p5dfmp9}}
	\caption{Orientation dynamics in $R_3$ ($\kappa=0.65$, $d_g=-0.9$) and $R_4$ ($\kappa=1.5$, $d_g=-0.9$). {Green, red and blue axes represent the direction of vectors $\hat{\mathbf{g}}$, $\hat{\mathbf{e}}$ and $\hat{\mathbf{g}}\times \hat{\mathbf{e}}$ respectively. For $d_g<0$ considered here, the stratification direction lies in the plane of red ($\hat{\mathbf{e}}$) and green ($\hat{\mathbf{g}}$) axes and the  viscosity increases in the positive $\hat{\mathbf{e}}$ and negative $\hat{\mathbf{g}}$ directions.} \label{fig:R3R4ProlateSpheroidRotation}}
\end{figure}

\begin{figure}
	\centering
	\subfloat{\includegraphics[width=0.49\textwidth]{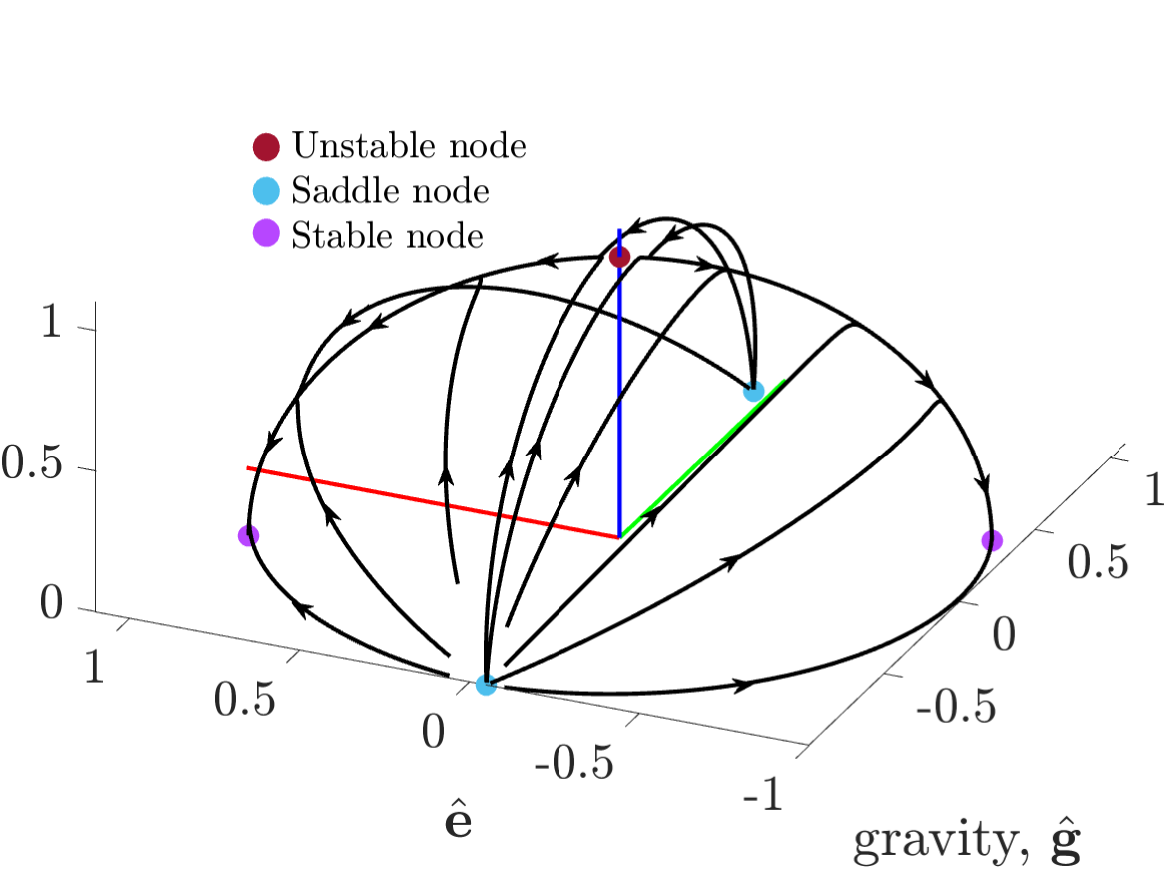}\label{fig:kapap3dfp9}}
	\subfloat{\includegraphics[width=0.49\textwidth]{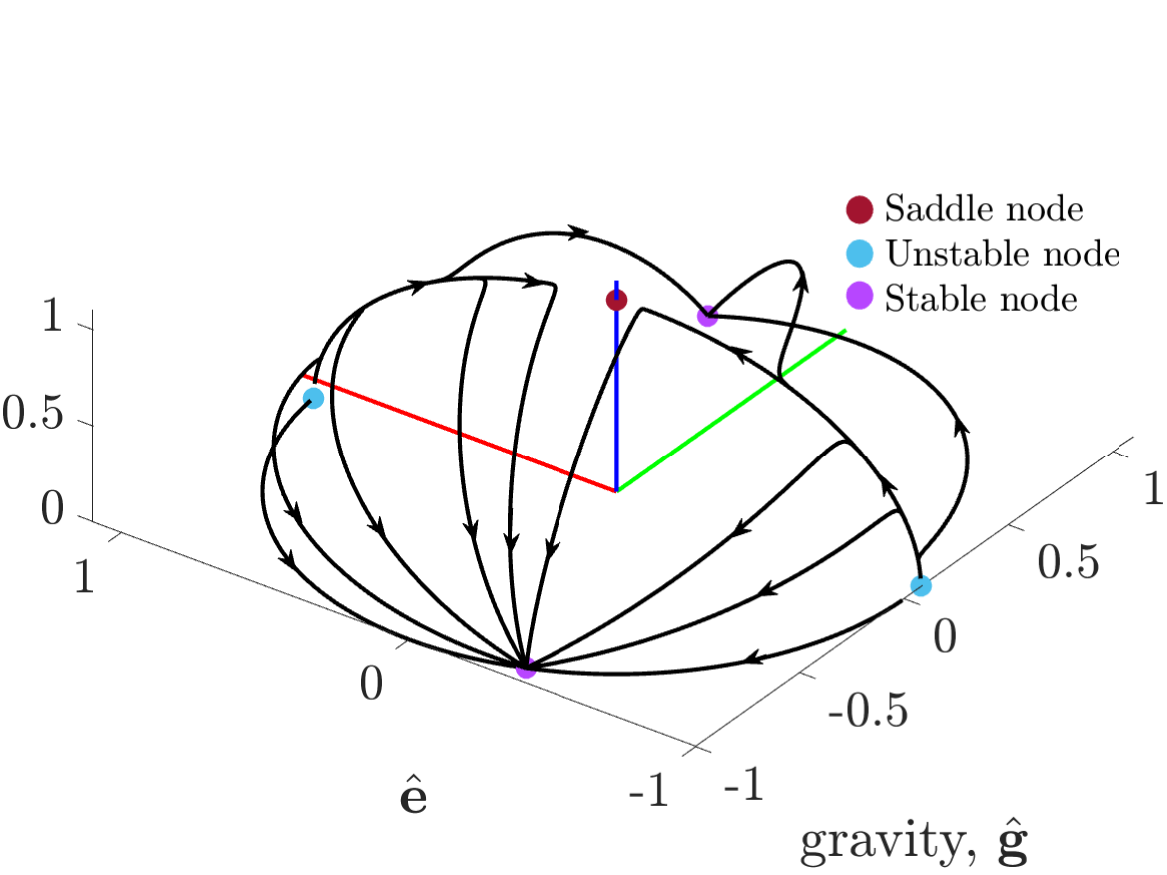}\label{fig:kapap1by3dfp9}}
	\caption{Orientation dynamics in $R_5$ ($\kappa=3.0$, $d_g=0.9$) and $R_6$ ($\kappa=1/3$, $d_g=0.9$).  {Green, red and blue axes represent the direction of vectors $\hat{\mathbf{g}}$, $\hat{\mathbf{e}}$ and $\hat{\mathbf{g}}\times \hat{\mathbf{e}}$ respectively. For $d_g>0$ considered here, the stratification direction lies in the plane of red ($\hat{\mathbf{e}}$) and green ($\hat{\mathbf{g}}$) axes and the viscosity increases in the positive $\hat{\mathbf{e}}$ and positive $\hat{\mathbf{g}}$ directions.}  \label{fig:R5R6ProlateSpheroidRotation}}
\end{figure}

Figures \ref{fig:L1SpheroidRotation} to \ref{fig:R5R6ProlateSpheroidRotation} show the particle trajectories in black, and axes $\hat{\mathbf{g}}$, $\hat{\mathbf{e}}$ and $\hat{\mathbf{g}}\times \hat{\mathbf{e}}$ are shown in green, red and blue respectively. These figures depict qualitatively different orientation (and hence the resulting translation) dynamics corresponding to regions $L_1$ (figures \ref{fig:L1SpheroidRotation} and \ref{fig:L1ProlateTranslation} for orientation and translation trajectories, respectively) and $R_i,i\in[1,6]$ discussed above. Trajectories in $R_1$ are illustrated in figure \ref{fig:R1SpheroidRotation} and those in $R_2$ in figures \ref{fig:R2ProlateSpheroidRotationTranslation} and \ref{fig:R2OblateSpheroidRotationTranslation}. Figure \ref{fig:R3R4ProlateSpheroidRotation} shows orientation trajectories in regions $R_3$ and $R_4$ and figure \ref{fig:R5R6ProlateSpheroidRotation} shows these trajectories in regions $R_5$ and $R_6$.

The oscillatory behavior of orientation dynamics within $\text{R}_1\cup \text{R}_2\cup L_1$ requires part of the viscosity variation to lie perpendicular to gravity, i.e. $|d_g|\ne 1$, and this region is centered around $\kappa=1$, i.e., a sphere. Any vector on a sphere can act as the orientation vector, which must necessarily undergo a neutral periodic orbit as a settling sphere always experiences the same stratification-induced torque, leading to continuous rotation. Therefore, in the $\kappa-d_g$ space plot of figure \ref{fig:PhaseDiagram}, each point on the vertical dashed line $\kappa=1$ denotes neutral or closed periodic orbits, with the rotation rate decreasing in magnitude as $|d_g|$ increases from 0. At $|d_g|=1$ ($\mathbf{d}\times\mathbf{g}=0$) the sphere experiences no stratification-induced rotation. The sphere line in figure \ref{fig:PhaseDiagram}, $\kappa=1$, acts as a bifurcation boundary for different possible behaviors upon changing $\kappa$.

When the viscosity variation is entirely perpendicular to gravity, i.e., along the dashed horizontal $d_g=0$ line, the condition $t_2t_3>0$ implies closed non-uniform, initial condition dependent periodic orbits for particle orientation with a time period $2\pi\eta_0^2/(\sqrt{t_2t_3}||\mathbf{g}||_2\beta)$. For spheroids, this occurs when $0.56 \lessapprox \kappa \lessapprox 2$ and is indicated as $L_1$ in figure \ref{fig:PhaseDiagram}. As discussed in appendix \ref{sec:Rotationatsmallbeta}, these closed orbits are analogous to the Jeffery orbits of a freely rotating spheroid in simple shear flow of a constant viscosity fluid. The GS plane in the stratification-induced rotation of a sedimenting spheroid is analogous to the shearing plane in Jeffery orbits.

Orientation trajectories for four different $\kappa$ in the regime $L_1$ are shown in figure \ref{fig:L1SpheroidRotation} (multiple curves in each plot for a $\kappa$ represents different initial orientations). For a sphere, the orientation trajectories are concentric circles parallel to the GS plane and centered around the $\hat{\mathbf{g}}\times \hat{\mathbf{e}}$ axis. For a prolate spheroid in this regime, $1<\kappa\lessapprox2$, the orientation trajectories deviate from circles and point downwards at the $\hat{\mathbf{g}}$ axis. On the contrary, for oblate particles within $L_1$ ($0.56\lessapprox \kappa \lessapprox1$) the orientation trajectories point downwards at the $\hat{\mathbf{e}}$ axis. Similar to the Jeffery orbits, for $\kappa\ne 1$ the particle spends different amounts of time in different parts of its orientation trajectory for sedimentation induced rotation within region $L_1$.

In particular, the particle spends more time (not shown) in the region of the orientation trajectory that points towards the GS plane (similar to Jeffery orbits pointing towards the shearing plane). Therefore, prolate spheroids in the $L_1$ regime spend more of the time with their axes of symmetry aligned with the gravity, $\hat{\mathbf{g}}$ axis, and oblate spheroids spend more time with their axes of symmetry aligned with the viscosity stratification, $\hat{\mathbf{e}}$ axis. In other words, during the majority of their orientation trajectory, the prolate spheroid's axis of symmetry and the oblate particle's face are aligned with gravity. This has profound implications for the sedimenting direction of the particle as shown in figure \ref{fig:L1ProlateTranslation}, where we compare the motion of a $\kappa=2.0$ particle initially placed at two different non-zero initial angles relative to gravity for unstratified and stratified fluids.

The dashed grey lines in each of the panels of figure \ref{fig:L1ProlateTranslation} show that for a uniform viscosity fluid, the particle sediments at a constant initial orientation-dependent angle relative to gravity. Therefore, along with vertical settling, a spheroid drifts horizontally. However, with a viscosity gradient perpendicular to gravity, the non-uniform periodic nature of the orientation trajectory ensures that the particle falls mostly along the gravity direction without drifting too far horizontally. There is an instantaneous drift, but it is centered about the initial location normal to gravity with the maximum drift bounded, because either the particle's axis of symmetry (prolate) or its face (oblate) is aligned with the gravity direction for the majority of the time during its orientation trajectory. The time period of the orientation trajectory reduces with increasing $\beta$, reducing the time spent in orientations other than when its axis of symmetry (prolate) or face (oblate) is aligned with gravity. Therefore, the translation trajectory of the particle's centroid becomes more confined as $\beta$, i.e., the magnitude of the viscosity gradient, increases (compare black curves in figure \ref{fig:kapa2df0epsp05trans} vs. \ref{fig:kapa2df0epsp5trans}). Similar behavior (not shown) is observed for an oblate spheroid where the orientation is mostly along the stratification direction (figure \ref{fig:kapap56df0orient}) which also leads to a translation direction that is more confined and aligned with gravity. This confinement effect is more pronounced for $\kappa$ close to 0.56 and 2 within the regime $L_1$ as closer to these $\kappa$, the particle's rotation rate is more non-uniform than for particles with $\kappa$ closer to 1.

When $d_g\ne0$, changing $\kappa$ from 1 causes a bifurcation in the orientation trajectories. While the boundaries between regions $R_1$ and $R_2$ ($\kappa=1$) are characterized by neutral closed orbits, spiraling trajectories are observed within regions $R_1$ and $R_2$, with the rate of spiraling increasing as $\kappa$ deviates from 1 in these regions. A similar bifurcation can also be observed as $d_g$ is altered from 0 when going from $L_1$ to $R_1$ and $R_2$. A prolate particle ($\kappa>1)$ spirals towards an orientation perpendicular to the GS plane if viscosity increases along gravity ($d_g>0$), i.e., the region $\text{R}_1={GS}_\text{spiral}^\perp$. Conversely, if $d_g<0$, it spirals towards the GS plane in the region $\text{R}_2={GS}_\text{spiral}^\parallel$. For an oblate particle, region ${GS}_\text{spiral}^\perp$ occurs if $d_g<0$ and ${GS}_\text{spiral}^\parallel$ if $d_g>0$. The switching of behaviors between prolate and oblate spheroids with the sign of $d_g$ is related to the reversed sign of $t_2-t_3$ (positive for oblate and negative for prolate spheroids) for these particles shown in figure \ref{fig:SedimentingTerms}. The orientation trajectories for a few starting orientations in region $R_1$ for both an oblate and prolate particle are shown in figure \ref{fig:R1SpheroidRotation}. The magnitudes of $\beta$ and $\mathbf{g}$ do not affect the shape of these trajectories, but they change the spiraling rate towards the $\hat{\mathbf{g}}\times \hat{\mathbf{e}}$ axis. Since the particle orientation is ultimately aligned normal to gravity, the particle settles parallel to gravity after the initial transient effects (not shown). The duration of the transient is inversely proportional to the rate of spiraling, which itself is proportional to $\beta||\mathbf{g}||$.
Figure \ref{fig:kapa1p5dfmp5} shows the orientation trajectories of a $\kappa=1.5$ spheroid within the $R_2$ region, where spiraling towards the GS plane is observed. A portion of each spiral in this figure points downwards towards the GS plane. Similar to the Jeffery orbits or the neutral orbits shown in figure \ref{fig:L1SpheroidRotation} for stratification-induced rotation of a sedimenting spheroid in region $L_1$, this downward pointing portion is the slowest part of the spiral. The projection of this bottleneck region in the GS plane (of the trajectories shown in figure \ref{fig:kapa1p5dfmp5}) is slightly misaligned with the gravity, $\hat{\mathbf{g}}$, axis for a $\kappa=1.5$ particle exemplifying prolate spheroids in $R_2$. For oblate spheroids in $R_2$, this projects close to, but slightly misaligned from, the $\hat{\mathbf{e}}$ axis (figure \ref{fig:kapap65dfp5}). When the particle orientation ultimately reaches the GS plane, it continues to tumble there, but in a non-uniform fashion. This has a direct consequence on the translation trajectories shown in figure \ref{fig:kapa1p5dfmp5Translationepsp05p005} and \ref{fig:kapap65dfp5Translationepsp05p005}. The bottleneck or slow region in each spiral leads to an independence of the sedimenting angle from the initial orientation compared to the constant viscosity case (grey lines). Irrespective of initial condition, a spheroid orients within the GS plane and due to the bottleneck, it falls at a similar average angle relative to gravity. The time period for each spiral is inversely proportional to $\beta$, and at higher $\beta$ the time spent outside of the bottleneck region is smaller, leading to straighter trajectories at larger $\beta$ shown in figures \ref{fig:kapa1p5dfmp5Translationepsp05p005} and \ref{fig:kapap65dfp5Translationepsp05p005}.

One notable difference appears between the sedimentation of prolate and oblate spheroids in $R_2$. This requires $d_g>0$ for oblate and $d_g<0$ for prolate spheroids, i.e., it requires stratification to be misaligned with gravity but viscosity to increase along gravity for oblate and decrease for prolate spheroids (figure \ref{fig:PhaseDiagram}). The direction of $\hat{\mathbf{e}}$ is perpendicular to $\hat{\mathbf{g}}$ and at least a part of viscosity increase is along $\hat{\mathbf{e}}$ (when $|d_g|\ne1$). Prolate particles in $R_2$ migrate towards the positive $\hat{\mathbf{e}}$ (figure \ref{fig:kapa1p5dfmp5Translationepsp05p005}) while the oblate particles towards the negative $\hat{\mathbf{e}}$ (figure \ref{fig:kapap65dfp5Translationepsp05p005}) axis. Therefore, the horizontal drift of prolate spheroids is towards higher viscosity fluid, while that of oblate spheroids is towards lower viscosity fluid in the region $R_2$.

Within $R_2$, as a particle is made less spherical, i.e., as $\kappa$ deviates further from one, the spiral points more downwards and, as mentioned above, the spiraling rate increases. Towards the edge of $R_2$ further from $\kappa=1$, the downwards pointing part of the spirals almost completely touches the GS plane such that at the edge between $R_2$ to $R_4$ in figure \ref{fig:PhaseDiagram} another bifurcation is observed. Hence, in $R_4$, spirals no longer exist, and instead, two fixed points appear on the GS plane as the GS plane bifurcates from a stable limit cycle to a stable sub-space, and the axis $\hat{\mathbf{g}}\times \hat{\mathbf{e}}$ bifurcates from an unstable spiral to an unstable node. The orientation trajectories of a $\kappa=1.5$ particle with $d_g=-0.9$ are shown in figure \ref{fig:kapap1p5dfmp9} and comparing them with figure \ref{fig:kapa1p5dfmp5} shows the bifurcation. The stable fixed point is closer to the $\hat{\mathbf{g}}$ axis for prolate spheroids, in continuation with the bottleneck region from $R_2$ for such particles. This alignment of the particle closer to gravity is also illustrated by a $p_g^{(0,2)}\approx 1$ (equation \eqref{eq:p_g}) for prolate spheroids in the region corresponding to $R_4$ in figure \ref{fig:SteadyBranches}. A similar topology of orientation trajectories is observed for oblate spheroids (not shown), but the location of the stable fixed point within the GS plane is more sensitive to the values of $\kappa$ and $d_g$ (as shown by the rapidly changing $p_g^{(0,2)}$ in the $R_4$ region for $\kappa<1$ particles than $\kappa>1$ in figure \ref{fig:SteadyBranches}). An oblate particle is aligned closer to $\hat{\mathbf{e}}$ axis near the $R_2-R_4$ boundary ($p_g^{(0,2)} \approx 0$ near this boundary in figure \ref{fig:SteadyBranches}), and as $\kappa$ reduces (such that the particle is more disc-like), it moves towards the $\hat{\mathbf{g}}$ axis (indicated in figure \ref{fig:SteadyBranches} by $p_g^{(0,2)}$ being closer to 1 than 0 upon reducing $\kappa$ for oblate particles in $R_4$). A similar effect is found upon increasing $d_g$ from 0 to 1 for oblate spheroids. The translation trajectories in region $R_4$ (not shown) follow a similar initial orientation independent trend as discussed above for $R_2$ for prolate and oblate spheroids, except that particles fall in a (perfectly) straight line as their orientation reaches a steady state in $R_4$.

The bifurcations that occur between $R_1$ and $R_3$ are similar to the $R_2$ to $R_4$ bifurcation just discussed. Region $R_1$, discussed earlier, is similar to $R_2$ but with spiraling away from the GS plane. Therefore, similar to the bifurcation from $R_2$ to $R_4$ discussed above, a bifurcation occurs at the edge between $R_1$ and $R_3$ in figure \ref{fig:PhaseDiagram} as the spiraling rate reaches infinity. Figure \ref{fig:kapapp65dfmp9} illustrates the orientation trajectories of a $\kappa=0.65$ particle with $d_g=0.9$, a case within $R_3$. In $R_3$, coming from $R_1$ (comparing figure \ref{fig:kapap65dfmp5} with \ref{fig:kapapp65dfmp9}), the GS plane changes from an unstable limit cycle to an unstable subspace with two fixed points (one saddle and the other unstable node), and the axis $\hat{\mathbf{g}}\times \hat{\mathbf{e}}$ bifurcates from a stable spiral node to a stable node. Since the ultimate orientation is perpendicular to gravity, the particle falls along gravity in $R_3$ (not shown).

When gravity and stratification are collinear, i.e., $|d_g|=1$, a spheroid will orient either parallel or perpendicular to gravity irrespective of its starting orientation. In either case, it sediments along the gravity direction. A prolate spheroid orients perpendicular to gravity if viscosity increases along gravity, i.e., $d_g=1$, and parallel if $d_g=-1$. An oblate spheroid shows the opposite trend relative to the sign of $d_g$. Regions where particles orient perpendicular to gravity are labeled as $L_2$ in figure \ref{fig:PhaseDiagram}. {This behavior is similar to that observed by \cite{anand2024sedimentation} and can also be deciphered for prolate spheroids from the formulae of \cite{gong2024active}.}

In the region $R_5\cup R_6$, demarcated by vertical lines at $\kappa\lessapprox.56$ and $\kappa\gtrapprox 2$ and horizontal lines for $d_g=1, \kappa>1$ and $d_g=-1, \kappa<1$, the $\hat{\mathbf{g}}\times \hat{\mathbf{e}}$ axis is a saddle node. Transitioning from $R_4$ to $R_6$ (comparing figures \ref{fig:kapap1p5dfmp9} and \ref{fig:kapap1by3dfp9}) involves a bifurcation where the saddle node on the GS plane becomes an unstable node, and the unstable node at the $\hat{\mathbf{g}}\times \hat{\mathbf{e}}$ axis becomes a saddle. This does not alter the topology around the stable fixed point on the GS plane. Similarly, transitioning from $R_3$ to $R_5$ (comparing figures \ref{fig:kapapp65dfmp9} and \ref{fig:kapap3dfp9}) causes the stable node at $\hat{\mathbf{g}}\times \hat{\mathbf{e}}$ axis to bifurcate into a saddle node, and the saddle node on the GS plane becomes a stable node without altering the topology around the unstable fixed point on the GS plane. Therefore, in both $R_5$ and $R_6$, the particle ultimately orients within the GS plane, but in a different fashion. In $R_5$, the stable direction of the saddle at $\hat{\mathbf{g}}\times \hat{\mathbf{e}}$ is sampled faster than its unstable direction, so the spheroids first approach a plane that includes the $\hat{\mathbf{g}}\times \hat{\mathbf{e}}$ axis before traversing towards the GS plane, as shown in figure \ref{fig:kapap3dfp9}. In $R_6$, the unstable eigenvalue of the saddle point at $\hat{\mathbf{g}}\times \hat{\mathbf{e}}$ has a greater magnitude than the stable eigenvalue, so the particle reaches its final location in the GS plane faster than in $R_5$ (figure \ref{fig:kapap1by3dfp9}). 

The contours of $p_g^{(0,2)}$ (equation \eqref{eq:p_g}) in figure \ref{fig:SteadyBranches} depict the final orientation of the spheroid's axis of symmetry relative to the gravity direction. As shown in this figure, for $d_g<0$, the final orientation in region $R_5\cup R_6$ is closer to $\hat{\mathbf{g}}$ for prolate and to $\hat{\mathbf{e}}$ for oblate spheroids. The final orientation is more sensitive to $d_g$ and $\kappa$ within $R_5\cup R_6$ for $d_g>0$, where both oblate and prolate spheroids orient closer to $\hat{\mathbf{g}}$ (yellow region in figure \ref{fig:SteadyBranches}) as $\kappa$ decreases. Increasing $d_g$, i.e., increasing the alignment between stratification and gravity, within $R_5\cup R_6$ for $d_g>0$ makes prolate and oblate spheroids orient towards $\hat{\mathbf{e}}$ (blue) and $\hat{\mathbf{g}}$ (yellow), respectively. The influence of stratification-induced rotation on sedimentation in $R_5$ and $R_6$ (not shown) is similar to that discussed earlier for $R_3$ and $R_4$. The particle falls ultimately in a straight, initial condition independent path, with its angle relative to gravity depending on the location of the stable fixed point on the GS plane, $p_g^{(0,2)}$.

At the intersection of the regions marked $R_5$ and $R_6$ when gravity is perpendicular to stratification, $d_g=0$, for particles with aspect ratio much different from a sphere \cite{anand2024sedimentation} also found a single stable orientation that is independent of the initial orientation. The value of $p_g^{(0,2)}$, from our calculations shown in figure \ref{fig:SteadyBranches} varies from about 0.998 to 0.931 for $2\lessapprox \kappa \lessapprox 10$ compares favorably with an equivalent value of about $|\cos(57\pi/74)|\approx 0.9\approx 0.94$ displayed in figure 14a of \cite{anand2024sedimentation}. However, for $0.1\lessapprox\kappa\lessapprox0.5$, we find $p_g^{(0,2)}$ to vary between 0.49 and 0.12, whereas for the same $\kappa$ range values displayed in figure 14b of \cite{anand2024sedimentation} vary between 0.5 and 0.25. Furthermore, \cite{anand2024sedimentation} find stable orientations for oblate spheroids with $\kappa$ as low as $0.01$, which is well within the $L_1$ (figure \ref{fig:PhaseDiagram}) region where neutral orbits are predicted by our calculations. Thus, the final orientation predicted by our calculations for large $\kappa$ spheroids agrees well with the previous investigations, but the orientation predicted for oblate spheroids is different. A future numerical investigation of oblate spheroids, akin to that conducted for prolate spheroids in section \ref{sec:LinearFlowsValidation} (black markers in figure \ref{fig:LinearFlowForcesSpheroid}), could serve as an independent validation of our findings.

\section{Freely suspended spheroids in linear flows with viscosity gradients}\label{sec:SpheroidsLinear}
\subsection{Spheres ($\kappa=1$)}\label{sec:SphereLinear}
A freely suspended sphere in a linear flow of uniform viscosity fluid simply rotates with the imposed fluid rotation, $\boldsymbol{\omega}_\infty=-0.5\boldsymbol{\epsilon}:\boldsymbol{\Omega}$, where $\boldsymbol{\Omega}$ is the anti-symmetric part of the imposed velocity gradient, and translates with the local velocity of the imposed flow, $\mathbf{u}_\text{fluid}$. Coupling between the translational and rotational motion due to viscosity stratification leads to the following particle motion,
\begin{align}
		\begin{split}&\mathbf{u}_\text{particle}=\mathbf{u}_\text{fluid}+\frac{ \beta l^2}{\eta_0}\mathbf{E}\cdot\mathbf{d}+\mathcal{O}(\beta^2).
\end{split}\end{align}

In a simple shear flow, with strain rate $\dot{\gamma}$, the velocity of a sphere relative to the viscosity stratified fluid is ${\dot{\gamma}\beta l^2}/({2\eta_0})\begin{bmatrix}d_2,&d_1,&0\end{bmatrix}$, where the directions 1, 2 and 3 are respectively in the flow, gradient and vorticity direction of the imposed flow. Thus, for magnitude $\beta$ and direction $\mathbf{d}=\begin{bmatrix}
	d_1,&d_2,&d_3
\end{bmatrix}$ of viscosity stratification, a sphere of radius $l$ can be moved across the flow streamlines with a speed ${\dot{\gamma}\beta l^2d_1}/({2\eta_0}).$

\begin{figure}
		\centering
		\subfloat{\includegraphics[width=0.33\textwidth]{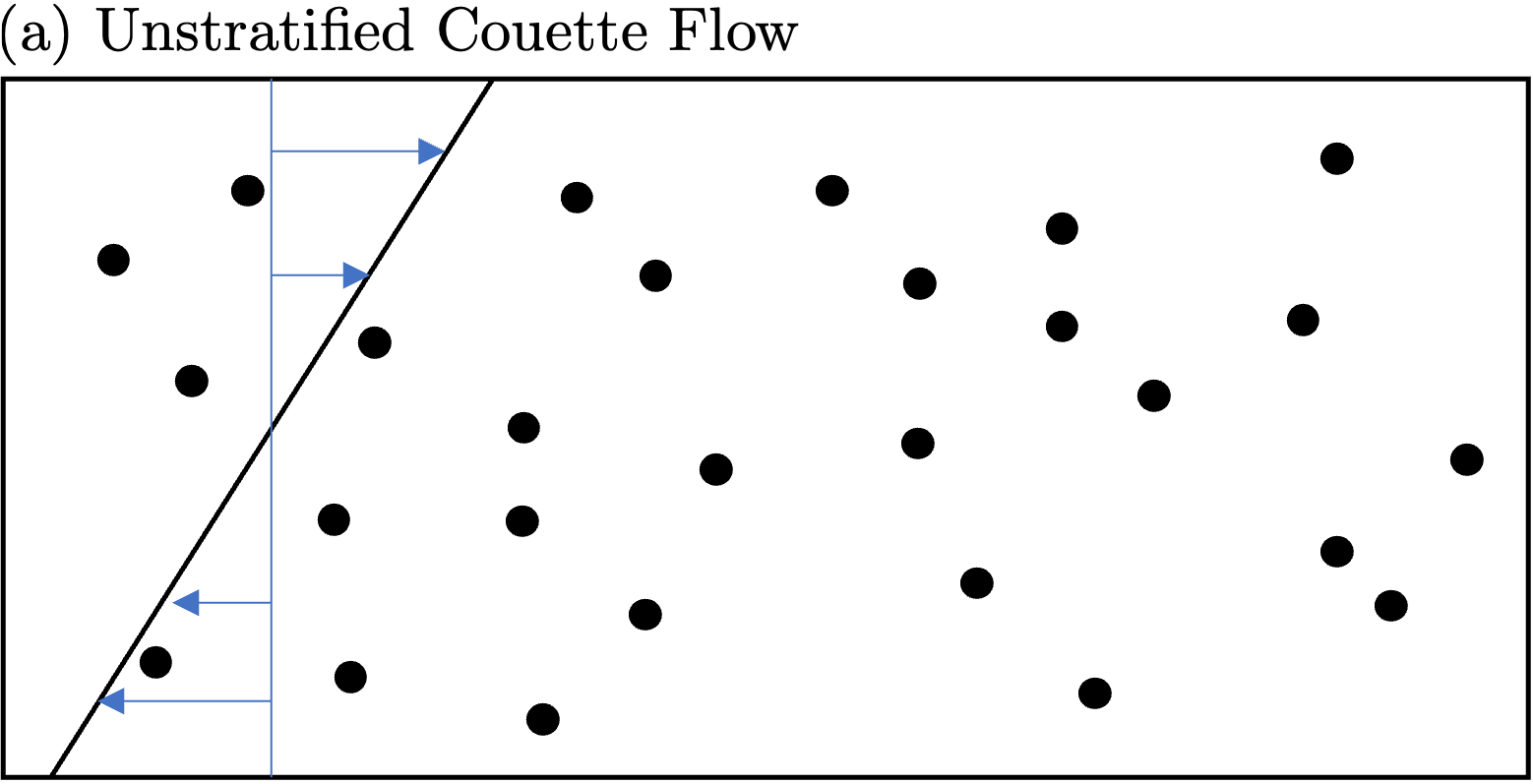} \label{UnstratifiedCouette}}
		\subfloat{\includegraphics[width=0.33\textwidth]{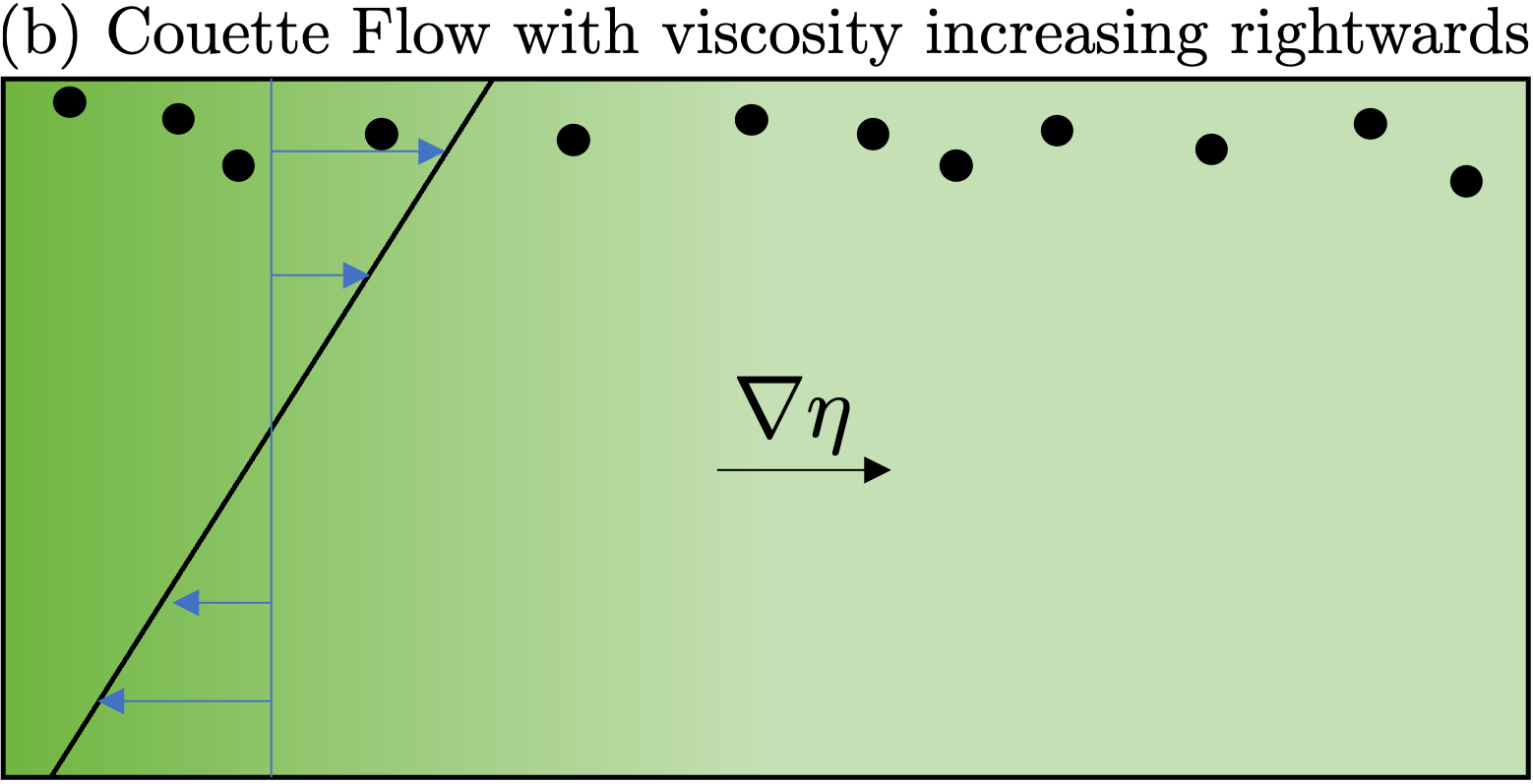}\label{StratifiedCouette} }
		\caption {Schematic of spheres dispersed in Couette flow of (a) unstratified and (b) stratified fluid with viscosity increase towards right. Based on the results of a single particle in an unbounded fluid, the particles migrate towards the wall that moves in the direction of increasing viscosity. \label{fig:Couette}}
	\end{figure}

\begin{figure}
		\centering
		\subfloat{\includegraphics[width=0.33\textwidth]{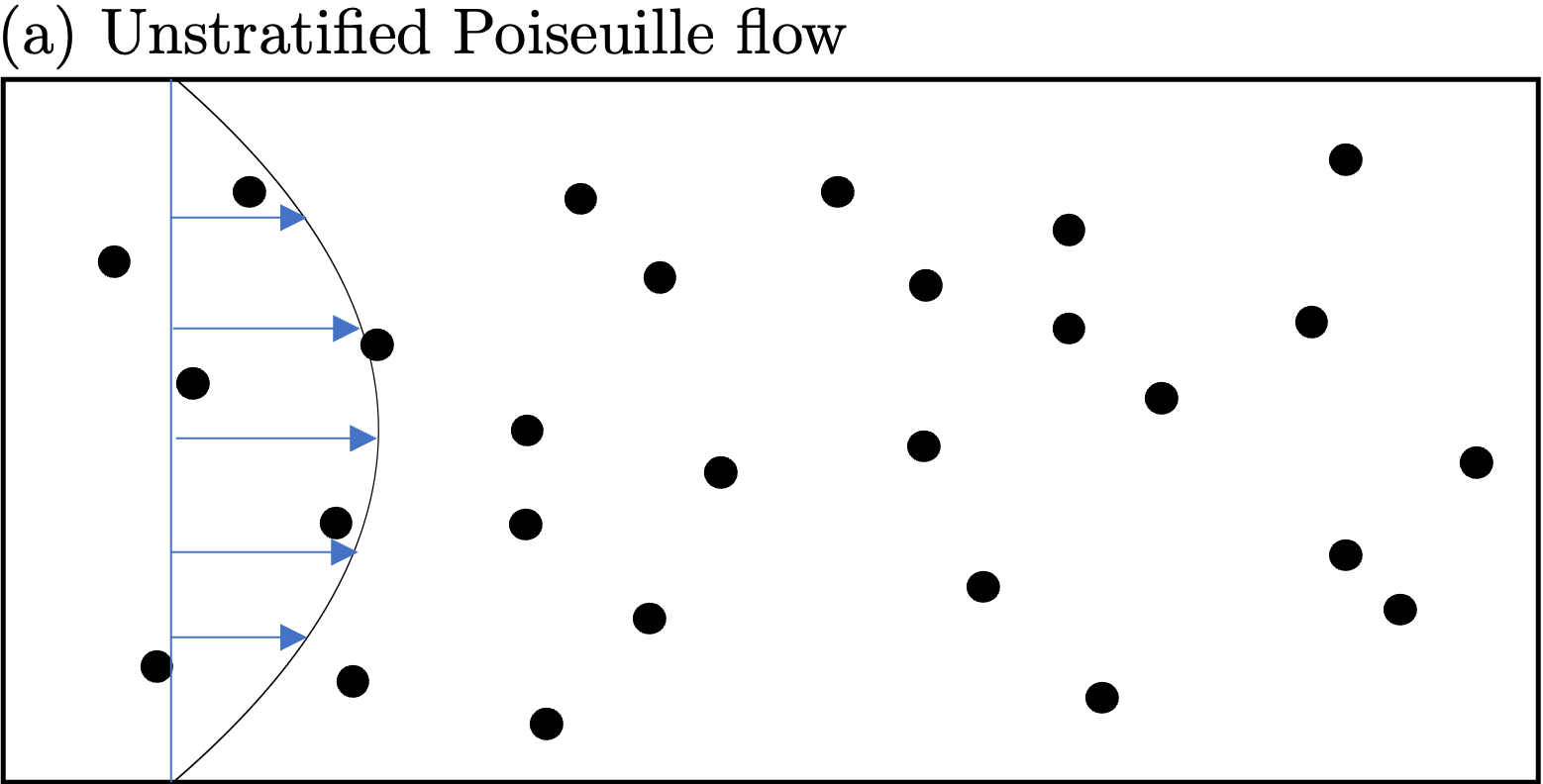} \label{UnstratifiedPoiseuille}}
		\subfloat{\includegraphics[width=0.33\textwidth]{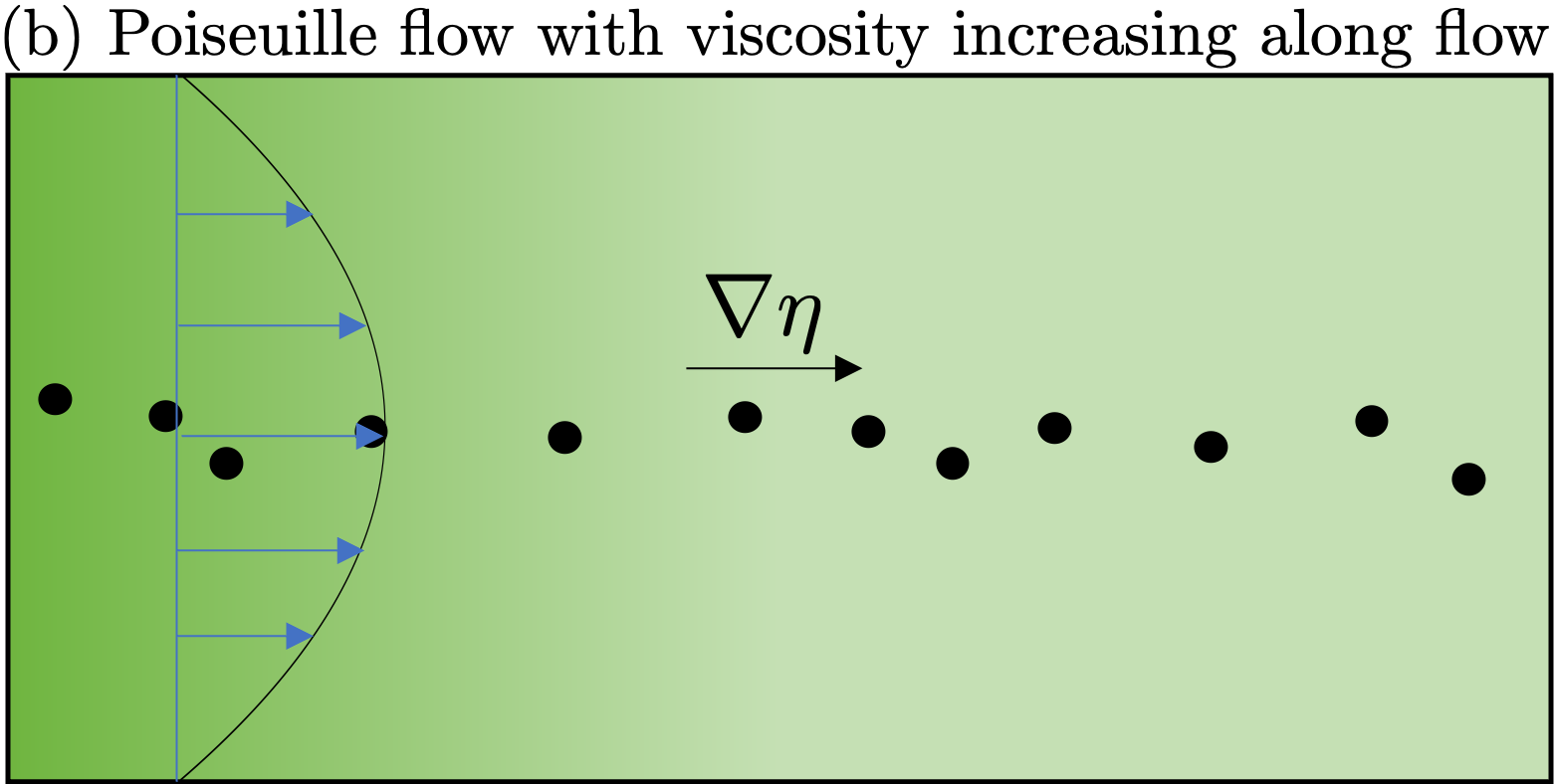}\label{StratifiedPoiseuille1} }
		\subfloat{\includegraphics[width=0.33\textwidth]{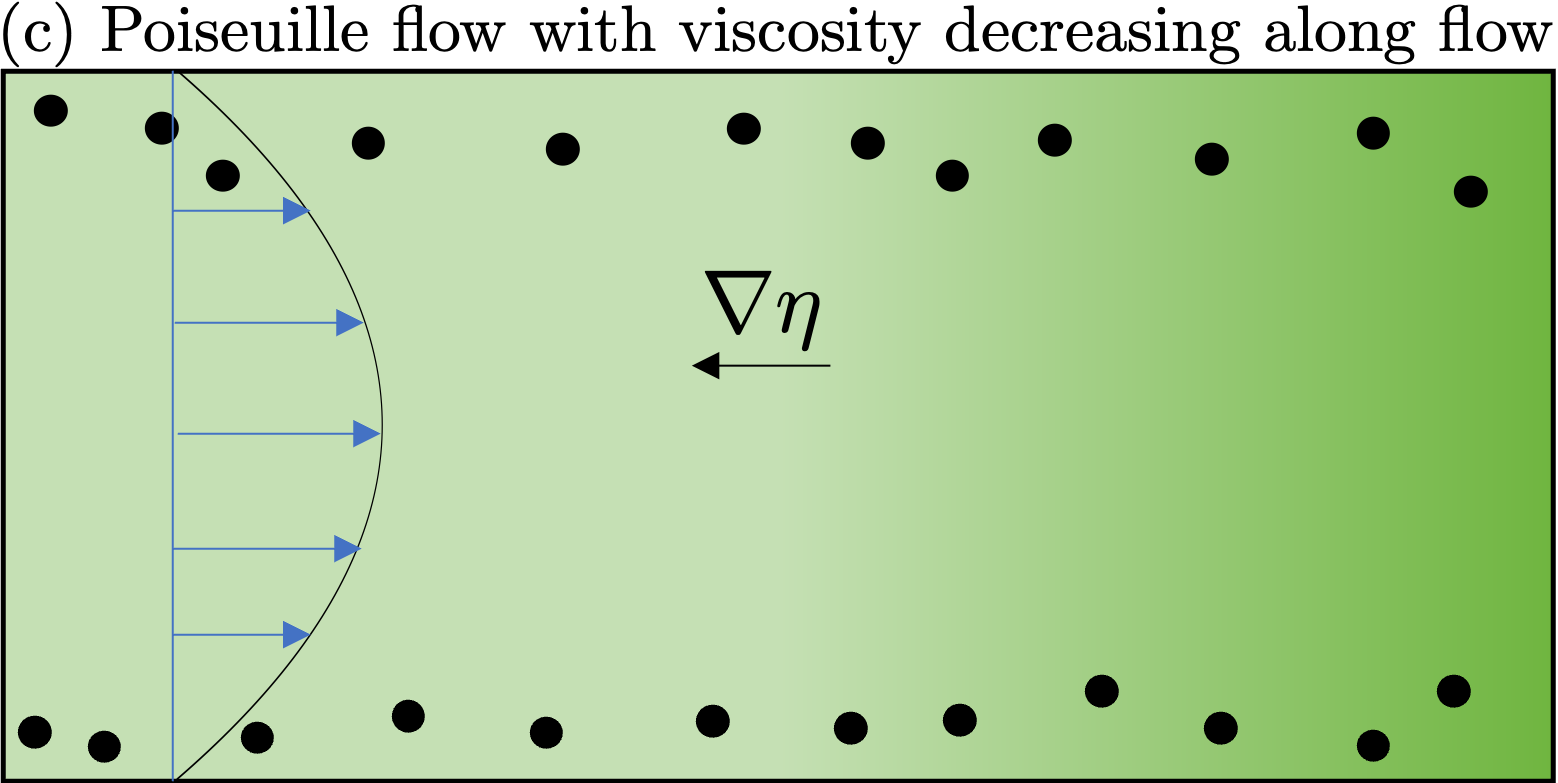}\label{StratifiedPoiseuille2} }
		\caption {Schematic of spheres dispersed in Poiseuille flow of (a) unstratified and stratified fluid with viscosity (b) increase and (c) decrease towards the flow direction. \label{fig:Poiseuille}}
	\end{figure}

Observing the effect of stratification in simple shear flow, we may conjecture this effect in particle-filled heated Couette and Poiseuille flows. If the viscosity is uniform, the position of the particles relative to the walls does not change (figures \ref{UnstratifiedCouette} and \ref{UnstratifiedPoiseuille}). If the inlet of a channel is at a different temperature than the outlet such that the viscosity increases along the channel length, in Couette flow, as schematically depicted in figure \ref{fig:Couette}, dispersed spheres will migrate towards the wall that moves in the same direction as the increasing viscosity. In the case of Poiseuille flow, if the viscosity increases along the flow direction, the particles migrate towards the center of the channel (figure \ref{StratifiedPoiseuille1}). If viscosity increases in the opposite direction, they move towards the walls due to the stratification-induced force and local shear (figure \ref{StratifiedPoiseuille2}). These hypotheses ignore inter-particle hydrodynamic interactions and assume the particles to be small enough such that locally they observe a simple shear flow in an unbounded fluid.
	
	
	In the case of uniaxial extensional flow, spheres move towards more viscous fluid if viscosity varies along the extensional axis, and towards less viscous fluid if viscosity varies along the compression axis. 
	n the case of uniaxial extensional flow, the stratification-induced relative velocity is $\beta l^2/\eta_0 \mathbf{E}\cdot\mathbf{d}+\mathcal{O}(\beta^2)$. Therefore, spheres move towards more viscous fluid if viscosity varies along the extensional axis, and towards less viscous fluid if viscosity varies along the compression axis.
	
	\subsection{Spheroids with $\kappa\ne 1$}\label{sec:SpheroidLinear}
\begin{figure}
		\centering
		\subfloat{\includegraphics[width=0.55\textwidth]{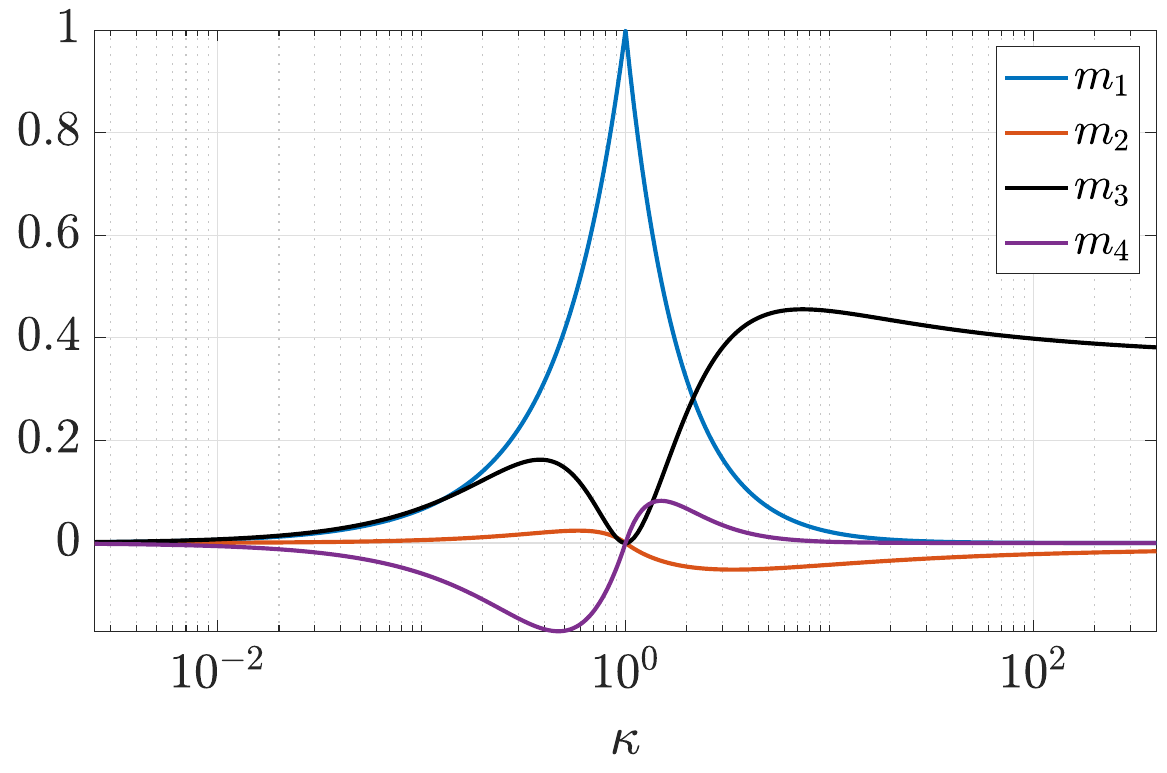}}
		\caption {Variation of coefficients, $m_i, i\in[1,4]$ in $\mathcal{O}(\beta)$ translation velocity (equation \eqref{eq:TranslCoeffs}) in linear flows around spheroids with aspect ratio $\kappa$ and major radius 1. \label{fig:LinearFlowTranslationCoefficientsSpheroid}}
	\end{figure}

	Next, we consider the effect of viscosity stratification on the translation trajectories of a spheroid with $\kappa\ne1$, where the particle's orientation also influences these dynamics (equation \eqref{eq:StratificationVelocity}), resulting in a more complex effect. The translation velocity of the particle relative to the fluid for a spheroid is given by
	\begin{equation}
		\mathbf{u}_\text{particle}-\mathbf{u}_\text{fluid}=\frac{\beta}{\eta_0} (m_1 \mathbf{E} + m_2 (\mathbf{p}\cdot\mathbf{E}\cdot\mathbf{p})\mathbf{I} +(m_2+m_3) (\mathbf{p}\cdot\mathbf{E}\cdot\mathbf{p})\mathbf{pp}+ m_4 [\mathbf{E}\cdot\mathbf{pp}+\mathbf{pp}\cdot\mathbf{E}])\cdot\mathbf{d}+\mathcal{O}(\beta^2),\label{eq:TranslCoeffs}
	\end{equation}
	where $m_i, i\in[1,4]$ as a function of $\kappa$ are displayed in figure \ref{fig:LinearFlowTranslationCoefficientsSpheroid}, and their analytical expressions are in equation \eqref{eq:StratLin} (the functional dependence of $m_i, i\in[1,4]$ on $f_1$, $f_3$, $q_1$, $q_3$ and $F_{ij}^{\Gamma_k},i,j\in[1,4], k\in[1,8]$ can be ascertained by comparing the RHS of the above equation with that of \eqref{eq:StratificationVelocity}). Here, only the first effects of the stratification-induced force on a rotating particle are accounted for, which arise from the rotation rate of the particle's axis of symmetry up to $\mathcal{O}(\beta)$. Thus, the particle rotates along Jeffery orbits (equation \eqref{eq:RotRateNewtonian}).
	
	\subsubsection{Spheroids in uniaxial extensional flow}
	Consider an extensional flow such that the imposed velocity gradient is $\Gamma_{ij}=\delta_{i1}\delta_{j1}-0.5(\delta_{i2}\delta_{j2}+\delta_{i3}\delta_{j3})=E_{ij},$ resulting in $\mathbf{u}_\text{fluid}=\mathbf{r}\cdot \mathbf{E}$. According to the particle rotation rate in a uniform viscosity fluid given by equation \eqref{eq:RotRateNewtonian}, a spheroid obtains a steady state orientation such that $\mathbf{E}\cdot\mathbf{p}=\alpha\mathbf{p}$, for a constant $\alpha$. Two possible values of $\alpha$ are +1 and -1/2, where one corresponds to a stable fixed point and the other to an unstable fixed point, depending upon the sign of $q_4/q_1$. In other words, a prolate spheroid orients with its major axis along the extensional axis ($\alpha=1$ is the stable case), and an oblate particle orients its face in a plane consisting of the extensional axis and one of the compressional directions ($\alpha=-1/2$ is the stable case). At this steady orientation, the spheroids move with the relative translation velocity,
	\begin{eqnarray}\begin{split}
			\mathbf{u}_\text{particle}-\mathbf{u}_\text{fluid}=\frac{\beta}{\eta_0 f_1}[m_1\mathbf{E}\cdot\mathbf{d}+ \alpha(m_2\mathbf{d}+(m_2+m_3+m_4)\mathbf{d}\cdot\mathbf{pp})]+\mathcal{O}(\beta^2),
		\end{split}
		\label{eq:ExtFlow}
	\end{eqnarray}
	where $\alpha=1$ for prolate and -1/2 for oblate spheroids. If the stratification $\mathbf{d}$ is directed along the extensional axis, the translation of the particles follows,
	\begin{equation}
		\mathbf{u}_\text{particle}-\mathbf{u}_\text{fluid}=\frac{\beta}{\eta_0 f_1}\begin{cases}
			(m_1+2m_2+m_3+m_4)\mathbf{d},\hspace{0.1in}\kappa>1\\
			(m_1-m_2/2)\mathbf{d},\hspace{0.1in}\kappa<1
		\end{cases}.\label{eq:Extdalongext}
	\end{equation}
	In the case where the stratification $\mathbf{d}$ is directed along the compressional axis,
	\begin{equation}
		\mathbf{u}_\text{particle}-\mathbf{u}_\text{fluid}=\frac{\beta}{\eta_0 f_1}\begin{cases}
			(-m_1/2+m_2)\mathbf{d},\hspace{0.1in}\kappa>1\\
			-1/2(m_1+2m_2+m_3+m_4)\mathbf{d},\hspace{0.1in}\kappa<1
		\end{cases}.\label{eq:ExtdalongComp}
	\end{equation}
	The coefficients of $\mathbf{d}$ inside the curly brackets in equation \eqref{eq:Extdalongext} are positive, and those in equation \eqref{eq:ExtdalongComp} are negative for all $\kappa$, with the largest magnitude occurring for a sphere (not shown). Hence, a viscosity gradient along the extensional axis makes prolate, oblate and spherical particles move towards more viscous fluid, and a viscosity gradient along a compression axis moves the spheroids towards the lower viscosity region.
	
	\subsubsection{Spheroids in simple shear flow}
	In simple shear flow with velocity gradient $\Gamma_{ij}=\delta_{i2}\delta_{j1}$, such that at the particle's center $u_{i,\text{fluid}}={r}_{2,\text{spheroid}}$, and considering just the Jeffery rotation of the spheroid's orientation, $\mathbf{p}=\begin{bmatrix}p_1&p_2&p_3 \end{bmatrix}^T$, the relative translation velocity is, \begin{align}\begin{split}
			u_{i,\text{particle}}-u_{i,\text{fluid}}=& \frac{\beta}{\eta_0} [0.5 m_1 (d_1\delta_{i2}+d_2\delta_{i1})+p_1p_2 (m_2d_i+(m_2+m_3)p_jd_jp_i) \\&+m_4((p_1\delta_{i2}+p_2\delta_{i1})p_kd_k+p_i(p_2d_1+p_1d_2))]+\mathcal{O}(\beta^2).
	\end{split}\end{align}
	The Jeffery orbits, or orientation trajectories, of spheroids in simple shear flow of a uniform viscosity fluid can be broadly classified into four types: (a) log-rolling, (b) tumbling, (c) flipping, and (d) wobbling orbits. The effect of stratification on the translation of a particle undergoing these orientation trajectories, along with the orbit descriptions, are discussed below.
	\begin{center}
		\textit{Log-rolling orbits}
	\end{center}
	A spheroid initially oriented along the vorticity direction (i.e. $p_1=p_2=0$ and $p_3=1$) does not change its orientation with time but simply rolls about its axis at angular velocity equal to half of the shear rate. This motion is thus referred to as log-rolling, where the translation velocity is
	\begin{equation}
		u_{i,\text{particle}}-u_{i,\text{fluid}}=0.5 m_1\frac{\beta}{\eta_0} ( (d_1\delta_{i2}+d_2\delta_{i1}))+\mathcal{O}(\beta^2).
	\end{equation}
	Since, $m_1$ does not change sign with $\kappa$ (figure \ref{fig:LinearFlowTranslationCoefficientsSpheroid}), the effect of stratification on a log-rolling spheroid is qualitatively similar to the motion of a sphere discussed in section \ref{sec:SphereLinear}. The magnitude of the stratification-induced velocity is largest for the sphere. In the case of a log-rolling spheroid, if stratification is along the vorticity direction of the imposed flow, i.e., $\mathbf{d}=\begin{bmatrix}
		0&0&1
	\end{bmatrix}$, the stratification-induced velocity is zero. Otherwise, the trajectory of the particle's centroid initially located at $\begin{bmatrix}0,&0,&0\end{bmatrix}^T$ moves along the parabola $0.5m_1 t \beta\begin{bmatrix} 0.5d_1t + d_2,& d_1, & 0\end{bmatrix}^T$. Hence, a particle placed at the origin moves along the flow direction if stratification is along the velocity-gradient direction of the imposed flow. For stratification along the flow direction, the particle is displaced along the gradient direction and then also swept along the flow direction by the imposed flow. Since $m_1>0$ for all spheroids (figure \ref{fig:LinearFlowTranslationCoefficientsSpheroid}), the particle is ultimately swept towards higher viscosity regions.
	
\begin{figure}
		\centering
		\subfloat{\includegraphics[width=0.45\textwidth]{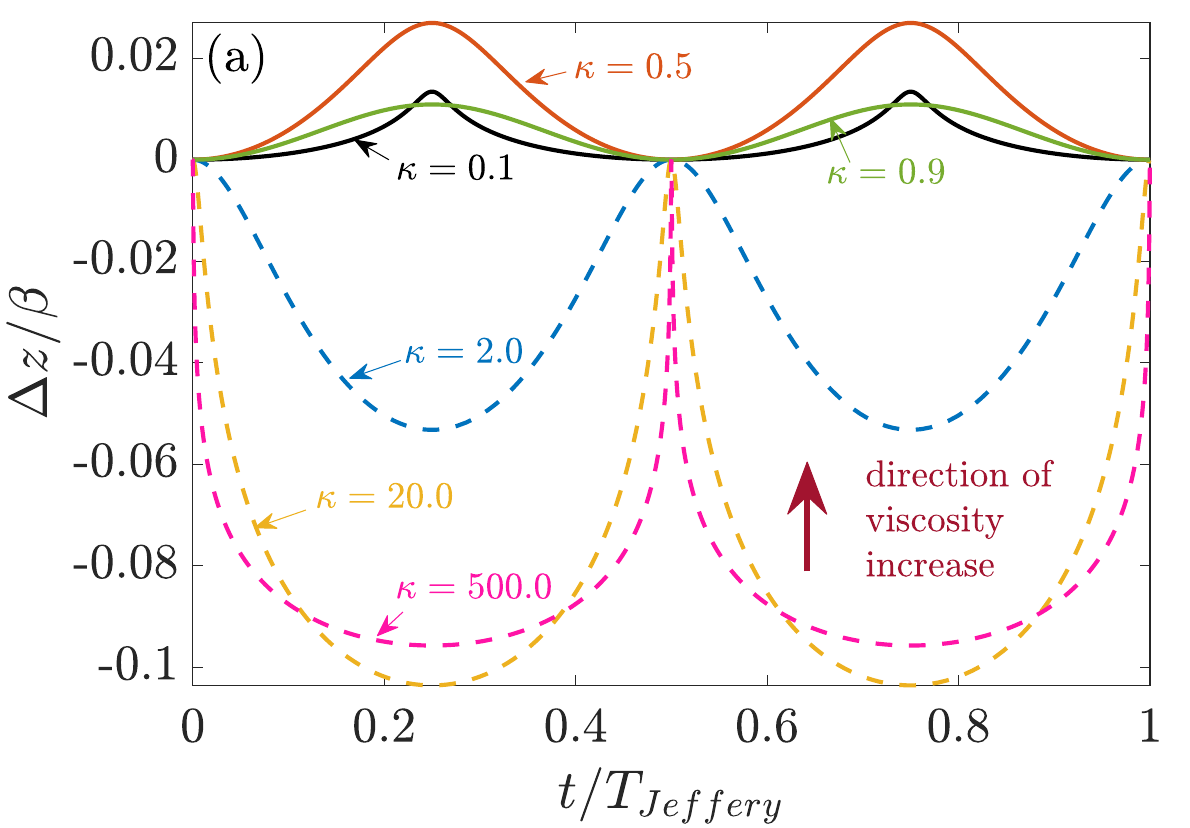}}\hspace{0.1in}
		\subfloat{\includegraphics[width=0.45\textwidth]{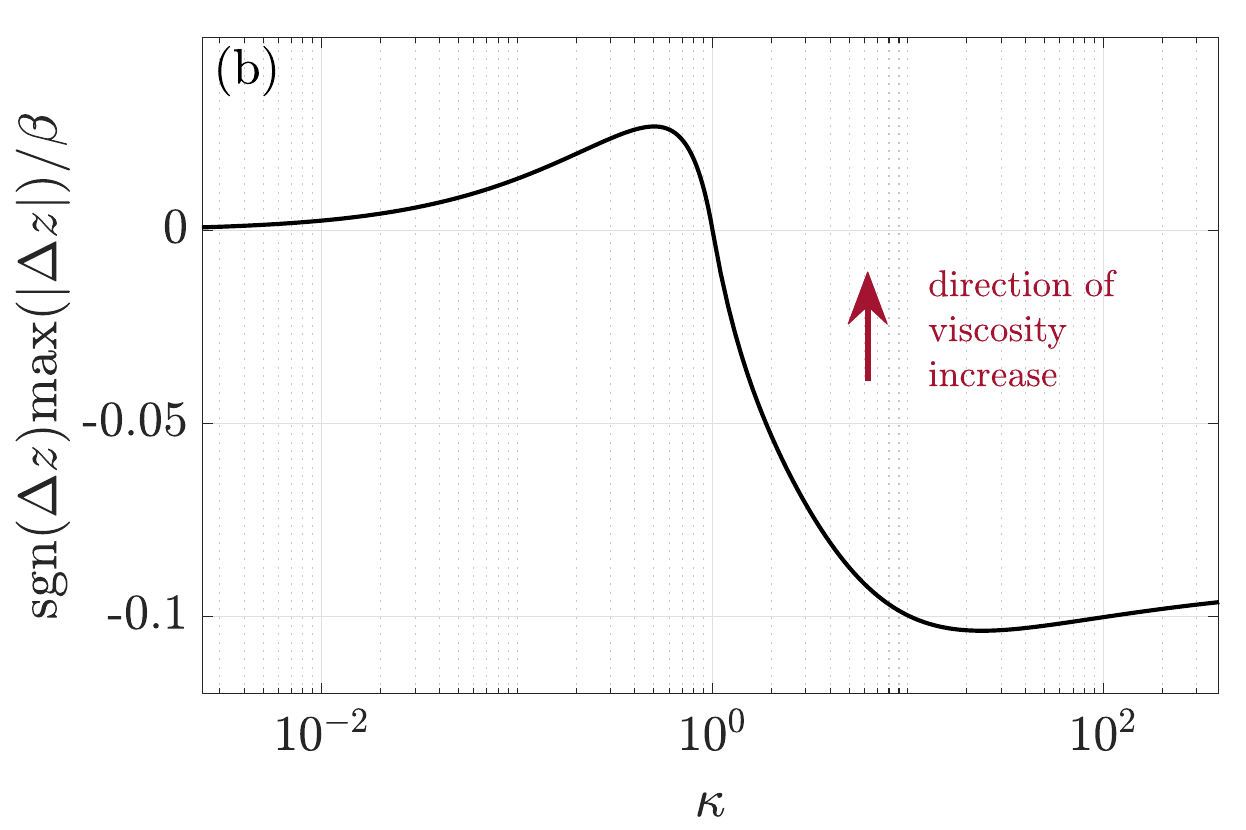}}
		\caption { (a) Excursion of spheroids (with major axis, $l=1$), $\Delta z/\beta$, tumbling in the flow-gradient (shearing) plane of simple shear flow along the vorticity direction when viscosity varies along the vorticity direction for different aspect ratio, $\kappa$. (b) Maximum excursion from the initial location as a function of $\kappa$. The largest magnitude of $\Delta z$ toward higher viscosity occurs for $\kappa\approx0.5$ and towards less viscous fluid for $\kappa\approx 20$. \label{fig:ViscosityinVorticityParticleinShearPlane}}
	\end{figure}

	\begin{center}
		\textit{Tumbling orbits}
	\end{center}
	A spheroid with initial orientation in the flow-gradient plane ($p_3=0$) remains there and continues to tumble in this tumbling orbit with a time period $T_\text{Jeffery}=2\pi(\kappa+1/\kappa)$ (normalized with the shear rate). The stratification-induced translation velocity of a tumbling spheroid is,
	\begin{equation}
		\begin{bmatrix} 	u_{1,\text{particle}}-u_{1,\text{fluid}}\\	u_{2,\text{particle}}-u_{2,\text{fluid}}\\	u_{3,\text{particle}}-u_{3,\text{fluid}}	\end{bmatrix}=\frac{\beta}{\eta_0} \begin{bmatrix}0.5 m_1 d_2 +m_2d_1p_1p_2+(p_1d_1+p_2d_2)(p_1^2p_2+m_4(p_1+p_2))\\
			0.5 m_1 d_1 +m_2d_2p_1p_2+(p_1d_1+p_2d_2)(p_1p_2^2+m_4(p_1+p_2))\\
			m_2d_3p_1p_2
		\end{bmatrix}.\label{eq:TumblingVelocity}
	\end{equation}
	Considering first the simpler case of stratification along the vorticity direction, i.e., $d_1=d_2=0, d_3=1$, the particle moves along the viscosity gradient direction with a rate $\beta p_1p_2m_2$. From figure \ref{fig:LinearFlowTranslationCoefficientsSpheroid}, we observe that $m_2$ changes sign at $\kappa=1$; therefore, at a particular orientation within the shearing plane, an oblate spheroid moves in the opposite direction to the prolate spheroid due to stratification effects. However, either particle's motion along the vorticity direction in this case is reversed when the particle is in the extensional quadrant ($p_1p_2>0$) compared to when it is in the compression quadrant ($p_1p_2<0$). Since a spheroid spends equal amounts of time in these quadrants, its centroid's translation velocity oscillates about the initial value. Therefore an oblate (prolate) particle, started with its centerline along the gradient direction, makes an excursion towards the higher (lower) viscosity region and periodically returns to its original location. A few examples of the particle's normalized displacement from its original location in the vorticity direction $\Delta z/\beta$ are shown in the left panel of figure \ref{fig:ViscosityinVorticityParticleinShearPlane}.
	
	Not only does $m_2$ decrease for aspect ratios further from the sphere ($\kappa\ll1$ and $\kappa\gg1$ in figure \ref{fig:LinearFlowTranslationCoefficientsSpheroid}), but for the majority of Jeffery orbits, a prolate spheroid spends most of the time oriented along the flow direction ($p_1=1$, $p_2=p_3=0$) and an oblate with its face in flow-vorticity plane ($p_2=1$, $p_1=p_3=0$), i.e., the orientations where the induced velocity $\beta p_1p_2m_2$ is zero. This implies that for $\kappa\ll 1$ and $\kappa\gg1$, the amplitude of the aforementioned oscillations in the position of the particle's centroid is smaller for these extreme values of $\kappa$. Oscillations are zero for a sphere, and hence an optimal $\kappa$ exists for both the $\kappa<1$ and $\kappa>1$ regimes where the particle oscillates with the largest amplitude. The right panel of figure \ref{fig:ViscosityinVorticityParticleinShearPlane} shows the $\kappa$ variation of the location of the particle's maximum excursion from its original location in the vorticity direction for a viscosity increase along the positive vorticity direction.
	
\begin{figure}
		\centering
		\subfloat{\includegraphics[width=0.45\textwidth]{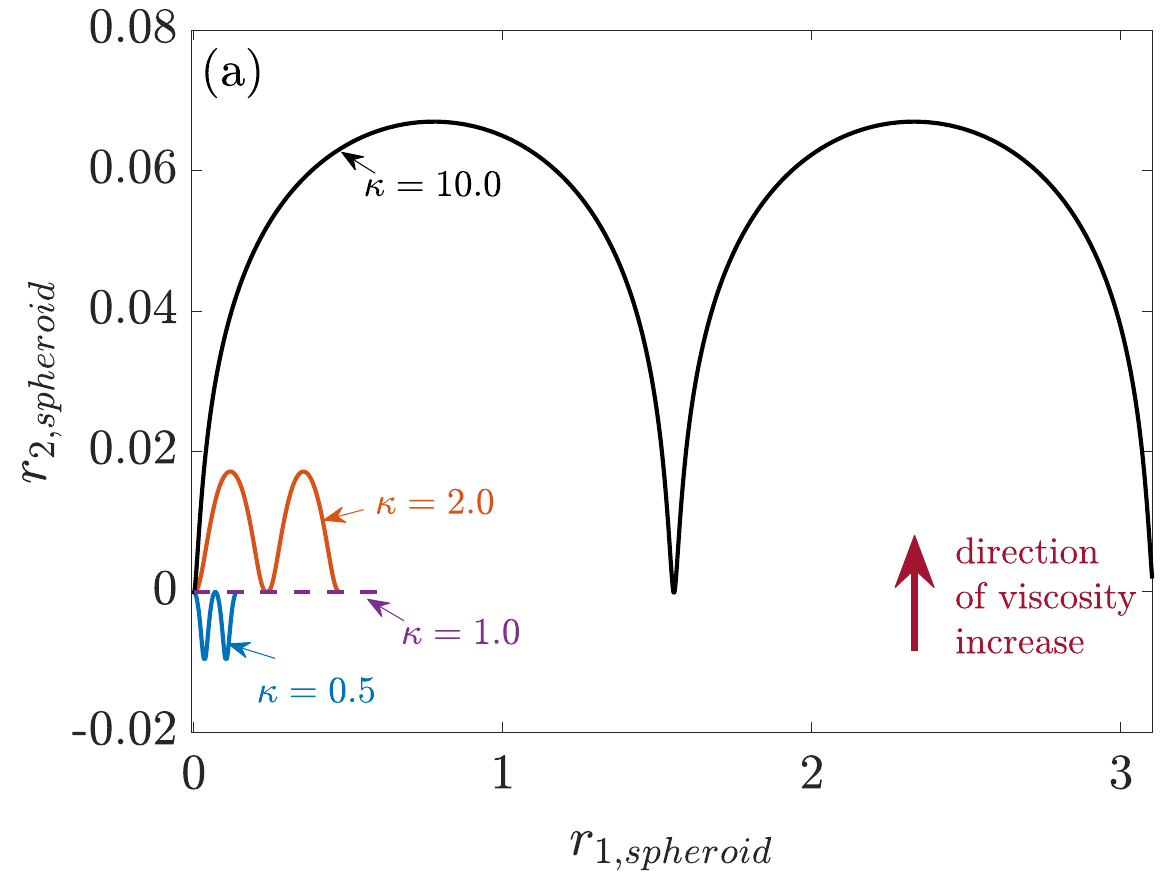}\label{fig:ViscosityinGradeientFlowParticleinShearPlanea} }\hspace{0.1in}\hspace{0.1in}
		\subfloat{\includegraphics[width=0.45\textwidth]{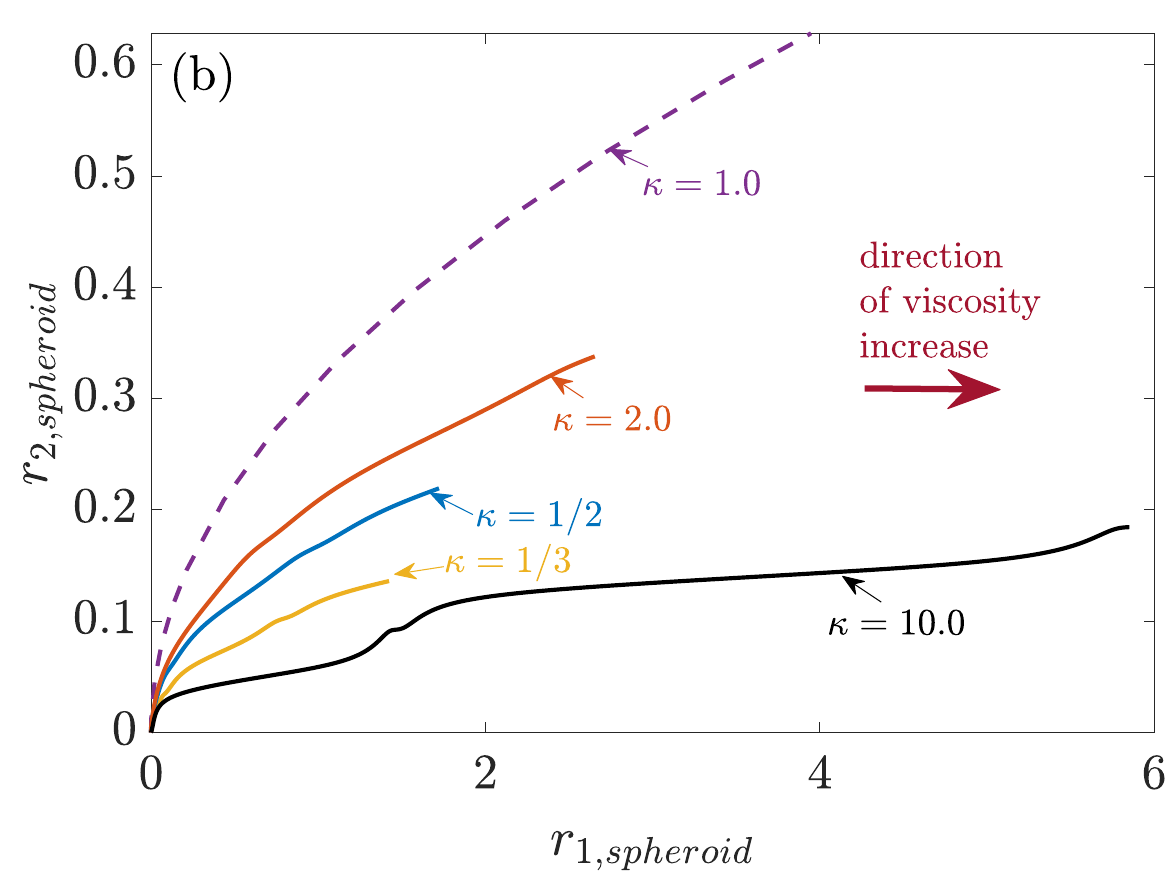}\label{fig:ViscosityinGradeientFlowParticleinShearPlaneb}}
		\caption {Translation trajectories for various spheroids (with major axis, $l=1$) when viscosity varies along (a) gradient direction and (b) flow direction. The viscosity gradient magnitude, $\beta=0.1$. \label{fig:ViscosityinGradeientFlowParticleinShearPlane}}
	\end{figure}

	Similar to the log-rolling case discussed above, for a particle in a tumbling orbit, if the stratification is in the flow-gradient plane (i.e., $d_3=0$) the particle will only translate within this plane as from equation \eqref{eq:TumblingVelocity} $d{r}_{3,\text{spheroid}}/dt|_\text{tumbling}=0$. However, unlike the log-rolling scenario, here a stratification purely in the gradient direction ($d_1=d_3=0$) will cause the particle to not only move along the flow but also along the gradient or the shearing direction. The trajectories of $\kappa=1/2, 1, 2$ and 10 particles over one respective Jeffery time period are shown in figure \ref{fig:ViscosityinGradeientFlowParticleinShearPlanea} when the viscosity increases along the gradient direction and $\beta=0.1$. Figure \ref{fig:ViscosityinGradeientFlowParticleinShearPlaneb} shows the trajectory for the same parameters, but with viscosity increasing along the flow direction.
	
\begin{figure}
		\centering
		\subfloat{\includegraphics[width=0.45\textwidth]{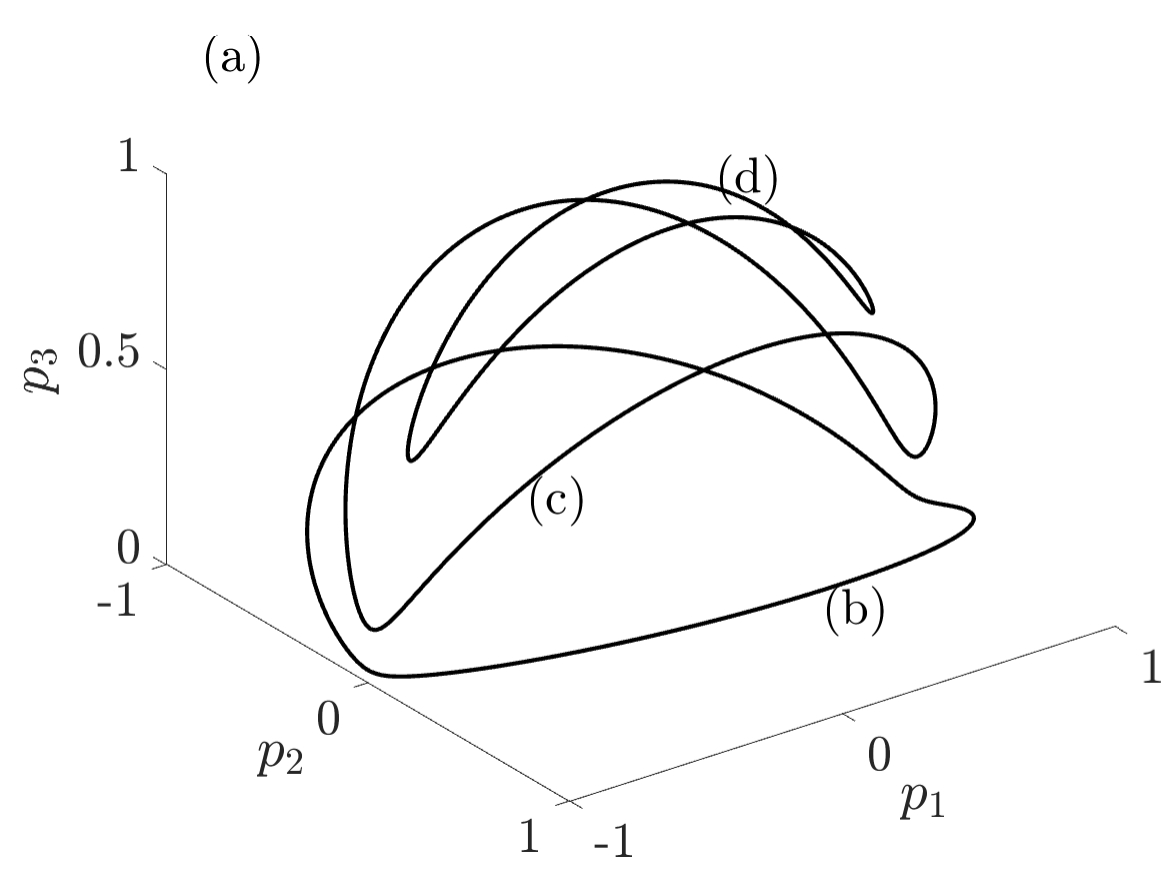}}
		\subfloat{\includegraphics[width=0.45\textwidth]{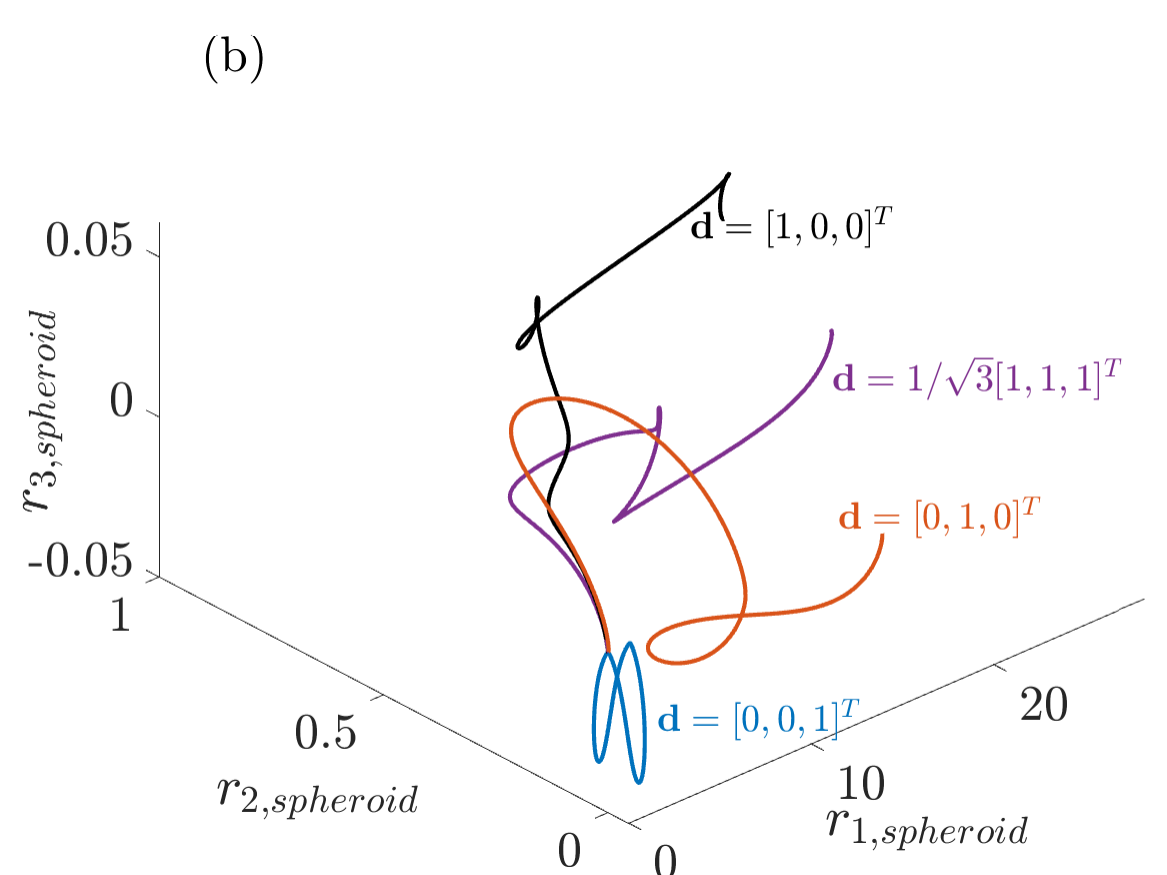}}\\
		\subfloat{\includegraphics[width=0.45\textwidth]{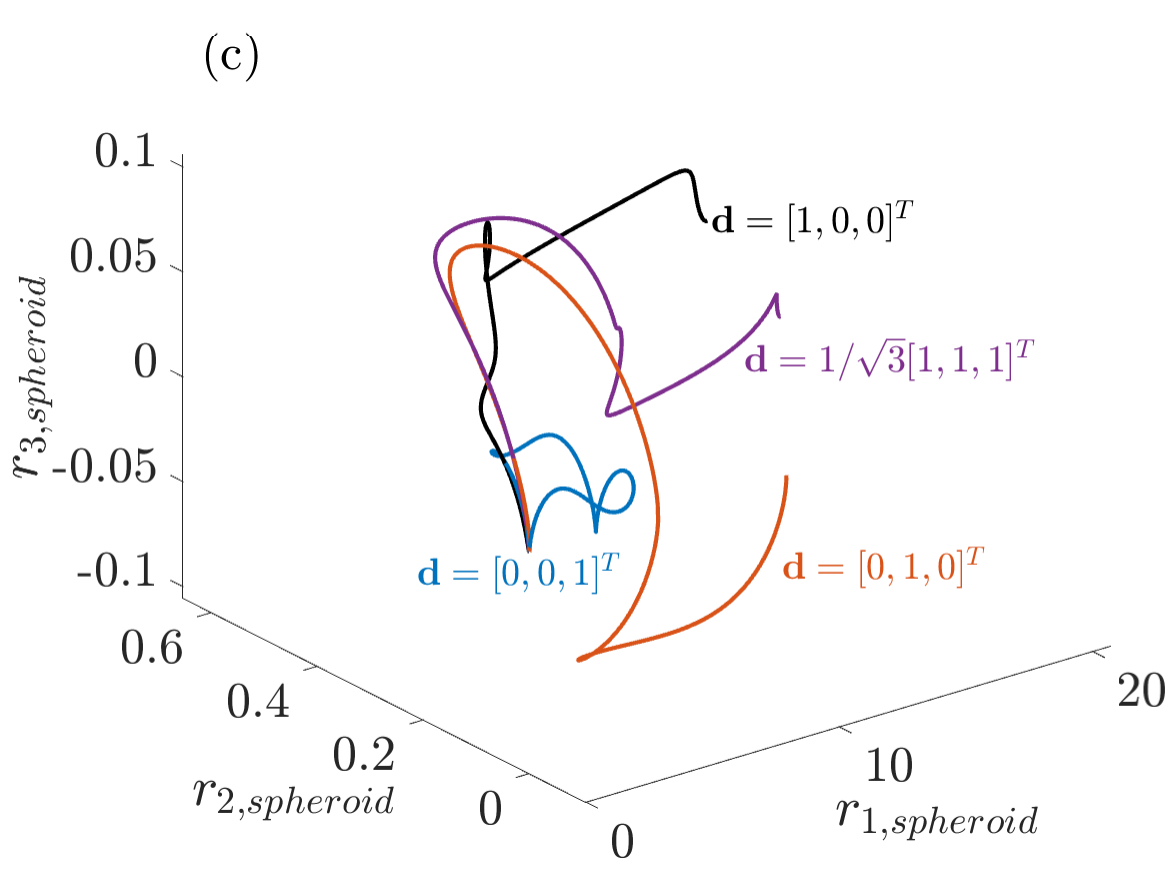}}
		\subfloat{\includegraphics[width=0.45\textwidth]{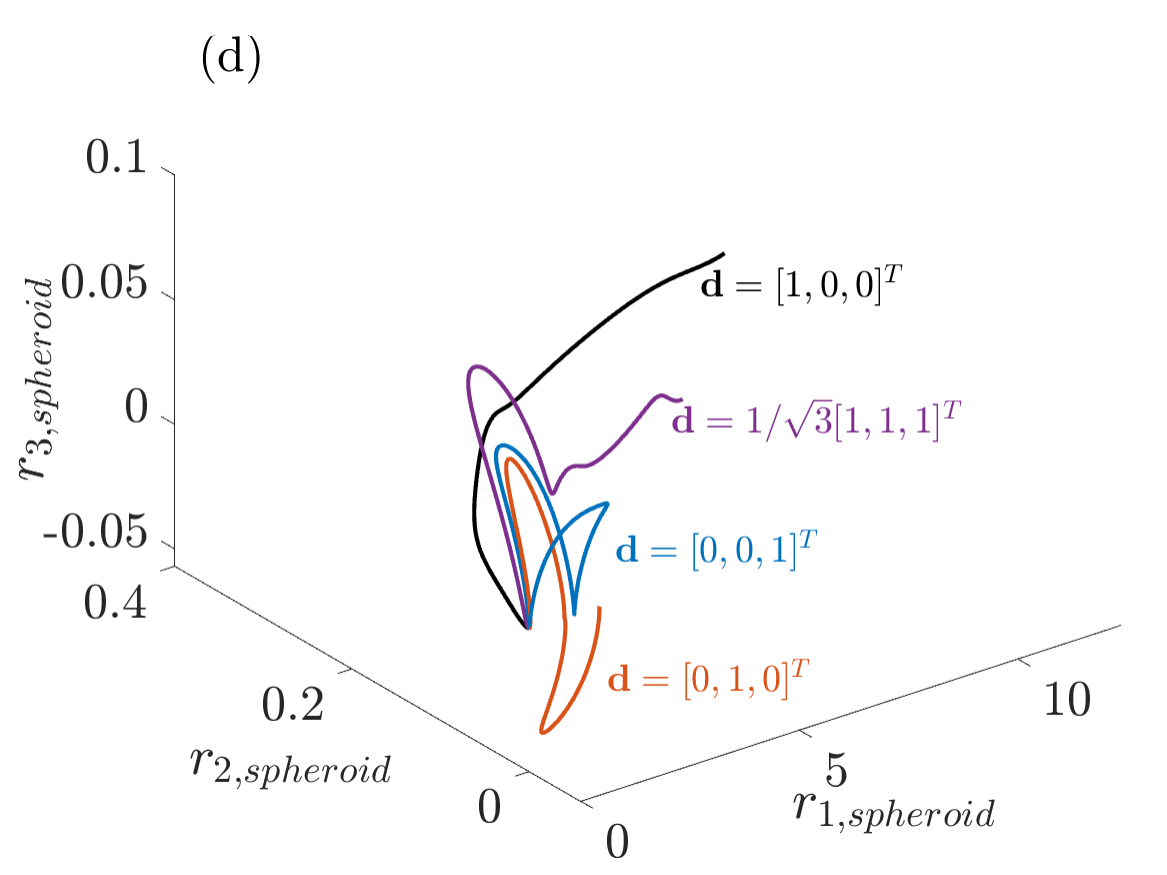}}
		\caption {Translation trajectories ((b), (c) and (d)) of a $\kappa=10$ spheroid with $\beta=0.1$ for different flipping (b, c) and wobbling (d) orientation trajectories shown in (a) and different directions of viscosity stratification, $\mathbf{d}$. \label{fig:FlippingandWobbling}}
	\end{figure}

	\begin{center}
		\textit{Flipping and wobbling orbits}
	\end{center}
	For an initial orientation close to but not on the flow-gradient plane, a spheroid with either large (prolate) or small (oblate) $\kappa$ spends most of its Jeffery orbit near the flow direction. Within a small time frame, it flips from one side of the gradient-vorticity plane to the other, and during this time, it traverses a larger three-dimensional orientation space. These are termed flipping orbits. For initial orientations close to but not on the vorticity axis, Jeffery orbits are also three-dimensional, but in these wobbling trajectories, the rotation of the particle throughout its orbit is more uniform than in the flipping orbits. Due to the three-dimensional rotation of the particle, shown in figure \ref{fig:FlippingandWobbling} for $\kappa=10$, the stratification-induced force translates the particle in a three-dimensional manner.
	
	In section \ref{sec:SphereSediment} we speculated that beyond the $\mathcal{O}(\beta)$ effects considered in this paper, a sedimenting sphere is expected to follow a curved settling trajectory due to the stratification induced force on the rotating particle. There the rotation induced at $\mathcal{O}(\beta)$ might be expected to induce a stratification force and modify the particle's translation motion at $\mathcal{O}(\beta^2)$. Similarly, here, for a spheroid in simple shear flow a stratification induced torque at $\mathcal{O}(\beta^2)$ may lead to a rotational motion of the particle that deviates from Jeffery orbits.
	
	\section{Conclusion and future directions}\label{sec:Conclusions}
	We have demonstrated the effect of {small} viscosity gradients in an inertia-less, incompressible fluid on the force, torque, and motion of a spheroid in various flow situations. The viscosity stratification (VS) induced torques and forces are obtained analytically through a combination of previously presented spheroidal harmonics \citep{dabade2015effects,dabade2016effect} and a generalized reciprocal theorem. In a uniform flow, where a particle in a constant viscosity fluid only experiences hydrodynamic drag, VS introduces a torque. In linear flows (such as simple shear, uniaxial extension, etc.), a spheroid may experience only a torque in a constant viscosity fluid, but VS leads to a force. Consequently, a freely sedimenting particle under gravity settles differently in a stratified fluid than in a uniform viscosity fluid. In a linear flow (such as a relative rotation between particle and fluid), while a spheroid does not translate relative to the local fluid with constant viscosity, the effect of variable viscosity breaks this symmetry as well. Therefore, the coupling between the imposed flow and the particle's rotation and translation due to VS leads to novel behavior even in the motion of the simplest spheroid, i.e., a sphere. It moves across the streamlines in simple shear flow and does not stay at the stagnation point of uniaxial extensional flow. Motion in simple shear may inspire particle sorting strategies based on controlling the viscosity of the fluid by altering temperature in microfluidics applications (figures \ref{fig:Couette} and \ref{fig:Poiseuille}). The effect of the VS force and torque is more profound for non-spherical spheroids, where the particle orientation plays an important role.
	
	If viscosity stratification is perpendicular to the free-stream velocity, a VS torque is induced on a fixed spheroid perpendicular to the flow-stratification plane. The torque that occurs when viscosity variation is along the spheroid's axis of symmetry and flow is perpendicular to it changes sign at $\kappa\approx 0.56$ and the torque when the stratification and flow directions are switched, changes sign at $\kappa\approx 2.0$. This sign change is the result of competition between the VS torque arising from $(\eta'/\eta_0)\boldsymbol{\sigma}^\text{Stokes}$, i.e., the Stokes stress acting in variable viscosity environment, and that from $\boldsymbol{\sigma}^\text{Stratified}$, i.e., the stress created due to modification of velocity and pressure by stratification. Near $\kappa=1$ (sphere), $(\eta'/\eta_0)\boldsymbol{\sigma}^\text{Stokes}$ dominates the torque generating mechanism, whereas $\boldsymbol{\sigma}^\text{Stratified}$ has a greater effect in the $\kappa\lessapprox0.56$ and $\kappa\gtrapprox2.0$ regimes. Within $\boldsymbol{\sigma}^\text{Stratified}$, the stratification-induced pressure is dominant, and contours of this variable, $p^\text{Stratified}$, as well as $\eta'p^\text{Stokes}$ (figure \ref{fig:StokesPresQ23U1} and \ref{fig:StokesPresQ12U3}), are used to illuminate the VS torque generation mechanism. Unlike the constant viscosity scenario, the pressure distribution is anti-symmetric in the directions perpendicular and parallel to the flow, leading to a torque (but still no force). A similar breaking in symmetry of pressure explains a VS force generated on a particle fixed in linear flows (figure \ref{fig:SchematicsLinearFlow}). Extra viscous stress in stratified fluid also plays a similar role to pressure in generating VS force and torque.
	
	A sphere settling under gravity in a fluid with a linear viscosity gradient (with magnitude $\beta$) {may} experience a horizontal drift along with vertical settling due to $\mathcal{O}(\beta^2)$ effects of viscosity stratification perpendicular to the gravity direction. For a spheroid with $\kappa\ne1$, novel settling dynamics arise at $\mathcal{O}(\beta)$. The types of orientation dynamics are illustrated using a two dimensional phase diagram (figure \ref{fig:PhaseDiagram}) in variables $\kappa$ and $d_g=\mathbf{d}\cdot\mathbf{g}/||\mathbf{g}||_2$ (the alignment of stratification, $\mathbf{d}$, and gravity, $\mathbf{g}/||\mathbf{g}||_2$). Depending upon $\kappa$ and $d_g$, a particle may obtain a stable steady state orientation in the gravity-stratification plane or on an axis perpendicular to it, spiral towards or away from the gravity-stratification plane, or rotate in non-uniform periodic orbits (in orientation space). Apart from periodic orbits, all the particle orientation trajectories approach a stable attractor in orientation space, thus showing an initial-orientation-independent behavior. A spheroid ultimately settles at a constant angle relative to gravity. This is in contrast to the motion in constant viscosity fluid, where the particle maintains its original orientation, and hence its settling angle is set by its initial condition (and $\kappa$). The sign switching discussed in the preceding paragraph plays a key role, as the orientation dynamics behavior is qualitatively most sensitive around $\kappa\lessapprox0.56$ and $\kappa\gtrapprox2.0$. Interestingly, the aspect ratios of \textit{Paramecium} and \textit{Escherichia coli} \citep{kreutz2012morphological,kaya2009characterization,liu2014real} are around 2.0 and micro-plastic population within the oceans is found to mostly have aspect ratio in the range $0.5<\kappa<2.5$ \citep{kooi2021characterizing}.
	
	Rotational-translation coupling due to viscosity stratification may affect the settling dynamics of a spheroid in two ways. Firstly, the VS torque changes the orientation of the particle's axis of symmetry, $\mathbf{p}$. This alters the sedimentation velocity, as the Newtonian sedimentation of a spheroid in a constant viscosity fluid depends on $\mathbf{p}$. Trivially, this does not affect sedimentation in the case of a sphere. Secondly, as the spheroidal particle rotates, the relative rotation between fluid and the particle leads to a stratification-induced force. For a freely settling particle, this second mechanism of altering the dynamics is activated only at higher orders in $\beta$ and requires continuous rotation of the particle. It is not accounted for in the present study at $\mathcal{O}(\beta)$, but may play an important role in larger viscosity gradients in an experiment.

	The dynamics of freely suspended spheroids is studied in two types of linear flows: simple shear and uniaxial extension. In uniaxial extension, a spheroid in a constant viscosity fluid orients its axis of symmetry along the extensional axis. In this orientation, a VS force is generated on the particle that moves it towards more viscous regions due to stratification along the extensional axis and lower viscosity fluid due to stratification along the compression axis. This can be explained simply by considering the $\eta'p^\text{Stokes}$ component of the extra stress in the stratified fluid. The pressure $p^\text{Stokes}$ acts to pull the particle surface outwards in the extensional direction on both sides of the particle, whereas it pushes the surface inwards in the compression direction. The variable component of viscosity $\eta'$ breaks this symmetry, thus translating the particle. In simple shear flow, in addition to the direction of stratification, the translation is influenced by the initial orientation of the spheroid and has a three dimensional structure if the particle is not initially oriented in the flow-gradient plane or the vorticity direction of the imposed flow. Similar to the curved settling trajectory of a freely sedimenting sphere, we speculate that the viscosity stratification will alter the orientational dynamics of a spheroid at $\mathcal{O}(\beta^2)$. 
	
	Our demonstration of VS induced torque and force, as well as the rotation-translation coupling in VS fluids, inspires further studies and applications. In liquids and gases, viscosity is determined by another factor or scalar, $s$, such as temperature \citep{kampmeyer1952temperature}, or concentration of a secondary species like salt \citep{jones1933viscosity}. While we have only considered linear spatial variation in viscosity, experimental observations indicate that viscosity dependence on these scalars is often non-linear \citep{ferreira2017viscosity}. Therefore, even if such scalars vary linearly in space, for example, in the case of a linear temperature change across a Couette cell, the spatial dependence of viscosity will be non-linear. The VS force on a spheroid experiencing relative uniform flow, or the VS torque on a spheroid experiencing linear flow, is zero because the relevant integrand is an odd function of position. However, as discussed in the beginning of section \ref{sec:GeneralParticle}, if the quadratic and higher-order spatial variations in the viscosity field are accounted for, a finite VS force in uniform flow and a VS torque in linear flows will be generated. Investigations of particle dynamics that incorporate more realistic viscosity properties such as by solving the scalar transport equation, perhaps using numerical techniques, are also useful extensions of the current work and are likely to reveal more novel particle dynamics that can be harnessed for practical applications.
	
	Based on the observation of the force exerted on a sphere in simple shear flow, we have conjectured the lateral migration of particles in Couette and Poiseuille flow in section \ref{sec:SpheroidsLinear}. These conjectures can be tested analytically or numerically by accounting for the finite size of the particle and its proximity to the wall, and they motivate future experiments and particle sorting applications.
	
	As a sphere rotates upon settling in a constant viscosity gradient, it will induce a velocity on another particle perpendicular to the vector joining their centers. While two identical spheres sediment with no relative velocity in a uniform viscosity fluid, the effects of particle-particle interaction in a stratified fluid may break this symmetry and cause them to approach and rotate around one another. This may be {further explored} analytically using the method of reflections {following the work of} \cite{ziegler2022hydrodynamic}, possibly elucidating yet another novel mechanism generated by viscosity variation. Lastly, as mentioned at the end of section \ref{sec:Intro} and at the beginning of section \ref{sec:GeneralParticle}, the dynamics of other fore-aft and axisymmetric particles such as cylinders, biconcave discs (red blood cells), bispherical objects, dumbbells, rings, etc., in a viscosity-stratified fluid for a variety of flows can be obtained using the equations presented in section \ref{sec:AbstractStratificationMoving}. This can be achieved after acquiring only limited data (non-zero force and torque components listed in table \ref{tab:NewtonianCoeffsStratifiedForces}) for the particle shape under consideration from a suitable numerical solver. Viscosity variation provides a new avenue to control the particle motion within liquids in engineering applications and must be accounted for to fully understand this motion in natural scenarios.
	
\noindent\small{\textbf{Acknowledgments:} We would like to express our gratitude to the anonymous reviewers for their valuable comments, which clarified the derivation of the stratification-induced force and torque using the reciprocal theorem.}\\
	\small{\textbf{Funding:} This work was supported by the U.S. Department of Energy, Office of Science, Advanced Scientific Computing Research (ASCR) Early Career Research Program. This paper describes objective technical results and analysis. Any subjective views or opinions that might be expressed in the paper do not necessarily represent the views of the U.S. Department of Energy or the United States Government. Sandia National Laboratories is a multimission laboratory managed and operated by National Technology and Engineering Solutions of Sandia, LLC, a wholly owned subsidiary of Honeywell International Inc., for the U.S. Department of Energy's National Nuclear Security Administration under contract DE-NA0003525. SAND-NO2024XXXJ.}\\
	\textbf{Competing Interests:} The authors report no conflict of interest.
	
	\bibliographystyle{jfm}
	\bibliography{MainDocument}
	\appendix
	\section{Stratification-induced force and torque obtained using a generalized reciprocal theorem}\label{sec:Reciprocal}
	In this appendix we present the derivation of the stratification-induced force and torque as a function of the Stokes flow fields, shown in equation \eqref{eq:StratificationForceTorque} in the main text. To apply the reciprocal theorem, it is convenient to write the stratified momentum equation \eqref{eq:StratificationProblem} in an equivalent form,
	\begin{equation}
		\nabla\cdot \widetilde{\mathbf{u}}^\text{Stratified}=0, \hspace{0.2in}\nabla\cdot\widetilde{\boldsymbol{\sigma}}^\text{Stratified}+2\beta\nabla\cdot[\eta'(\mathbf{e}^\text{Stokes}-\mathbf{E}_\infty)]=0,\label{eq:StratificationProblemMod}
	\end{equation} with boundary conditions,
	\begin{equation}
		\widetilde{\mathbf{u}}^\text{Stratified}=0, \text{on the particle surface, and as }\mathbf{x}\rightarrow\mathbf{x}_\text{out}.\label{eq:StratificationProblemModBC}
	\end{equation}
	Here, \begin{eqnarray}
		&\widetilde{\boldsymbol{\sigma}}^\text{Stratified}=\boldsymbol{\sigma}^\text{Stratified}-\boldsymbol{\sigma}^\text{Stratified}_\infty-\beta\frac{\eta'(\mathbf{x})}{\eta_0}p^\text{Stokes}\mathbf{I}, \\
		&\widetilde{\boldsymbol{\sigma}}^\text{Stratified}=-\widetilde{p}^\text{Stratified}\mathbf{I}+(\eta_0+\beta\eta'(\mathbf{x}))[\nabla\widetilde{\mathbf{u}}^\text{Stratified}+(\nabla\widetilde{\mathbf{u}}^\text{Stratified})^T]-\beta\frac{\eta'(\mathbf{x})}{\eta_0}p^\text{Stokes}\mathbf{I},\\
		&\widetilde{\mathbf{u}}^\text{Stratified}=\mathbf{u}^\text{Stratified}-\mathbf{u}^\text{Stratified}_\infty,\\
		& \widetilde{p}^\text{Stratified}={p}^\text{Stratified}-{p}^\text{Stratified}_\infty.\end{eqnarray}
	The force and torque generated on the particle by $\widetilde{\boldsymbol{\sigma}}^\text{Stratified}$ are given by
	\begin{equation}
		\widetilde{\mathbf{f}}^\text{Stratified}= \int_{r_p}\text{dS}\hspace{0.05in}(\mathbf{n}\cdot\widetilde{\boldsymbol{\sigma}}^\text{Stratified}), \hspace{0.2in}	\widetilde{\mathbf{q}}^\text{Stratified}= \int_{r_p}\text{dS}\hspace{0.05in}\mathbf{x}\times(\mathbf{n}\cdot\widetilde{\boldsymbol{\sigma}}^\text{Stratified}).\label{eq:StratifiedForceTorqueDirect}
	\end{equation}
	In terms of $\widetilde{\boldsymbol{\sigma}}^\text{Stratified}$, the total fluid stress (equation \eqref{eq:TotalStressDecomposed}) is,
	\begin{equation}
		\boldsymbol{\sigma}=\boldsymbol{\sigma}^\text{Stokes}+2\beta \eta'\mathbf{e}^\text{Stokes}+\boldsymbol{\sigma}^\text{Stratified}_\infty+\widetilde{\boldsymbol{\sigma}}^\text{Stratified}.\label{eq:TotalStressDecomposed2}
	\end{equation}
	We can analytically obtain the force and torque due to the first two terms once the viscosity profile is known. For the force and torque arising from remaining components we use the regular perturbation in $\beta$ followed by a generalized reciprocal theorem.
	
	{Performing a regular perturbation in $\beta$, we may express, \begin{eqnarray}
			\widetilde{\mathbf{u}}^\text{Stratified}=(\widetilde{\mathbf{u}}^\text{Stratified})^{(0)}+\beta(\widetilde{\mathbf{u}}^\text{Stratified})^{(1)}+\mathcal{O}(\beta^2),\\
			\widetilde{p}^\text{Stratified}=(\widetilde{p}^\text{Stratified})^{(0)}+\beta(\widetilde{p}^\text{Stratified})^{(1)}+\mathcal{O}(\beta^2).
		\end{eqnarray}
		From the stratified momentum equation \eqref{eq:StratificationProblemMod} at the leading order in $\beta$, $(\widetilde{\mathbf{u}}^\text{Stratified})^{(0)}=0$ and $(\widetilde{p}^\text{Stratified})^{(0)}=0$. Thus, 
		\begin{align}\begin{split}
				\widetilde{\boldsymbol{\sigma}}^\text{Stratified}&=\beta (\widetilde{\boldsymbol{\sigma}}^\text{Stratified})^{(1)}+\mathcal{O}(\beta^2)\\&= \beta\big[-(\widetilde{p}^\text{Stratified})^{(1)}\mathbf{I}+\eta_0[(\nabla\widetilde{\mathbf{u}}^\text{Stratified})^{(1)}+((\nabla\widetilde{\mathbf{u}}^\text{Stratified})^{(1)})^T]\big]+\mathcal{O}(\beta^2),
			\end{split}
		\end{align}
		leading to 
		\begin{equation}
			\nabla\cdot (\widetilde{\mathbf{u}}^\text{Stratified})^{(1)}=0, \hspace{0.2in}\nabla\cdot(\widetilde{\boldsymbol{\sigma}}^\text{Stratified})^{(1)}+\nabla\cdot[\eta'(\mathbf{e}^\text{Stokes}-\mathbf{E}_\infty)]=0.\label{eq:StratificationProblemMod2}
	\end{equation} }
	
	Let there be an auxiliary Stokes problems defined around the particle under consideration, such that
	\begin{equation}
		\nabla \cdot\mathbf{B}=0, \nabla \cdot\mathbf{b}=0, B_{ijk}=-\delta_{ij}q_k+\frac{\partial b_{jk}}{\partial x_i}+\frac{\partial b_{ik}}{\partial x_j},\label{eq:AuxEqn}
	\end{equation}
	with boundary conditions,
	\begin{equation}
		b_{ij}={b}^\mathbf{f}_{ij} \text{ or } {b}^\mathbf{q}_{ij}, \text{on the particle surface, and } b_{ij} =0, \text{as } |\mathbf{x}|\rightarrow \mathbf{x}_\text{out},\label{eq:AuxBoundary}
	\end{equation}
	where ${b}^\mathbf{f}_{ij}={\delta}_{ij}$ and ${b}^\mathbf{q}_{ij}=\epsilon_{ijk}r_k$ are 2-tensors. 
	The physical relevance of this auxiliary velocity field is as follows. For the problem when $b_{ij}={b}^\mathbf{f}_{ij}$, a vector $b_{ij}\cdot U_j$ is the Stokes velocity field around the particle translating with $U_j$ in a quiescent fluid. Similarly, for the problem with $b_{ij}={b}^\mathbf{q}_{ij}$ as particle surface condition, a vector $b_{ij}\cdot \omega_j$ corresponds to the velocity disturbance created by a particle rotating with angular velocity $-\omega_j$ in a quiescent fluid (i.e. a surface velocity $\boldsymbol{\omega}\times \mathbf{x}$). Using $\nabla \cdot \mathbf B=0$ and from the symmetry of $B_{lki}$ and {$\widetilde{\sigma}^\text{Stratified}_{lk}$} about the $l$ and $k$ indices we obtain,
	\begin{equation}
		(\widetilde{\sigma}^\text{Stratified}_{lk})^{(1)}\frac{\partial b_{ki}}{\partial r_l}=\frac{\partial B_{lki} (\widetilde{u}^\text{Stratified}_k)^{(1)}}{\partial r_l},\label{eq:AuxBoundary2}
	\end{equation}
	which, using the chain rule and writing the volume integral in the fluid domain, bounded by the particle surface on the inside and the outer boundary at $\mathbf{x}_\text{out}$, leads to
	\begin{equation}{\int_\text{Fluid} \text{dV}\hspace{0.1in}\frac{\partial}{\partial r_l}[(\widetilde{\sigma}^\text{Stratified}_{lk})^{(1)}b_{ki}-B_{lki}(\widetilde{u}^\text{Stratified}_k)^{(1)}]=\int_\text{Fluid} \text{dV}\hspace{0.1in}b_{ki}\frac{\partial(\widetilde{\sigma}_{lk}^\text{Stratified})^{(1)}}{\partial r_l}.}\label{eq:ReciprocalDerivation}\end{equation}
	Using the divergence theorem, the left side of the above equation can be written as the sum of two surface integrals,
	$\int_{|\mathbf{x}|\rightarrow|\mathbf{x}_\text{out}|} \text{dS}\hspace{0.1in}n_l[(\widetilde{\sigma}^\text{Stratified}_{lk})^{(1)}b_{ki}-B_{lki}(\widetilde{u}^\text{Stratified}_k)^{(1)}]-
	\int_{\mathbf{x}_\text{p}} \text{dS}\hspace{0.1in}n_l[(\widetilde{\sigma}^\text{Stratified}_{lk})^{(1)}b_{ki}-B_{lki}(\widetilde{u}^\text{Stratified}_k)^{(1)}],$
	where the surface normal $n_l$ points into the fluid (away from the particle) on the particle surface and outwards on the outer boundary.
	A particle that exerts a force (or force dipole) on the fluid produces a velocity which decays as $1/r$ (or $1/{r}^2$) in the far field, i.e., at $\mathbf{x}_\text{out}$. Hence, the velocities {$\widetilde{\mathbf{u}}^\text{Stratified}$} and $\mathbf{b}$ scale as $1/r$ and $1/{r}^2$, respectively, and the stresses {$\widetilde{\boldsymbol{\sigma}}^\text{Stratified}$} and $\mathbf{B}$ scale as $1/{r}^2$ and $1/{r}^3$. Therefore, the first surface integral (at $|\mathbf{x}|\rightarrow\infty$) vanishes (this is the reason for redefining the stratified momentum balance and associated variables). In confined domains, i.e., when $\mathbf{x}_\text{out}$ is a solid no-slip or a periodic boundary the integral is more straightforwardly shown to vanish, since $\mathbf{b}=0$ and $\widetilde{\mathbf{u}}=0$ on solid walls and the terms cancel on the periodic boundaries. Depending on whether ${b}^\mathbf{f}_{ij} \text{ or } {b}^\mathbf{q}_{ij}$ is used in the boundary condition in equation \eqref{eq:AuxBoundary}, the surface integral at the particle surface, $\beta\int_{\mathbf{x}_\text{p}} \text{dS}\hspace{0.1in}n_l(\widetilde{\sigma}^\text{Stratified}_{lk})^{(1)}b_{ki}$ is either part of force or torque. Therefore, upon using equation \eqref{eq:StratificationProblemMod2} and $(u^\text{Stratified}_k)^{(1)}=0$ on the surface we obtain
	\begin{equation}
		\int_{\mathbf{x}_\text{p}} \text{dS}\hspace{0.1in}n_l(\widetilde{\sigma}^\text{Stratified}_{lk})^{(1)}b_{ki}=2\int_\text{Fluid} \text{dV}\hspace{0.1in}b_{ki}\frac{\partial(\eta'({e}_{lk}^\text{Stokes}-{E}_{lk}))}{\partial r_l}.\label{eq:ReciprocalDerivationInter}
	\end{equation}
	Thus, the net force acting on a particle placed at the origin in a fluid with viscosity $\eta=\eta_0(t)+\beta \eta'(\mathbf{x},t)$ (such that $\eta'(0,t)=0$) is,
	\begin{equation}
		\mathbf{f}=	\eta_0\mathbf{f}^\text{Stokes}+2\beta \int_{r_p}\text{dS}\hspace{0.05in}\eta'\mathbf{n}\cdot\mathbf{e}^\text{Stokes}+\int_{r_p}\text{dS}\hspace{0.05in}\mathbf{n}\cdot\boldsymbol{\sigma}_\infty^\text{Stratified}+2\beta\int_\text{Fluid} \text{dV}\hspace{0.1in}\nabla\cdot(\eta'(\mathbf{e}^\text{Stokes}-\mathbf{E}_\infty))\cdot\mathbf{b}^\mathbf{f}{+\mathcal{O}(\beta^2)}.
	\end{equation}
	Noting that,
	$\int_{|\mathbf{x}|\rightarrow|{x}_\text{out}|}\text{dS}\hspace{0.05in}{b}_{ki}(\eta'{n}_l({e}^\text{Stokes}_{lk}-{E}_{lk}))\rightarrow0$,
	and using $\nabla\cdot\mathbf{b}=0$ (equation \eqref{eq:AuxEqn}), $\nabla\cdot(\boldsymbol{\sigma}_\infty^\text{Stratified}+\eta'\mathbf{E}_\infty)=0$ and the Gauss divergence theorem, we find the net force (and similarly torque) on the particle to be,
	\begin{align}
		\begin{split}
			&\mathbf{f}(t)=	\eta_0(t)\mathbf{f}^\text{Stokes}(t)-2\beta\int_\text{Fluid} \mathbf{dx}\hspace{0.1in}\eta'(\mathbf{x},t)(\mathbf{e}^\text{Stokes}(\mathbf{x};t)-\mathbf{E}_\infty(t)):\nabla\mathbf{b}^\mathbf{f}(\mathbf{x}){+\mathcal{O}(\beta^2)},\\
			&\mathbf{q}(t)=	\eta_0(t)\mathbf{q}^\text{Stokes}(t)-2\beta\int_\text{Fluid} \mathbf{dx}\hspace{0.1in}\eta'(\mathbf{x},t)(\mathbf{e}^\text{Stokes}(\mathbf{x};t)-\mathbf{E}_\infty(t)):\nabla\mathbf{b}^\mathbf{q}(\mathbf{x}){+\mathcal{O}(\beta^2)}.
		\end{split}
	\end{align}
	The above formulae are repeated in equation \eqref{eq:StratificationForceTorque} in the main text.
	
	\section{Further analysis of the rotation of a freely sedimenting particle}\label{sec:Rotationatsmallbeta}
	The orientation of a freely settling fore-aft and axisymmetric particle under gravity is governed by equation \eqref{eq:rotationratefull}, i.e., 
	\begin{equation}	
		\dot{\mathbf{p}}=	-\frac{\beta }{\eta_0^2}\cdot\Big[ t_3(\mathbf{p}\cdot \mathbf{d}){\mathbf{g}}-t_2 (\mathbf{p} \cdot{\mathbf{g}})\mathbf{d}\Big]\cdot(\mathbf{I}-\mathbf{pp})+\mathcal{O}(\beta^2).\label{eq:rotatateg}
	\end{equation}
	The dynamical system of the $\mathcal{O}(\beta)$ rotation of the particle given in equation \eqref{eq:rotatateg} is amenable to further analysis discussed in this section.
	
	\noindent\textbf{Case 1: Gravity and stratification are collinear, $d_g=\pm1$ }\\
	First consider the case when $\mathbf{g}=\pm||\mathbf{g}||\mathbf{d}$, i.e., gravity and stratification are collinear. Here equation \eqref{eq:rotatateg} simplifies to
	\begin{equation}
		\dot{\mathbf{p}}=	-d_g(t_3-t_2)\frac{\beta }{\eta_0^2}||\mathbf{g}||(\mathbf{p}\cdot \mathbf{d})(\mathbf{d}-(\mathbf{d}\cdot\mathbf{p})\mathbf{p}).
	\end{equation}
	The two equilibrium orientations are (a) $\mathbf{p}\parallel \mathbf{d}$ (i.e. $\mathbf{p}= \mathbf{d}$, $(\mathbf{d}\cdot\mathbf{p})\mathbf{p}=\mathbf{p}$), and, (b) $\mathbf{p}\perp \mathbf{d}$ (i.e. $\mathbf{p}\cdot \mathbf{d}=0$). The orientation space is a unit sphere, and if one of the equilibrium locations (a) or (b) is stable, the other is unstable.
	
	To analyze the stability, consider $\mathbf{p}\parallel \mathbf{d}$, i.e., $(\mathbf{d}\cdot\mathbf{p})\mathbf{p}=\mathbf{p}$, with $\mathbf{p}=\mathbf{d}+\boldsymbol{l}$ such that $|\boldsymbol{l}|\ll1$ is a small perturbation. The perturbation dynamics are given by
	\begin{equation}
		\dot{\boldsymbol{l}}= \frac{\beta }{\eta_0^2}||\mathbf{g}||d_g(t_3-t_2) (\boldsymbol{l}+ (\mathbf{d}\cdot\boldsymbol{l})\mathbf{d}).\label{eq:ColinearStability}
	\end{equation}
	Hence, the equilibrium point $\mathbf{p}=\mathbf{d}$ is unstable, i.e., $\dot{\boldsymbol{l}}= A {\boldsymbol{l}}$, with $A>0$, if $s=d_g[t_3-t_2]>0$ and stable i.e., $\dot{\boldsymbol{l}}= - A {\boldsymbol{l}}$, if $s<0$. In other words, for viscosity increasing in the direction of gravity ($d_g>0$), the particle will approach the axis $\mathbf{p}=\mathbf{d}$ if $[t_3-t_2]<0$ and the plane $\mathbf{p}\perp\mathbf{d}$ if $[t_3-t_2]>0$. Alternatively, for viscosity decreasing in the gravity direction, the particle will settle towards the plane $\mathbf{p}\perp\mathbf{d}$ if $[t_3-t_2]<0$ and the axis $\mathbf{p}=\mathbf{d}$ if $[t_3-t_2]>0$.
	
	\noindent\textbf{Case 2: Non-collinear gravity and stratification}\\
	Richer orientation dynamics are found when $\mathbf{g}$ and $\mathbf{d}$ are not collinear, i.e. $|d_g|<1$. Expressing the rotation rate in the basis defined by $\hat{\mathbf{e}}$, $\hat{\mathbf{g}}$ and $\hat{\mathbf{g}}\times\hat{\mathbf{e}}$ (equations \eqref{eq:CoordinateSystem1} and \eqref{eq:CoordinateSystem2}) allows us to obtain the following fixed points ($\dot{\mathbf{p}}=\mathbf{0}$) of the dynamical system of equation \eqref{eq:rotatateg},
	\begin{eqnarray}
		\mathbf{p}^{(0,1)}=\pm \hat{\mathbf{g}}\times\hat{\mathbf{e}}, p_g=p_e=0,\\
		\mathbf{p}^{(0,2)}= \pm (p_g^{(0,2)}\hat{\mathbf{g}}-\sqrt{1-(p_g^{(0,2)})^2}\hat{\mathbf{e}}),\\
		\mathbf{p}^{(0,3)}=\pm( p_g^{(0,3)}\hat{\mathbf{g}}+ \sqrt{1-(p_g^{(0,3)})^2}\hat{\mathbf{e}}),
	\end{eqnarray}
	where
	\begin{align}\begin{split}
			&p_g^{(0,2)}= \sqrt{\frac{d_g^2 (t_2+t_3)+d_g \sqrt{d_g^2 (t_2+t_3)^2-4 t_3 t_2}-2 t_3}{2(t_2-t_3)}},\\
			&p_g^{(0,3)}= \sqrt{\frac{d_g^2 (t_2+t_3)-d_g \sqrt{d_g^2 (t_2+t_3)^2-4 t_3 t_2}-2 t_3}{2(t_2-t_3)}}.\label{eq:FixedPointsVortStrat}
	\end{split}\end{align}
	One of the fixed points, $\mathbf{p}^{(0,1)}$, is where the particle orients perpendicular to the GS plane. Two other branches of fixed points, $\mathbf{p}^{(0,2)}$ and , $\mathbf{p}^{(0,3)}$, are for particle orientation within the GS plane. Here, if the fixed points exist in the orientation space (i.e. $\mathbf{p}^{(0,2)}, \mathbf{p}^{(0,3)} \in \mathbb{R}$), $\mathbf{p}^{(0,2)}$ is stable and $\mathbf{p}^{(0,3)}$ is unstable within the plane. Globally, the nature of these fixed points depends on if the plane is stable or unstable. These fixed points might not exist as depending upon $t_2$, $t_3$ and $d_g$ non-real values and values greater than 1 of $p_g^{(0,2)}$ and $p_g^{(0,3)}$ are possible. In that case, the GS plane is a (stable, or unstable) limit cycle or a neutral orbit.
	
	Consider the stability of the first fixed point, $\mathbf{p}^{(0,1)}=\hat{\mathbf{g}}\times\hat{\mathbf{e}}$, through the linearization of the dynamical system of equation \eqref{eq:rotatateg} at this fixed point. The reduced dynamics close to $\mathbf{p}=\mathbf{p}^{(0,1)}$ projected in the GS plane are,
	\begin{equation}
		\dot{\widetilde{\mathbf{p}}}=\frac{\beta }{\eta_0^2}||\mathbf{g}||\begin{bmatrix}
			-d_g(t_3-t_2)&-\sqrt{1-d_g^2}t_3\\
			\sqrt{1-d_g^2}t_2&0
		\end{bmatrix}\widetilde{\mathbf{p}}.
	\end{equation}
	The eigenvalues of this reduced system are,
	\begin{equation}
		\gamma^{\mathbf{p}^{(0,1)}}=\frac{1}{2}\frac{\beta }{\eta_0^2}||\mathbf{g}||\Bigg( -d_g(t_3-t_2)\pm\sqrt{d_g^2(t_3+t_2)^2-4t_2t_3}\Bigg).
	\end{equation}
	When,
	\begin{equation}
		d_g^2(t_3+t_2)^2-4t_2 t_3=(d_g(t_3-t_2))^2+4(d_g^2-1)t_2t_3<0 \label{eq:SpiralingConditions}
	\end{equation}
	the fixed point $\mathbf{p}^{(0,1)}$ is a spiral (stable spiral when $d_g(t_3-t_2)>0$ and unstable if $d_g(t_3-t_2)<0$). Since, $d_g^2\le 1$, this condition is never satisfied when,
	\begin{equation}
		t_2t_3<0.
	\end{equation}
	In other words, the particle's orientation dynamics may behave in an oscillatory manner near $\mathbf{p}^{(0,1)}$ when $t_2t_3>0$. Within this regime, the spiraling/ oscillatory behavior requires,
	\begin{equation}
		|d_g|<|d_g|_\text{spiral}, |d_g|_\text{spiral}= \frac{2\sqrt{t_2t_3}}{t_3+t_2}.
	\end{equation}
	Among the cases when $\mathbf{p}^{(0,1)}$ is not a spiral, it is a saddle node for all $d_g$ when $t_2 t_3<0$. Here, if $d_g(t_3-t_2)>0$, the unstable eigenvalue of the saddle has a greater magnitude than its stable direction and vice versa for $d_g(t_3-t_2)<0$. When $t_2 t_3>0$ and $|d_g|>|d_g|_\text{spiral}$, $\mathbf{p}^{(0,1)}$ is a stable node if $d_g(t_3-t_2)>0$, and an unstable node if $d_g(t_3-t_2)<0$. When $\mathbf{p}^{(0,1)}$ is an unstable spiral or node, i.e., $d_g(t_3-t_2)<0$, the particle approaches the GS plane irrespective of the initial orientation. From equations \eqref{eq:FixedPointsVortStrat}, one can notice that no fixed points exist on the GS plane (i.e., $p_g^{0,2}$ and $p_g^{0,3}$ are not real numbers) when the spiraling conditions in equation \eqref{eq:SpiralingConditions} are satisfied. Hence, the GS plane is a limit cycle when $\mathbf{p}^{(0,1)}$ is a spiral. A stable (unstable) spiral at $\mathbf{p}^{(0,1)}$ corresponds to an unstable (stable) limit cycle at the GS plane. When $\mathbf{p}^{(0,1)}$ is an unstable node, $\mathbf{p}^{(0,2)}$ is a stable node and $\mathbf{p}^{(0,3)}$ is a saddle node. When $\mathbf{p}^{(0,1)}$ is a stable node, $\mathbf{p}^{(0,2)}$ is a saddle node and $\mathbf{p}^{(0,3)}$ is an unstable node.
	
	When the stratification direction is perpendicular to gravity, $d_g=0$, and $t_2t_3>0$, a unique rotational dynamics behavior occurs. This regime corresponds to a neutral periodic orbit in the GS plane and a center at the $\mathbf{p}^{(0,1)}$ axis which indicates that between the GS plane and the $\hat{\mathbf{g}}\times\hat{\mathbf{e}}$ axis, the orientation trajectories may be such that the particle will continue to rotate in a non-uniform periodic orbit depending on the initial orientation. This is reminiscent of the Jeffery orbits of axisymmetric particles with $||q_4/q_1||_2\le1$ in a simple shear flow of uniform viscosity fluid. We will now explore this analogy in further detail.
	
	The rotational dynamics of equation \eqref{eq:rotatateg} in the $\theta-\phi$ system ($p_1=\sin(\theta)\cos(\phi)$, $p_2=\sin(\theta)\sin(\phi)$, and $p_3=\cos(\theta)$) are expressed through the equations
	\begin{align}\begin{split}
			&\frac{d \theta}{d t}=\frac{\beta}{\eta_0^2}||\mathbf{g}|| (t_2- t_3) \cos (\phi) \sin (\theta) \cos (\theta) (\mathbf{d}\cdot\mathbf{g} \cos (\phi)+\sqrt{1-\mathbf{d}\cdot\mathbf{g}^2} \sin (\phi))\\
			&\frac{d \phi}{d t}=\frac{\beta}{\eta_0^2}||\mathbf{g}|| (-\mathbf{d}\cdot\mathbf{g} (t_2- t_3) \sin (\phi) \cos (\phi)+\sqrt{1-\mathbf{d}\cdot\mathbf{g}^2} [t_2 \cos ^2(\phi)+ t_3 \sin ^2(\phi)]).
	\end{split}\end{align}
	When $\mathbf{d}\cdot\mathbf{g}=0$, the particle dynamics can be written as,
	\begin{equation}
		\frac{d \theta}{d t}=\frac{\beta}{\eta_0^2}||\mathbf{g}|| (t_2- t_3) \cos (\phi) \sin (\theta) \cos (\theta) \sin (\phi),\hspace{0.2in}\frac{d \phi}{d t}=\frac{\beta}{\eta_0^2}||\mathbf{g}|| ( t_2 \cos ^2(\phi)+ t_3 \sin ^2(\phi)).\label{eq:JeffLikeViscGradSed}
	\end{equation}
	This equation is similar to the rotation dynamics of spheroids in simple shear flow with shear rate $\dot{\gamma}$ of a uniform viscosity fluid,
	\begin{align}\begin{split}
			&\frac{d \theta^\text{Jeffery}}{d t}=\dot{\gamma} \frac{\kappa^2-1}{\kappa^2+1}\cos (\phi^\text{Jeffery}) \sin (\theta^\text{Jeffery}) \cos (\theta^\text{Jeffery}) \sin (\phi^\text{Jeffery}),\\
			&\frac{d \phi^\text{Jeffery}}{d t}= \dot{\gamma} \Big(\frac{\kappa^2}{\kappa^2+1}\cos ^2(\phi^\text{Jeffery})+ \frac{1}{\kappa^2+1} \sin ^2(\phi^\text{Jeffery})\Big).
	\end{split}\end{align}
	The closed-form solution of these equations, i.e. the Jeffery orbits, are
	\begin{equation}
		\tan(\phi^\text{Jeffery})=\kappa\tan\Bigg(\frac{\dot{\gamma}t}{\kappa+1/\kappa}\Bigg),\hspace{0.05in}
		\tan(\theta^\text{Jeffery})=C\frac{1 }{[\kappa^2\cos^2(\phi^\text{Jeffery})+\sin^2(\phi^\text{Jeffery})]^{1/2}}.
	\end{equation}
	For our case (equation \eqref{eq:JeffLikeViscGradSed}), with $\mathbf{d}\cdot\mathbf{g}=0$, the closed form solutions are
	\begin{equation}
		\tan(\phi(t))=\sqrt{\frac{t_2}{t_3}}\tan\Bigg(\frac{\beta}{\eta_0^2}|| \mathbf{g}||\sqrt{t_2t_3}t\Bigg),\hspace{0.05in} \tan(\theta(t))
		=C \frac{1}{[ \frac{t_2}{t_3}\cos^2(\phi(t))+\sin^2(\phi(t))]^{1/2}}.
	\end{equation}
	We can observe from these solutions as well that there is a non-uniform periodic solution only when $t_2t_3>0$.
	
	\section{Effect of stratification on a sphere}\label{sec:AppendixSphereCase}
	In the results presented in section \ref{sec:SpheroidResultsFix}, the stratification-induced force and torque on the spheroids are obtained by integrating equation \eqref{eq:StratificationForceTorque} using the Stokes fields represented by the spheroidal harmonics formulation \citep{dabade2015effects,dabade2016effect}. A spheroid is a sphere in the limit of particle aspect ratio approaching one. Hence, the forces and torques on a sphere can also be obtained by considering the spheroidal particle's expression in the appropriate limit. However, the relevant expressions for the sphere can be obtained directly through the simpler formula of Stokes flow fields around a spherical particle. Not only is this an easier calculation, but it provides a source of validation for the use of the spheroidal harmonics formulation. The following particle tensors are relevant for the sphere of radius $l$,
	\begin{align}&\begin{split}
			{b}_{ik}^\mathbf{f}=\frac{3 l}{4 |\mathbf{x}|}\Big( {\delta}_{ik}+\frac{{x_ix_k}}{|\mathbf{x}|^2}\Big)-\frac{l^3}{4|\mathbf{x}|^3}\Big(-{\delta_{ik}}+3\frac{{x_i x_k}}{|\mathbf{x}|^2}\Big), 	\hspace{0.2in}{b}_{ij}^\mathbf{q}=\epsilon_{ijk}x_k\frac{l^3}{|\mathbf{x}|^3}.
	\end{split}\end{align}
	A sphere of radius, $l$, placed in a Stokes flow with a uniform imposed fluid velocity $\mathbf{u}_\text{flow}$ has the following pressure and velocity
	\begin{align}
		&{u}^\text{Stokes}_i=\Bigg[{\delta_{ij}}-\frac{3 l}{4 |\mathbf{x}|}\Big( {\delta}_{ij}+\frac{{x_ix_j}}{|\mathbf{x}|^2}\Big)+\frac{l^3}{4|\mathbf{x}|^3}\Big(-{\delta_{ij}}+3\frac{{x_ix_j}}{|\mathbf{x}|^2}\Big) \Bigg]{u}_{j,\text{flow}}, p^\text{Stokes}=p_\infty-\frac{3l}{2}\frac{{u}_{i,\text{flow}}{x}_i}{|\mathbf{x}|^3}.
	\end{align}
	The velocity and pressure field around a sphere fixed in a linear flow, $\nabla \mathbf{u}_\infty=\mathbf{E}+\boldsymbol{\Omega}$, with $ \mathbf{u} _\infty=\mathbf{x}\cdot \nabla \mathbf{u}_\infty$ is
	\begin{align}
		&{u}^\text{Stokes}_i= {u} _{i,\infty}+\bigg(-\frac{l^5}{|\mathbf{x}|^5}\delta_{ik}x_j+\frac{5}{2}\Big(\frac{l^5}{|\mathbf{x}|^7}-\frac{l^3}{|\mathbf{x}|^5}\Big)x_ix_jx_k\bigg) E_{jk}-\frac{l^3}{|\mathbf{x}|^3}x_j\Omega_{ji},\label{eq:StokesVelocity}
		\\& p^\text{Stokes}=p_\infty-\frac{5l^3}{|\mathbf{x}|^5}E_{jk}x_jx_k.\label{eq:StokesPressure}
	\end{align}
	The force on a fixed sphere due to linear stratification in a uniform flow is zero since $\mathbf{b}^\mathbf{f}\cdot(2\mathbf{e}^\text{Stokes}+\nabla p^\text{Stokes}\mathbf{x})\cdot\mathbf{d}$ is an odd function of $\mathbf{n}$. There is however a finite stratification-induced torque since $\mathbf{b}^\mathbf{q}\cdot(\mathbf{e}^\text{Stokes}+\nabla p^\text{Stokes}\mathbf{x})\cdot\mathbf{d}$ is an even function of $\mathbf{n}$. Evaluating equation \eqref{eq:StratificationForceTorque} using the Stokes fields defined in this section we find the stratification-induced torque on a sphere fixed in uniform flow with velocity, $\mathbf{u}_\text{flow}$ to be,
	\begin{equation}
		\mathbf{q}^\text{Stratified}_{\mathbf{u}_\text{flow}}=2\pi l^3 \beta \mathbf{d}\times\mathbf{u}_\text{flow}, \label{eq:SphereUniformTorque}
	\end{equation}
	and, the stratification-induced force on a sphere fixed in linear flow with velocity gradient $\boldsymbol{\Gamma}=\mathbf{E}+\boldsymbol{\Omega}$ ($\mathbf{E}$ and $\boldsymbol{\Omega}$ being the symmetric and anti-symmetric parts of $\boldsymbol{\Gamma}$) is,
	\begin{equation}
		\mathbf{f}^\text{Stratified}_{\boldsymbol{\Gamma}}= {2\pi l^3}\beta(3\mathbf{E}-\boldsymbol{\Omega}) \cdot\mathbf{d}.\label{eq:SphereForceStratLin}
	\end{equation}
	
	\section{Force and torque expressions for a fixed spheroid}\label{sec:ExpressionsExtra}
	In the main text, we use the constant viscosity and stratification-induced forces and torques on a fixed spheroid in various flows. While we discussed their trends with the particle aspect ratio, $\kappa$, these unwieldy expressions are presented in this section. We consider a prolate spheroid with an aspect ratio $\kappa={\xi_0}/{\sqrt{\xi_0^2-1}}$ and major axis length $2l=2d\xi_0$ ($d$ is the focal length). When the expressions for an oblate spheroid are not presented, these can be found from the corresponding prolate formulae through the transformation $\xi_0\rightarrow \sqrt{1-\xi_0^2}$ and $d\rightarrow-\sqrt{-1} d$. An oblate particle has aspect ratio $\kappa=\sqrt{\xi_0^2-1}/\xi_0$ and major axis length $2l=2d\xi_0$. The expressions for force and torque on a fixed spheroid in constant viscosity fluids, stratification-induced torques in uniform flow, and stratification-induced force in linear flows are presented in different subsections below. The stratification-induced torques in uniform flow are compared with those calculated by \cite{anand2024sedimentation}. 
	\subsection{Constant viscosity fluid}
	The forces and torques acting on a fixed prolate spheroid in a uniform flow and linear flow of a fluid with unit constant viscosity (the coefficients in the columns two and three of table \ref{tab:NewtonianCoeffsStratifiedForces}) are,
	\begin{eqnarray}
		\begin{split}
			&f_1= \frac{16 \pi d}{\xi_0+(3-\xi_0^2)\coth^{-1}(\xi_0)},\hspace{0.2in}f_3=\frac{8 \pi d}{-\xi_0+(1+\xi_0^2)\coth^{-1}(\xi_0)},\\
			&q_1= \frac{16 \pi d^3}{3} \frac{1-2\xi_0^2}{\xi_0-(\xi_0^2+1)\coth^{-1}(\xi_0)}, \hspace{0.1in}q_3= \frac{16 \pi d^3}{3} \frac{\xi_0^2-1}{\xi_0-(\xi_0^2-1)\coth^{-1}(\xi_0)},\hspace{0.1in}q_4= \frac{q_1}{{2\xi_0^2-1}}.
		\end{split}
	\end{eqnarray}
	For a prolate spheroid, the Bretherton ratio, $\frac{q_4}{q_1}$, appearing in the rotation rate equation \eqref{eq:RotRateNewtonian} of a particle freely suspended in a linear flow of a constant viscosity fluid is 
	\begin{equation}
		\frac{q_4}{q_1}= \frac{1}{{2\xi_0^2-1}}=\frac{\kappa^2-1}{\kappa^2+1}.
	\end{equation}
	In the limit $\kappa\rightarrow1$ or $\xi_0\rightarrow\infty$, i.e., in the limit of a sphere we obtain, the familiar expressions,
	\begin{equation}
		\lim_{\xi_0\rightarrow\infty}(f_1)=\lim_{\xi_0\rightarrow\infty}(f_3)=6\pi l ,
		\lim_{\xi_0\rightarrow\infty}(q_1)=\lim_{\xi_0\rightarrow\infty}(q_3)=8\pi l^3,
		\lim_{\xi_0\rightarrow\infty}(q_4)=0,
	\end{equation}
	where $l$ is the sphere's radius.
	\subsection{Uniform flow of stratified fluids}\label{sec:SpheroidFixUniformAppend}
	As indicated in the fourth column of table \ref{tab:NewtonianCoeffsStratifiedForces}, no additional force is induced by stratification on a spheroid in uniform flow (or on a sedimenting particle). However, stratification does lead to an extra torque and the coefficients for a prolate spheroid are,
	\begin{eqnarray}\begin{split}
			&Q_{32}^{U_1}=\frac{16\pi d^3}{3}\frac{\xi_0^2-1}{-\xi_0+(\xi_0^2-3)\coth^{-1}(\xi_0)},\\
			&Q_{23}^{U_1}=\frac{8\pi d^3}{3}\Big(\frac{1}{\xi_0+(3-\xi_0^2)\coth^{-1}(\xi_0)}-\frac{\xi_0^2}{\xi_0-(1+\xi_0^2)\coth^{-1}(\xi_0)}\Big),\\
			&Q_{12}^{U_3}= \frac{4\pi d^3}{3}\frac{-5 \xi_0^3+(5 \xi_0^4-2 \xi_0^2-3) \coth ^{-1}(\xi_0)+3 \xi_0}{(\xi_0-(\xi_0^2+1) \coth ^{-1}(\xi_0))^2}.
		\end{split}\label{eq:Q32Q12}
	\end{eqnarray}
	Based on the decomposition of stratification-induced torque introduced in equations \eqref{eq:ForceTorqueA} and \eqref{eq:ForceTorqueB}, the torques on a prolate spheroid due to $(\eta'/\eta_0)\boldsymbol{\sigma}^\text{Stokes}$ are,
	\begin{equation}
		{Q_{23}^{U_1}}_\text{A}=\frac{16\pi d^3}{3}\frac{\xi_0^2}{\xi_0+(3-\xi_0^2)\coth^{-1}(\xi_0)},{Q_{12}^{U_3}}_\text{A}= \frac{8\pi d^3}{3}\frac{1- \xi_0^2}{\xi_0-(\xi_0^2+1) \coth ^{-1}(\xi_0)}. \label{eq:DecompAB}
	\end{equation}
	The other components due to $\boldsymbol{\sigma}^\text{Stratified}$ are obtained as ${Q_{23}^{U_1}}_\text{B}={Q_{23}^{U_1}}-{Q_{23}^{U_1}}_\text{A}$ and ${Q_{12}^{U_3}}_\text{B}= {Q_{12}^{U_3}}- {Q_{12}^{U_3}}_\text{A}$ from equations \eqref{eq:Q32Q12} and \eqref{eq:DecompAB}.
	\begin{figure}
		\centering
		\subfloat{\includegraphics[width=0.5\textwidth]{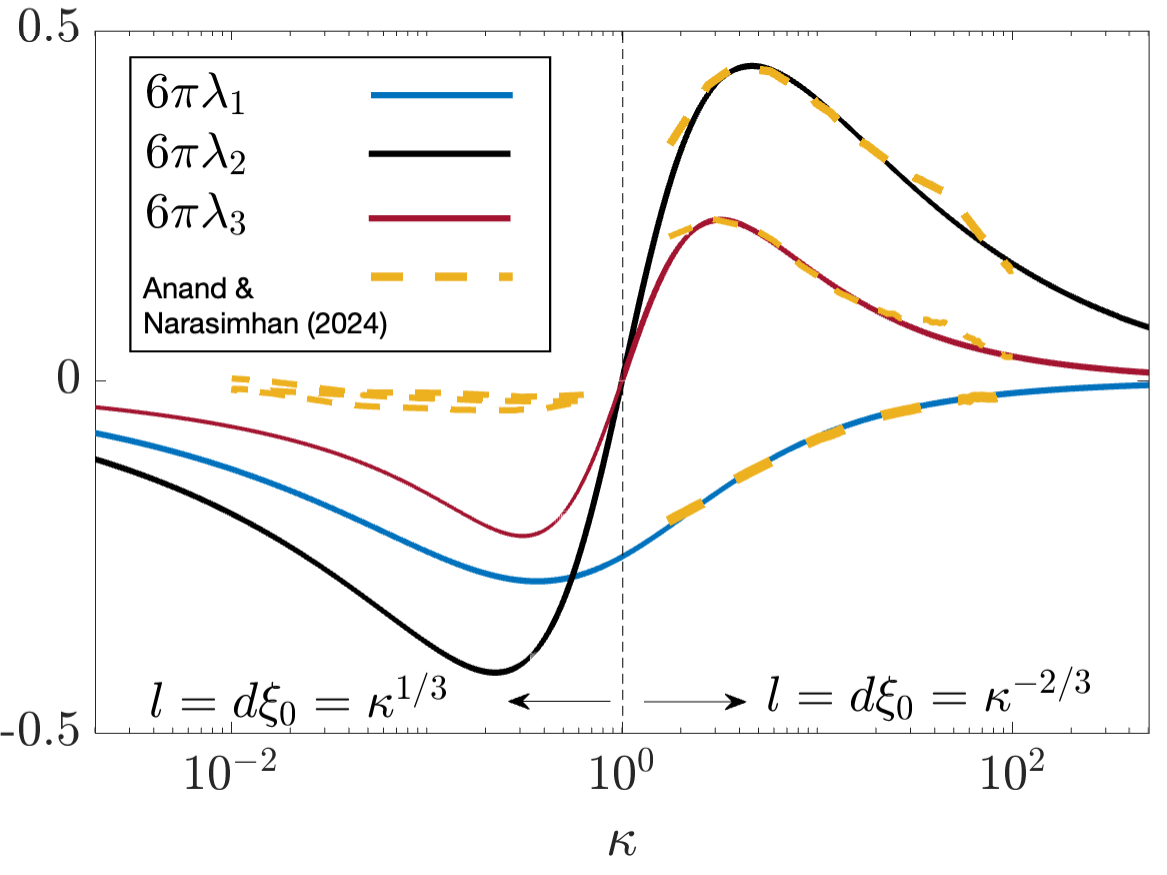}}
		\caption {Comparison of $\kappa$ variation of parameters in the angular velocity equation \eqref{eq:angvelgPurdue} with that from \cite{anand2024sedimentation}. The major axis of the spheroid, $l=d\xi_0$, with aspect ratio $\kappa$ is chosen to be $\kappa^{-2/3}$ for prolate and $\kappa^{1/3}$ for oblate spheroids, conforming to the length scale used in the comparison paper.\label{fig:ComparePurdue}}
	\end{figure}
	The sedimenting particle's rotation at $\mathcal{O}(\beta)$ (equation \eqref{eq:rotatateg}) depends upon $t_i,i\in[1,3]$ (shown in equation \eqref{eq:t1t2t3}). For a prolate spheroid,
	\begin{align}\begin{split}
			&t_1=-\frac{\xi_0^4}{16 d \pi }(\xi_0+\coth^{-1}(\xi_0)(1-\xi_0^2)),\\
			&t_2= \frac{\xi_0^4}{32 d \pi(2\xi_0^2-1)}(3 \xi_0-5 \xi_0^3+(5 \xi_0^4-2 \xi_0^2-3) \coth ^{-1}(\xi_0)),\\
			&t_3=\frac{\xi_0^4}{32 d\pi(2\xi_0^2-1)}(\xi_0^3-\xi_0+(-\xi_0^4+4 \xi_0^2+1) \coth ^{-1}(\xi_0)),\label{eq:t1t2t3Prolate}
	\end{split}\end{align}
	and for an oblate spheroid,
	\begin{align}\begin{split}
			&t_1=\frac{(\xi_0^2-1)^2}{16d\pi} (\sqrt{\xi_0^2-1}-\xi_0^2 \cot ^{-1}(\sqrt{\xi_0^2-1})),\\
			&t_2=\frac{(\xi_0^2-1)^2} {32 d\pi (2 \xi_0^2-1)}((5 \xi_0^2-2) \sqrt{\xi_0^2-1}+(8-5 \xi_0^2) \xi_0^2 \cot ^{-1}(\sqrt{\xi_0^2-1})),\\
			&t_3=\frac{(\xi_0^2-1)^2}{32 d\pi (2 \xi_0^2-1)} ((\xi_0^4+2 \xi_0^2-4) \cot ^{-1}(\sqrt{\xi_0^2-1})-\xi_0^2 \sqrt{\xi_0^2-1}).\label{eq:t1t2t3Oblate}
	\end{split}\end{align}
	
	\subsubsection{Comparison with \cite{anand2024sedimentation}}
	As mentioned in section \ref{sec:Intro}, \cite{anand2024sedimentation} recently studied the sedimentation of spheroids in fluids with linearly varying viscosity. The angular velocity from equation \eqref{eq:angvelg} can also be expressed as,
	\begin{equation}
		\boldsymbol{\omega}_\text{particle} =\frac{\beta}{\eta_0^2}[ \lambda_1\mathbf{d}\times\mathbf{g}+\lambda_2(\mathbf{p}\cdot\mathbf{d})\mathbf{g}\times\mathbf{p}+\lambda_3(\mathbf{p}\cdot\mathbf{g})\mathbf{d}\times\mathbf{p}].\label{eq:angvelgPurdue}
	\end{equation}
	In the \cite{anand2024sedimentation} study, the parameters $\lambda_i, i\in[1,3]$ are evaluated through the numerical integration of the volume integral obtained via a reciprocal theorem (equation \eqref{eq:StratificationForceTorque} in the present study and equation 2.7 of \cite{anand2024sedimentation}). Since equations \eqref{eq:angvelg} and \eqref{eq:angvelgPurdue} are equivalent, these parameters are related to $t_i, i\in[1,3]$ defined earlier in equations \eqref{eq:t1t2t3Prolate} and \eqref{eq:t1t2t3Oblate} as
	\begin{equation}
		\lambda_1=t_1, \lambda_2=t_1+t_3, \text{ and }\lambda_3=-t_1-t_2.\label{eq:EquivalencewithPurdue}
	\end{equation}
	A comparison of the analytically obtained $\lambda_i$ values and that through graph digitization of the data presented in figure 6 of \cite{anand2024sedimentation} is shown in figure \ref{fig:ComparePurdue}. The $\lambda_1$ values for prolate spheroids match for the two studies except at the smallest $\kappa$ presented in the comparison work. The parameters, $\lambda_2$ and $\lambda_3$ for prolate spheroids match at moderate $3 \lessapprox \kappa \lessapprox 25$ but values from our analytical expressions deviate from the data of \cite{anand2024sedimentation} outside this range. Some of the results presented in discussions within section \ref{sec:SpheroidsSedimenting} are qualitatively consistent between two studies because the signs of the parameters $\lambda_i, i\in[1,3]$ for a given $\kappa$ (or equivalently $t_i, i\in[1,3]$) are the same in both studies. The magnitude of $\lambda_i, i\in[1,3]$ are much larger in our study and show a qualitatively different variation with $\kappa$ for oblate spheroids. Since the scaling used in figure \ref{fig:ComparePurdue} is such that each spheroid has the same volume as a sphere of radius 1, i.e., $4\pi/3$, the values of $\lambda_i, i\in[1,3]$ for a spherical particle must be the same whether this limit, $\kappa\rightarrow 1$, is approached from the right (prolate) or left (oblate). Unlike the data from \cite{anand2024sedimentation}, our expressions satisfy this requirement, as shown in figure \ref{fig:ComparePurdue}. The violation of this requirement can also be observed by comparing the two plots in figure 6 of the comparison paper \citep{anand2024sedimentation}.
	
	\subsection{Linear flows}\label{sec:LinearFlowsAppend}
	The presence of linear viscosity stratification leads to an extra force on an axisymmetric and fore-aft symmetric particle. In the particle reference frame, this stratification-induced force is $\mathbf{F}_\text{strat}^\text{body}\cdot\mathbf{d}^\text{body}$, where the non-zero components of $\mathbf{F}_\text{strat}^\text{body}$ for a fore-aft and axisymmetric particle fixed in the various linear flows defined in \eqref{eq:BasisFlows} are listed in the fourth column of table \ref{tab:NewtonianCoeffsStratifiedForces}. For a prolate spheroid, the different non-zero components of $\mathbf{F}_\text{strat}^\text{body}$ are
	\begin{align}\begin{split}
			&F_{11}^{\Gamma_1}=-\frac{8 \pi d^3 (\xi_0^2-1) (-3 \xi_0^3+(3 \xi_0^4-6 \xi_0^2+11) \coth ^{-1}(\xi_0)+5 \xi_0)}{3 ((\xi_0^2-3) \coth ^{-1}(\xi_0)-\xi_0) (-3 \xi_0^3+3 (\xi_0^2-1)^2 \coth ^{-1}(\xi_0)+5 \xi_0)},\\
			&F_{11}^{\Gamma_2}=\frac{4 \pi d^3 (-3 \xi_0^3+3 (\xi_0^2-1)^2 \coth ^{-1}(\xi_0)+5 \xi_0)}{3 ((\xi_0^2-3) \coth ^{-1}(\xi_0)-\xi_0) ((3 \xi_0^2-1) \coth ^{-1}(\xi_0)-3 \xi_0)},\\
			&F_{33}^{\Gamma_2}=\frac{4}{3} \pi d^3 (\frac{\xi_0^2}{(\xi_0^2+1) \coth ^{-1}(\xi_0)-\xi_0}+\frac{1}{(3 \xi_0^2-1) \coth ^{-1}(\xi_0)-3 \xi_0}),\\
			&F_{13}^{\Gamma_4}=\frac{8 \pi d^3 \Bigg(\begin{matrix}
					-3 \xi_0^5+5 \xi_0^3+\coth ^{-1}(\xi_0) (6 \xi_0^6-8 \xi_0^4-(\xi_0-1) (\xi_0+1) \\(3 \xi_0^4+1) \xi_0 \coth ^{-1}(\xi_0)+4 \xi_0^2+2)-2 \xi_0
				\end{matrix}\Bigg)}{ \Bigg(\begin{matrix}(3(\xi_0^2-3) \coth ^{-1}(\xi_0)-\xi_0)((\xi_0^2+1) \coth ^{-1}(\xi_0)-\xi_0)\\ (-3 \xi_0^2+3 (\xi_0^2-1) \xi_0 \coth ^{-1}(\xi_0)+2) \end{matrix}\Bigg)},\\
			&F_{31}^{\Gamma_4}=\frac{ 4 \pi d^3\Bigg(\begin{matrix} (\xi_0 (\xi_0^2-3) (3 \xi_0^2-2)+(\xi_0^2-1) \coth ^{-1}(\xi_0) \\(-6 \xi_0^4+14 \xi_0^2+(3 \xi_0^4-6 \xi_0^2-5) \xi_0 \coth ^{-1}(\xi_0)+6))\end{matrix}\Bigg)} {3 (-3 \xi_0^2+3 (\xi_0^2-1) \xi_0 \coth ^{-1}(\xi_0)+2) (\xi_0-(\xi_0^2+1) \coth ^{-1}(\xi_0))^2},\\
			&F_{12}^{\Gamma_6}=\frac{16 \pi d^3 (\xi_0^2-1)}{3 (\xi_0^2-3) \coth ^{-1}(\xi_0)-3 \xi_0},\\
			&F_{31}^{\Gamma_7}=\frac{4 \pi d^3 (\xi_0 (5 \xi_0^2-3)+(-5 \xi_0^4+2 \xi_0^2+3) \coth ^{-1}(\xi_0))}{3 (\xi_0-(\xi_0^2+1) \coth ^{-1}(\xi_0))^2},\\
			&F_{23}^{\Gamma_8}=\frac{8}{3} \pi d^3 (\frac{\xi_0^2}{\xi_0^2 (-\coth ^{-1}(\xi_0))+\xi_0-\coth ^{-1}(\xi_0)}+\frac{1}{(\xi_0^2-3) \coth ^{-1}(\xi_0)-\xi_0}).
		\end{split} \label{eq:StratLin}
	\end{align}
	The parameters used in the $\mathcal{O}(\beta)$ translation velocity (equation \eqref{eq:TranslCoeffs}) for a prolate spheroid are
	\begin{eqnarray}\begin{split}
			&m_1=\frac{d^2 (\xi_0^2-1) (-3 \xi_0^3+(3 \xi_0^4-6 \xi_0^2+11) \coth ^{-1}(\xi_0)+5 \xi_0)}{6 (-3 \xi_0^3+3 (\xi_0^2-1)^2 \coth ^{-1}(\xi_0)+5 \xi_0)},\\
			&m_2=\frac{d^2 ((\xi_0^2-1) \coth ^{-1}(\xi_0) ((3 \xi_0^4+6 \xi_0^2-1) \coth ^{-1}(\xi_0)-2 \xi_0 (3 \xi_0^2+2))+(3 \xi_0^2-5) \xi_0^2)}{6 (-3 \xi_0^3+3 (\xi_0^2-1)^2 \coth ^{-1}(\xi_0)+5 \xi_0) ((3 \xi_0^2-1) \coth ^{-1}(\xi_0)-3 \xi_0)},\\
			&m_3=-\frac{\begin{matrix}d^2 (\xi_0 (3 \xi_0^4-8 \xi_0^2+4)+\xi_0 (3 \xi_0^6-30 \xi_0^4+37 \xi_0^2-10) \coth ^{-1}(\xi_0)^2\\+(-6 \xi_0^6+38 \xi_0^4-34 \xi_0^2+4) \coth ^{-1}(\xi_0))\end{matrix}}{3 (2 \xi_0^2-1) (-3 \xi_0^2+3 (\xi_0^2-1) \xi_0 \coth ^{-1}(\xi_0)+2) ((3 \xi_0^2-1) \coth ^{-1}(\xi_0)-3 \xi_0)},\\
			&m_4=\frac{d^2 \begin{matrix}(9 \xi_0^7-30 \xi_0^5+31 \xi_0^3+(\xi_0^2-1)^2 \coth ^{-1}(\xi_0) \\(-18 \xi_0^4+90 \xi_0^2+(9 \xi_0^5-78 \xi_0^3+45 \xi_0) \coth ^{-1}(\xi_0)-22)-10 \xi_0)\end{matrix}}{6 (2 \xi_0^2-1) (-3 \xi_0^2+3 (\xi_0^2-1) \xi_0 \coth ^{-1}(\xi_0)+2) (-3 \xi_0^3+3 (\xi_0^2-1)^2 \coth ^{-1}(\xi_0)+5 \xi_0)}.
	\end{split}\end{eqnarray}
	For a sphere , $\lim\limits_{\kappa\rightarrow1}(m_1)=l^2,\lim\limits_{\kappa\rightarrow1}(m_2)=\lim\limits_{\kappa\rightarrow1}(m_3)=\lim\limits_{\kappa\rightarrow1}(m_4)=0.$
\end{document}